\documentclass[fleqn]{lmcs}
\pdfoutput=1
\usepackage[utf8]{inputenc}

\usepackage{lastpage}
\lmcsdoi{22}{1}{20}
\lmcsheading{}{\pageref{LastPage}}{}{}%
{Sep.~13,~2023}{Mar.~06,~2026}{}

\usepackage[british]{babel}
\usepackage{graphicx,amsfonts,amssymb,microtype,mathdefs,array,mathrsfs,amsthm,relsize}
\usepackage[section]{thmdefs2}
\usepackage[hidelinks]{hyperref}

\let\sc\scshape

\makeatletter
\newcommand\m@thsm@ller[2]{\mbox{\relscale{0.91}$\m@th#1#2$}}
\let\smaller\undefined
\DeclareRobustCommand\smaller[1]{\relax\ifmmode{\mathpalette\m@thsm@ller{#1}}\else{\relscale{0.91}#1}\fi}
\makeatother

\newlist{enuma}{enumerate}{10}
\setlist[enuma]{label={\normalfont(\alph*)}}
\newlist{enumr}{enumerate}{10}
\setlist[enumr]{label={\normalfont(\roman*)}}
\newlist{enumn}{enumerate}{10}
\setlist[enumn]{label={\normalfont(\arabic*)}}

\DeclareRobustCommand*{\dom}{\qopname\relax o{dom}}
\DeclareRobustCommand*{\rng}{\qopname\relax o{rng}}

\newcommand*{\id}{\mathrm{id}}
\newcommand*{\suc}{\mathrm{suc}}
\newcommand*{\val}{\mathrm{val}}
\newcommand*{\term}{\mathrm{term}}
\newcommand*{\Flat}{\mathrm{flat}}
\newcommand*{\sing}{\mathrm{sing}}
\newcommand*{\dist}{\mathrm{dist}}
\newcommand*{\union}{\mathrm{union}}
\newcommand*{\pt}{\mathrm{pt}}
\newcommand*{\fin}{\mathrm{fin}}
\newcommand*{\wilke}{\mathrm{wilke}}
\newcommand*{\thin}{\mathrm{thin}}
\newcommand*{\reg}{\mathrm{reg}}
\newcommand*{\TA}{\mathrm{TA}}
\newcommand*{\op}{\mathrm{op}}
\newcommand*{\MSO}{\smaller{\mathrm{MSO}}}
\newcommand*{\FO}{\smaller{\mathrm{FO}}}
\newcommand*{\Id}{\mathrm{Id}}
\newcommand*{\one}{\mathsf{1}}
\newcommand*{\emptyseq}{\smaller{\langle\rangle}}
\newcommand*{\?}{\kern .08em}
\DeclareRobustCommand*{\Belowseg}{\mathord\Downarrow}
\DeclareRobustCommand*{\Aboveseg}{\mathord\Uparrow}
\newcommand*{\mVert}{\mathrel\Vert}
\newcommand*{\llangle}{\langle\!\langle}
\newcommand*{\rrangle}{\rangle\!\rangle}

\newcommand\upqed{\vskip-\baselineskip\vskip-\belowdisplayskip}

\begin{document}
\title{The Expansion Problem for Infinite Trees}
\author{Achim Blumensath\lmcsorcid{0009-0006-6315-1019}}
\address{Masaryk University Brno}
\email{blumens@fi.muni.cz}

\begin{abstract}
We study Ramsey-like theorems for infinite trees and similar combinatorial tools.
As an application we consider the expansion problem for tree algebras.
\end{abstract}

\maketitle

\section{Introduction}   

While the theory of languages of infinite trees is well-established by now,
it is far less developed than other formal language theories.
Further progress in this direction is currently hampered by our lack of understanding
of the combinatorial properties of infinite trees. In particular, for many purposes
the currently known Ramsey type theorems for trees are simply not strong enough.
What would be needed instead are, for instance, analogues of Simon factorisation trees for
infinite trees.
Such Ramsey arguments are ubiquitous in the study of languages of infinite objects.
For instance, in the theory of $\omega$-words they appear
in the original complementation proof for B\"uchi automata (for a modern account see,
e.g.,~\cite{Thomas90}), when expanding a Wilke algebra to an $\omega$-semigroup
(see, e.g.,~\cite{PerrinPin04}), or in the more recent work on distance automata
and boundedness problems (see, e.g.,~\cite{Simon90,Simon94,AbdullaKY08,Colcombet21}).

There are at least two persistent problems when trying to extend our repertoire
of combinatorial tools to infinite trees.
The first one concerns the step from arbitrary trees to regular ones\?:
while many arguments only work if the considered trees are regular,
the only known method of reducing a given tree to an equivalent regular one
is based on automata, which are not applicable in all contexts.
For instance, in~\cite{BojanczykId09} all proofs work exclusively with regular trees,
and only at the very end the authors transfer their results to arbitrary trees
(which was possible in this particular case since the languages under consideration were regular
and therefore uniquely determined by which regular trees they contain).

The second problem concerns trees that are highly-branching.
Many of the known tools from the theory of $\omega$-semigroups can be generalised
to trees that are thin, i.e., that have only countably many infinite branches.
But all attempts to extend them to (at least some) non-thin trees have failed so far.
For example, in~\cite{BojanczykIdSk13} the authors only consider languages of
thin trees since their methods do not apply to non-thin ones.
Later in~\cite{BilkowskiSkrzypczak13}, they then successfully adapted their approach
to study unambiguous languages of non-thin trees,
utilising the fact that trees in unambiguous languages are in a certain sense governed by
their thin prefixes.

The only combinatorial methods known so far that work well even in light of the above issues
are those based on automata and games since, in a certain sense, games provide a way
to reduce a problem concerning the whole tree into one only involving a single branch.
(In fact, one of the motivations for this paper stems from a wish to deeper understand
how exactly this is achieved, in particular during the translation of a formula into an
automaton. Theorem~\ref{Thm: reduction to Txthin} below might be considered to give a partial
answer.)
Unfortunately, there are many questions that resist to being phrased in
automata-theoretic or game-theoretic terms.

In this article we start by quickly reviewing the existing techniques to study combinatorial
properties of infinite trees, presenting them in the unifying language of
tree algebras from~\cite{Blumensath20}. We then take a look at several new approaches.
We determine how far they carry and what the problems are that prevent us from continuing.
Our contributions are mainly conceptual. We raise many open questions, but provide few answers.
None of the results below are very deep and several remain partial.
The main purpose of the article is to draw attention to a problem I~consider central
for further progress.

It seems that such progress will likely not come from abstract considerations but by
working on concrete problems. The recently developed algebraic approaches
to languages of infinite trees \cite{BojanczykId09,Blumensath11c,Blumensath13a,BojanczykIdSk13,
BilkowskiSkrzypczak13,BojanczykKlin18,BojanczykCaPlSk19,Blumensath20,Blumensath21,Blumensath23}
seems to provide many opportunities to test our combinatorial tools.
Our focus will therefore be on a particular application from this domain,
one that we call the \emph{Expansion Problem\?:} the problem of whether a given algebra
whose product is defined only for some trees can be expanded to one whose product is
defined everywhere, analogously to the expansion of a Wilke algebra (where the product
is defined only for ultimately periodic words) to an $\omega$-semigroup (where we can
multiply arbitrary $\omega$-words).
This problem turns out to be a good test bed for the various approaches we consider.
We solve it in some special cases, but none of our approaches is strong enough
to solve the general case.

The overview of the article is as follows.
We start in Section~\ref{Sect:tree algebras} with setting up the algebraic framework
we will be working in.
Section~\ref{Sect:Ramsey} contains a brief survey of the existing Ramsey Theorems for trees.
The Expansion Problem is defined in Section~\ref{Sect:dense}, where we also recall
some tools from~\cite{Blumensath21} to prove uniqueness of expansions.
The main technical part of the article are Sections
\ref{Sect:evaluations}~and~\ref{Sect:labellings}, which contain two tools
to study expansions. The first one are so-called \emph{evaluations,} which are a weak
form of a Simon tree, the second one are \emph{consistent labellings,} which are somewhat
similar to automata.
The final two sections (\ref{Sect:unambiguous}~and~\ref{Sect:branch-continuous})
contain two applications. The first one recalls results of~\cite{BilkowskiSkrzypczak13}
about a characterisation of unambiguous languages in terms of consistent labellings,
while the second one uses consistent labellings to define classes of tree algebras with unique
expansions.

Finally, let us highlight the concrete contributes of this article. (All terminology will
be defined in the respective section below.)
\begin{itemize}
\item We streamline and generalise the definitions of two combinatorial tools from the
  literature\?: evaluations~\cite{Puppis10,CartonColcombetPuppis18} and consistent
  labellings~\cite{BilkowskiSkrzypczak13}.
\item We prove the existence of expansions for $\MSO$-definable $\bbT^\reg$-algebras in
  Theorem~\ref{Thm: unique MSO-definable expansion of Tree algebras}.
\item We prove the existence of certain evaluations in Theorems
  \ref{Thm: definable algebras have condensations}~and~\ref{Thm: reduction to Txthin}.
\item We solve the expansion problem for thin trees in Section~\ref{Sect:thin trees},
  and the expansion problems for deterministic and (partially for) branch-continuous
  tree algebras in Section~\ref{Sect:branch-continuous}.
\end{itemize}

\section{Tree algebras}   
\label{Sect:tree algebras}

We start with a brief introduction to the algebraic framework we will be working in.
A~more detailed account can be found in~\cite{Blumensath20,Blumensath21,BlumensathLN2}
(in increasing order of abstractness).
Let us fix notation and conventions. For $n < \omega$, we set $[n] := \{0,\dots,n-1\}$.
We denote tuples $\bar a = \langle a_0,\dots,a_{n-1}\rangle$ with a bar.
The empty tuple is~$\emptyseq$.
The \emph{range} of a function $f : A \to B$ is the set $\rng f := f[A]$.
We denote the disjoint union of two sets by $A + B$, and we denote the union of two functions
$f : A \to B$ and $f' : A' \to B$ by $f + f' : A + A' \to B$.

Let us quickly recall some material from the theory of $\omega$-semigroups (see, e.g.,
\cite{PerrinPin04} for an introduction). An \emph{$\omega$-semigroup} is a two-sorted structure
$\langle S_1,S_\omega\rangle$ with three products
\begin{align*}
  {}\cdot{} : S_1 \times S_1 \to S_1\,,\quad
  {}\cdot{} : S_1 \times S_\omega \to S_\omega\,,\quad
  \pi : (S_1)^\omega \to S_\omega
\end{align*}
satisfying several associative laws. A~\emph{Wilke algebra} is a two-sorted structure
$\langle S_1,S_\omega\rangle$ with two products and an $\omega$-power operation
\begin{align*}
  {}\cdot{} : S_1 \times S_1 \to S_1\,,\quad
  {}\cdot{} : S_1 \times S_\omega \to S_\omega\,,\quad
  {-}^\omega : S_1 \to S_\omega
\end{align*}
again satisfying several associative laws. The laws for the $\omega$-power are
\begin{align*}
  (ab)^\omega = a(ba)^\omega \qtextq{and} (a^n)^\omega = a^\omega,
  \quad\text{for all } a,b \in S_1 \text{ and } 0 < n < \omega\,.
\end{align*}
One can show via a Ramsey argument that every finite Wilke algebra has a unique
expansion to an $\omega$-semigroup.

To make this article accessible to a wider audience, we have tried to keep the
category-theoretic prerequisites at a minimum.
Let us briefly recall some standard notions.
A \emph{functor}~$\bbF$ is an operation that maps every sorted set~$A$ to some
sorted set~$\bbF A$, and every function $f : A \to B$ between such sets to a function
$\bbF f : \bbF A \to \bbF B$ such that $\bbF$~preserves identity maps and composition
of functions.
A \emph{natural transformation} $\tau : \bbF \Rightarrow \bbG$ between two functors
$\bbF$~and~$\bbG$ is a family $\tau = (\tau_A)_A$ (indexed by all sorted sets~$A$)
of functions $\tau_A : \bbF A \to \bbG A$ satisfying
\begin{align*}
  \tau_A \circ \bbF f = \bbG f \circ \tau_B\,,
  \quad\text{for all } f : A \to B\,.
\end{align*}
Usually, we omit the index~$A$ and simply write~$\tau$ instead of~$\tau_A$.

To model ranked trees, we work in a many-sorted setting where the sort of a tree
represents the set of \emph{variables} or \emph{holes} appearing in it.
Hence, we fix a countably infinite set~$X$ of variables and use the set
$\Xi := \PSet_\fin(X)$ of finite subsets of~$X$ as sorts.
In addition, for technical reasons we equip the labels of our trees with an ordering.
Hence, we will work with \emph{partially ordered $\Xi$-sorted sets} which are
families $A = (A_\xi)_{\xi \in \Xi}$ where each component $A_\xi$~is partially ordered.
A~function $f : A \to B$ between two such sets is then a family $f = (f_\xi)_{\xi \in \Xi}$
of monotone functions $f_\xi : A_\xi \to B_\xi$.
In the following we will for simplicity use the term \emph{sorted set} for
`partially ordered $\Xi$-sorted set' and the term \emph{function} for a function
between such sets.
Sometimes it is convenient to identify a sorted set $A = (A_\xi)_{\xi \in \Xi}$ with
its disjoint union $A = \sum_{\xi \in \Xi} A_\xi$.
Then a function $f : A \to B$ corresponds to a sort-preserving and order-preserving
function between the corresponding disjoint unions.

Given a sorted set~$A$ an \emph{$A$-labelled tree} is a possibly infinite tree~$t$
where the vertices are labelled by elements of~$A$ and the edges by variables from~$X$
in such a way that a vertex with a label $a \in A_\xi$ of sort~$\xi$ has exactly one outgoing
edge labelled by~$x$, for every $x \in \xi$ (and no other edges).
We identify such a tree with a function $t : \dom(t) \to A$, where $\dom(t)$~is the set of
vertices of~$t$. (We consider $\dom(t)$ to be a sorted set where $v \in \dom(t)$ has the same
sort as its label $t(v)$.) As usual, we identify the vertices of~$t$ with finite sequences of
directions. Since, in our case, we can take the variables for directions,
this turns $\dom(t)$ into a prefix-closed subset of~$X^*$.
Using this identification, we can write the root of~$t$ as~$\emptyseq$.
If there is an $x$-labelled edge from a vertex~$u$ to~$v$, we call~$v$ the \emph{$x$-successor}
of~$u$. We denote it by $\suc_x(v)$.
A~\emph{branch} of~$t$ is a maximal path starting at the root.
\pagebreak[5]
\begin{Def}
Let $A$~be a sorted set.

    \begin{enuma}
\item We set $\bbT^\times A := (\bbT^\times_\xi A)_{\xi \in \Xi}$ where
$\bbT^\times_\xi A$~denotes the set of all $(A + \xi)$-labelled trees~$t$ (where the elements
in~$\xi$ are assumed to have sort~$\emptyset$) satisfying the following conditions.
\begin{itemize}
\item Every variable $x \in \xi$ appears at least once in~$t$.
\item The root of~$t$ is not labelled by a variable.
\end{itemize}
We order the elements of $\bbT^\times A$ by
\begin{align*}
  s \leq t \quad\defiff\quad
  \dom(s) = \dom(t) \qtextq{and} s(v) \leq t(v) \text{ for all vertices } v\,.
\end{align*}

\item For a function $f : A \to B$, we denote by $\bbT^\times f : \bbT^\times A \to \bbT^\times B$
the function applying~$f$ to every label of the given tree (leaving the variables unchanged).

\item For $t \in \bbT^\times_\xi A$, we denote by $\dom_0(t) \subseteq \dom(t)$ the set of all
vertices that are not labelled by a variable. \qedhere
\end{enuma}
\end{Def}

We need the following two operations on trees.
\begin{Def}
Let $A$~be a sorted set.

    \begin{enuma}
\item The \emph{singleton operation} $\sing : A \to \bbT^\times A$
maps every letter $a \in A_\xi$, to the tree $\sing(a)$ consisting of
the root with label~$a$ attached to which is one leaf with label~$x$, for every
$x \in \xi$.

\item The \emph{flattening operation} $\Flat : \bbT^\times\bbT^\times A \to \bbT^\times A$
is a generalisation of term substitution.
It takes a tree~$t$ labelled by trees $t(v) \in \bbT^\times A$ and combines them into
a single tree as follows (see Figure~\ref{Fig:flat}).
\end{enuma}
\begin{figure}
\centering
\includegraphics{Expansion-1.mps}
\caption{The flattening operation\?: $t$ and $\Flat(t)$\label{Fig:flat}}
\end{figure}
\begin{itemize}
\item We take the disjoint union of all trees $t(v)$, for $v \in \dom(t)$
  (where, if $t(v) = x \in X$ is a variable and not a tree,
   we treat $t(v)$ as a $1$-vertex tree whose root is labelled~$x$)\?;
\item from each component~$t(v)$ such that $t(v)$~is a proper tree and not just a variable,
  we delete every vertex labelled by a variable $x \in X$\?;
\item we redirect every edge of $t(v)$ leading to such a deleted vertex to the root
  of~$t(u_x)$, where $u_x$~is the $x$-successor of~$v$ in~$t$\?; and
\item we unravel the resulting graph into a tree.
  \qedhere
\end{itemize}
\end{Def}

\begin{Rem}
The triple $\langle\bbT^\times,\Flat,\sing\rangle$ forms what is called a \emph{monad}
in category-theoretical language, which means that
$\Flat : \bbT^\times\bbT^\times \Rightarrow \bbT^\times$ and
$\sing : \Id \Rightarrow \bbT^\times$ are natural transformations satisfying
the following three equations.
\begin{align*}
  \Flat \circ \sing &= \id\,, \\
  \Flat \circ \bbT^\times\sing &= \id\,, \\
  \Flat \circ \bbT^\times\Flat &= \Flat \circ \Flat\,.
\end{align*}
\upqed
\end{Rem}

\begin{Def}
Let $t \in \bbT^\times A$ be a tree.

    \begin{enuma}
\item
The \emph{tree order}~$\preceq$ is the ordering on~$\dom(t)$ defined by
\begin{align*}
  u \preceq v \quad\defiff\quad u \text{ lies on the path from the root to } v\,.
\end{align*}

\item A \emph{factorisation} of~$t$ is a tree $T \in \bbT^\times\bbT^\times A$ with $\Flat(T) = t$.
We call each tree $T(v)$, for $v \in \dom_0(T)$, a \emph{factor} of~$t$.

\item Given vertices $u$~and $\bar v = (v_x)_{x \in \xi}$ of~$t$
such that $\bar v$~forms an antichain (with respect to~$\preceq$)
and $u \prec v_x$, for all~$x$, we define
\begin{align*}
  [u,\bar v) := \set{ w \in \dom(t) }{ u \preceq w \text{ and } v_x \npreceq w\,,
                      \text{ for all } x }\,.
\end{align*}
We call~$\xi$ the \emph{sort} of $[u,\bar v)$.
\end{enuma}

We denote by $t[u,\bar v)$ the restriction of $t : \dom(t) \to A$ to the set
$[u,\bar v) \cup \bar v$ (where $v_x$~is labelled by the variable~$x$ while all other vertices
have the same label as in~$t$).
We call $t[u,\bar v)$ the \emph{factor} of~$t$ \emph{between} $u$~and~$\bar v$.
In the special case where $\xi = \emptyset$, we obtain the \emph{subtree} of~$t$ \emph{rooted}
at~$u$, which we usually denote by~$t|_u$.
\end{Def}
\begin{Rem}
    \begin{enuma}
\item When identifying the vertices of a tree with words in~$X^*$,
the tree order~$\preceq$ is just the prefix ordering.

\item A factor $t[u,\bar v)$ may contain additional variables besides those at the
vertices~$\bar v$. More precisely, we have
$t[u,\bar v) \in \bbT^\times_{\xi \cup \zeta} A$ where $\xi$~is the sort of
$[u,\bar v)$ and $\zeta$~is the set of those variables of~$t$ that appear at some
vertex $w \in [u,\bar v)$. \qedhere
\end{enuma}
\end{Rem}

There are several special classes of trees we are interested in below.
\begin{Def}
    \begin{enuma}
\item
A \emph{submonad} of~$\bbT^\times$ is a functor~$\bbT^0$ such that
\begin{itemize}
\item $\bbT^0 A \subseteq \bbT^\times A$, for every sorted set~$A$,
\item $\bbT^0 f = \bbT^\times f \restriction \bbT^0 A$, for every function $f : A \to B$, and
\item $\bbT^0 A$~is closed under $\Flat$~and~$\sing$, that is,
  \begin{alignat*}{-1}
    \Flat(t) &\in \bbT^0 A\,, &&\quad\text{for all } t \in \bbT^0\bbT^0 A\,,\\
    \sing(a) &\in \bbT^0 A\,, &&\quad\text{for all } a \in A\,.
  \end{alignat*}
\end{itemize}
We write $\bbT^0 \subseteq \bbT^\times$ to denote this fact.

\item Similarly, given two submonads $\bbT^0,\bbT^1 \subseteq \bbT^\times$, we write
$\bbT^0 \subseteq \bbT^1$ if $\bbT^0 A \subseteq \bbT^1 A$, for all~$A$.

\item We are particularly interested in the following submonads.
$\bbT$~denotes the subset of all \emph{linear trees,} i.e., trees where each variable
appears exactly once. $\bbT^\fin$~denotes the subset of \emph{finite linear trees,}
$\bbT^\reg$~the subset of \emph{regular linear trees,} $\bbT^\thin$~the subset of
\emph{thin linear trees,} i.e., trees with only countably many infinite branches,
and $\bbT^\wilke := \bbT^\thin \cap \bbT^\reg$ the subset of all trees that are thin and regular.
The corresponding classes of non-linear trees are denoted by $\bbT^{\times\fin}$,
$\bbT^{\times\reg}$, etc. \qedhere
\end{enuma}
\end{Def}
\begin{Rem}
Note that $\bbT^{\times\thin}$ and $\bbT^{\times\wilke}$ do not form submonads of~$\bbT^\times$,
since they are not closed under $\Flat$. This is different for $\bbT^\thin$ and $\bbT^\wilke$,
which are in fact submonads of~$\bbT$.
\end{Rem}

In algebraic language theory one uses algebras (usually finite ones) to recognise languages.
In our setting these algebras take the following form.
\pagebreak[5]
\begin{Def}
Let $\bbT^0 \subseteq \bbT^\times$.

    \begin{enuma}
\item
A \emph{$\bbT^0$-algebra} $\frakA = \langle A,\pi\rangle$ consists of a sorted set~$A$
and a \emph{product} $\pi : \bbT^0 A \to A$ satisfying
\begin{align*}
  \pi \circ \sing = \id
  \qtextq{and}
  \pi \circ \bbT^0\pi = \pi \circ \Flat\,.
\end{align*}
The first equation is called the \emph{unit law,} the second one the \emph{associative law.}

\item A $\bbT^0$-algebra~$\frakA$ is \emph{finitary} if it is finitely generated and
every domain~$A_\xi$ is finite.

\item A \emph{morphism} between two $\bbT^0$-algebras $\frakA = \langle A,\pi\rangle$ and
$\frakB = \langle B,\pi\rangle$ is a function $\varphi : A \to B$ commuting with the respective
products in the sense that
\begin{align*}
  \varphi \circ \pi = \pi \circ \bbT^0\varphi\,.
\end{align*}

\item The \emph{free $\bbT^0$-algebra} generated by a sorted set~$\Sigma$ is the
algebra $\bbT^0\Sigma := \langle\bbT^0\Sigma,\Flat\rangle$.

\item A $\bbT^0$-algebra~$\frakA$ \emph{recognises} a language $K \subseteq \bbT^0_\xi\Sigma$
if there exists a morphism $\eta : \bbT^0\Sigma \to \frakA$ with
$K = \eta^{-1}[P]$, for some $P \subseteq A_\xi$. \qedhere
\end{enuma}
\end{Def}

\begin{Exam}
The following algebra $\frakA = \langle A,\pi\rangle$ recognises the language~$K$ of
all trees $t \in \bbT^\times_\emptyset\{a,b\}$ that contain at least one letter~$a$.
For each sort~$\xi$, we use two elements $0_\xi,1_\xi$. Hence,
\begin{align*}
  A_\xi := \{0_\xi,1_\xi\}\,.
\end{align*}
The product is defined by
\begin{align*}
  \pi(t) := \begin{cases}
              1 &\text{if } t \text{ contains the label } 1\,, \\
              0 &\text{otherwise}\,.
            \end{cases}
\end{align*}
Then $K = \varphi^{-1}(1_\emptyset)$, where $\varphi : \bbT^\times\{a,b\} \to A$ is the morphism
mapping $\sing(a)$~to~$1$ and $\sing(b)$~to~$0$.
\end{Exam}

We will often use the usual term notation for trees and elements in an algebra.
That is, for $s \in \bbT^\times_\xi A$ and a $\xi$-tuple~$\bar r$ of trees and/or variables,
we denote by $s(\bar r)$ the tree obtained from~$s$ by replacing every variable $x \in \xi$
by the tree~$r_x$. Similarly, for an algebra~$\frakA$, an element $a \in A_\xi$,
and a $\xi$-tuple~$\bar b$ of elements and/or variables, we denote by $a(\bar b)$ the
product of the tree $s(\bar r)$ where $s := \sing(a)$ and $r_x := \sing(b_x)$
(or $r_x := b_x$, if $b_x$~is a variable).

A complication of the theory of infinite trees is the fact that some finitary
$\bbT^\times$-algebras recognise non-regular languages~\cite{BojanczykKlin18}.
For this reason we have to consider a smaller class of algebras.
\begin{Def}
    \begin{enuma}
\item We denote \emph{first-order logic} by $\FO$ and \emph{monadic second-order logic} by $\MSO$.

\item
Let $\bbT^0 \subseteq \bbT^\times$.
A $\bbT^0$-algebra~$\frakA$ is \emph{$\MSO$-definable} if it is finitary and there exists
a finite set $C \subseteq A$ of generators of~$\frakA$ with the following property\?:
for every $a \in A$, there exists an $\MSO$-formula~$\varphi_a$ such that
\begin{align*}
  t \models \varphi_a \quad\defiff\quad \pi(t) \geq a\,,
  \quad\text{for all } t \in \bbT^0 C\,.
\end{align*}
If all formulae~$\varphi_a$ belong to~$\FO$, we call $\frakA$ \emph{$\FO$-definable.} \qedhere
\end{enuma}
\end{Def}
\begin{Exam}
The algebra from the previous example is $\FO$-definable.
(The formulae $\varphi_0$~and~$\varphi_1$ only have to check whether or not the given tree
contains the label~$1$.)
\end{Exam}

Using this notion we obtain the following characterisation
(for proofs see Theorems 3.3~and~3.4 of~\cite{Blumensath20}\?; or more generally
Theorem~9.4 and Corollary~9.7 of~\cite{Blumensath21}).
\begin{Thm}\label{Thm: definable algebras}
Let $\bbT^0 \subseteq \bbT^\times$.
    \begin{enuma}
\item A finitary\/ $\bbT^0$-algebra\/~$\frakA$ is\/ $\MSO$-definable if, and only if,
  every language recognised by\/~$\frakA$ is regular.
\item A language $K \subseteq \bbT^0\Sigma$ is regular if, and only if, it is
  recognised by some\/ $\MSO$-definable\/ $\bbT^0$-algebra.
\end{enuma}
\end{Thm}
The definition above is not very enlightening as it is basically just a restatement
of the preceding theorem. Although a~more algebraic characterisation has been found
in~\cite{Blumensath20}, a simpler one would be appreciated.
In particular, it would be nice to find a system of inequalities axiomatising
the class of $\MSO$-definable algebras.
\begin{Open}
Find a concrete description of a system of inequalities that axiomatises
the class of $\MSO$-definable $\bbT^\times$-algebras.
\end{Open}
\noindent
By general arguments we know that such a system of inequalities exists,
although it might be infinite and the terms in the inequalities are in general profinite
(see~\cite{Blumensath21} for the details).

\section{Partition theorems for trees}   
\label{Sect:Ramsey}

Let us start with a brief overview of the existing partition theorems for trees,
followed by some remarks on how they might be extended and how they might not.
The seminal partition theorem for trees is the one by Milliken.
\begin{Def}
Let $t \in \bbT_\emptyset\Sigma$ be a tree.

    \begin{enuma}
\item The \emph{level} of a vertex $v \in \dom(t)$ is the number of vertices~$u$
with $u \prec v$. We denote it by~$\abs{v}$.

\item Let $\one$~be a set containing exactly one element of each sort.
A \emph{strong embedding} of a tree $s \in \bbT_\emptyset\one$ into~$t$
is a function $\varphi : \dom(s) \to \dom(t)$ such that, for all $u,v \in \dom(s)$,
\begin{align*}
 u \preceq v         &\quad\iff\quad \varphi(u) \preceq \varphi(v)\,, \\
 \suc_x(u) \preceq v &\quad\iff\quad \suc_x(\varphi(u)) \preceq \varphi(v)\,, \\
 \abs{u} = \abs{v}   &\qtextq{implies} \abs{\varphi(u)} = \abs{\varphi(v)}\,.
\end{align*}
\upqed
\end{enuma}
\end{Def}
\begin{Thm}[Milliken~\cite{Milliken79}]
Let $t \in \bbT A$ be an infinite tree without leaves, $C$~a finite set of colours, $m < \omega$,
and let $S_m \subseteq \bbT_\emptyset\one$ be the set of all finite trees such that every leaf
has level~$m$.
Given a function~$\lambda$ mapping every strong embedding $s \to t$ with $s \in S_m$
to some colour in~$C$, there exist a strong embedding $\varphi : h \to t$ and a colour $c \in C$
such that $h \in \bbT_\emptyset\one$ is a tree without leaves and
\begin{align*}
  \lambda(\varphi \circ \psi) = c\,,
  \quad\text{for all strong embeddings } \psi : s \to h \text{ with } s \in S_m\,.
\end{align*}
\end{Thm}

For our purposes it is sufficient to consider embeddings $s \to t$ with $s \in S_1$,
which correspond to strongly embedded factors $t[u,\bar v)$ of~$t$.
The main limitation of the theorem is that it does not give us any information about factors
$[u,\bar v)$ whose end-points are not strongly embedded.
For a stronger statement we need additional assumptions on the labelling.
For instance, for labellings of finite words, there is the Factorisation Tree Theorem of
Simon~\cite{Simon90} which states that, if the labelling is additive
(i.e., the colours form a semigroup), we can recursively factorise the given word into
homogeneous parts.
This theorem has been adapted to trees by Colcombet~\cite{Colcombet21} as follows.

\begin{Def}
Let $t \in \bbT A$ be a tree and $\frakS$~a semigroup.

    \begin{enuma}
\item
An \emph{additive $\frakS$-labelling} of~$t$ is a function~$\lambda$ that maps every edge~$e$
of~$t$ to a semigroup element $\lambda(e) \in S$.
Each such function can be extended to all non-empty finite paths~$p = (v_i)_i$ by setting
\begin{align*}
  \lambda(p) := \prod_i \lambda(v_i,v_{i+1})\,.
\end{align*}
In case $\frakS$~is an $\omega$-semigroup, we can extend the notation $\lambda(p)$ to
infinite paths using the same definition as above.
Finally, for vertices $u \prec v$, we will also use the notation
\begin{align*}
  \lambda(u,v) := \lambda(p)\,, \quad\text{where $p$ is the path from $u$ to } v\,.
\end{align*}

\item Given a function $\sigma : \dom(t) \to [k]$, we define a binary relation~$\sqsubset_\sigma$
on $\dom(t)$ (see Figure~\ref{Fig: Ramseyan split}) by
\begin{align*}
  x \sqsubset_\sigma y
  \quad\defiff\quad
  &x \prec y\,,\quad
  \sigma(x) = \sigma(y)\,,
  \quad\text{and} \\[1ex]
  &\sigma(z) \leq \sigma(x)\,, \quad\text{for all } x \preceq z \preceq y\,.
\end{align*}
As usual, $\sqsubseteq_\sigma$~denotes the reflexive version of~$\sqsubset_\sigma$.
\begin{figure}
\centering
\includegraphics{Expansion-2.mps}
\caption{A function~$\sigma$ (along a single branch of~$t$) and the corresponding relation
$\sqsubseteq_\sigma$, indicated via the grey bars.}\label{Fig: Ramseyan split}
\end{figure}

\item A \emph{weak Ramseyan split} of an $\frakS$-labelling~$\lambda$
is a function $\sigma : \dom(t) \to [k]$ such that
\begin{align*}
  \lambda(x,y) = \lambda(x,y) \cdot \lambda(x',y')\,,
  \quad
  &\text{for all } x \sqsubset_\sigma y \text{ and } x' \sqsubset_\sigma y' \\
  &\text{such that } y \sqsubseteq_\sigma y' \text{ or } y' \sqsubseteq_\sigma y\,.
\end{align*}
\upqed
\end{enuma}
\end{Def}

\begin{Thm}[Colcombet]\label{Thm: existence of Ramseyan splits}
Let\/ $\frakS$~be a finite semigroup.
There exists a number $N < \omega$ such that every\/ $\frakS$-labelling~$\lambda$ of some
tree~$t$ has a weak Ramseyan split $\sigma : \dom(t) \to [N]$ with $\sigma(\emptyseq) = N-1$.
Furthermore, this split is\/ $\MSO$-definable.
\end{Thm}

As an example of how to apply this theorem, let us mention the following result
from~\cite{Blumensath13a} (Theorem~4.4) that can be used to turn arbitrary trees
into regular ones while preserving an edge labelling.
\begin{Def}
Let $\frakS$~be an $\omega$-semigroup and $\lambda$~an $\frakS$-labelling of some tree~$t$.
We write
\begin{align*}
  \lim \lambda := \bigset{ \lambda(\beta) }{ \beta \text{ a branch of } t }\,.
\end{align*}
\upqed
\end{Def}
\begin{Thm}\label{Thm: regular tree with same labelling limit}
Let $\frakS$~be a finite $\omega$-semigroup.
For every $\frakS$-labelling~$\lambda$ of a tree~$t$, there exists a regular tree~$t_0$
and a regular $\frakS$-labelling~$\lambda_0$ of~$t_0$ such that
$\lim \lambda_0 = \lim \lambda$.
\end{Thm}
\noindent
The proof consists in fixing a weak Ramseyan split of~$\lambda$ and using it to
replace certain subtrees of~$t$ by back-edges. The unravelling of the resulting graph is the
desired regular tree~$t_0$.

As a second application let us see how to use Ramseyan splits to evaluate
products in a $\bbT$-algebra.
\begin{Def}
Let $\frakA$~be a $\bbT^0$-algebra, for some $\bbT^0 \subseteq \bbT^\times$.

    \begin{enuma}
\item
The \emph{canonical edge labelling}~$\lambda$ of a tree $t \in \bbT^0 A$ is defined by
\begin{align*}
  \lambda(u,v) := \pi(t[u,v))\,, \quad\text{for } u \prec v\,.
\end{align*}

\item
Let $L$~be either $\FO$ or $\MSO$, and let $\lambda$~be a given type of labelling,
i.e., a function assigning to each tree~$t$ some labelling~$\lambda_t$
(which may be an edge labelling or a vertex labelling).
We say that $\frakA$~is $L$-definable \emph{with respect to}~$\lambda$
if $\frakA$~is finitary and, for every finite $C \subseteq A$ and every $a \in A_\xi$,
there exists an $L$-formula~$\varphi_a$ such that
\begin{align*}
  \langle t,\lambda_t\rangle \models \varphi_a
  \quad\iff\quad
  \pi(t) \geq a\,,
  \quad\text{for all } t \in \bbT^0_\xi C\,,
\end{align*}
where
$\langle t,\lambda_t\rangle :=
  \bigl\langle \dom(t), {\preceq}, (\suc_x)_x, (P_c)_{c \in C + \xi}, (R_s)_{s \in S}\bigr\rangle$
denotes the usual encoding of~$t$ as a relational structure
with unary predicates~$P_c$ for the labels of~$t$ and additional relations~$R_s$
(either unary or binary) for the labelling~$\lambda_t$.
In case $\lambda_t$~is an edge labelling, we assume that the relations~$R_s$ contain
the labels for all pairs of vertices $u \prec v$, not only those where $v$~is a successor of~$u$.
(This is important in the case where $L = \FO$.) \qedhere
\end{enuma}
\end{Def}
\begin{Rem}
Note that the canonical edge labelling of a tree $t \in \bbT_\xi A$ is an $\frakS$-labelling
for the $\omega$-semigroup $\frakS = \langle S_1,S_\omega\rangle$ with elements
\begin{align*}
  S_1 := \sum_{\eta \subseteq \xi} A_{\eta + \{z\}}
  \qtextq{and}
  S_\omega := \sum_{\eta \subseteq \xi} A_\eta\,,
\end{align*}
where $z \in X$ is some fixed variable with $z \notin \xi$ that we use to define
the semigroup product.
\end{Rem}

\begin{Prop}\label{Prop: product FO-definable with canonical edge labelling}
Let $\bbT^0 \subseteq \bbT^\times$.
Every finitary $\bbT^0$-algebra is $\FO$-definable with respect to the canonical edge labelling.
\end{Prop}
\begin{proof}
Fix $a \in A_\xi$ and $C \subseteq A$.
The formula~$\varphi_a$ checks whether the given tree~$t$ has a leaf.
If this is the case it picks one, say~$v$, and then checks whether
\begin{align*}
  \lambda(\emptyseq,v) \cdot t(v) \geq a\,.
\end{align*}
Otherwise, for every sort~$\zeta$ used by some element of~$C$,
we fix some variable $z \in \zeta$.
If we start at the root of~$t$ and follow the successors labelled by one of these
chosen variables, we obtain an infinite branch of~$t$. This branch is $\FO$-definable.
The formula~$\varphi_a$ checks that the branch contains an infinite sequence
$v_0 \prec v_1 \prec\dots$ of vertices such that
\begin{align*}
  &\lambda(v_i,v_k) = \lambda(v_0,v_1)\,, \quad\text{for all } i < k < \omega\,, \\
  &\lambda(\emptyseq,v_0) \cdot \lambda(v_0,v_1)^\omega \geq a\,.
\end{align*}
This fact can be expressed in first-order logic using a trick of
Thomas~\cite[Lemma~1.4]{Thomas81}.
\end{proof}

We can improve this result by encoding the corresponding edge labelling by a vertex labelling.
\begin{Def}
Let $L$~be either $\FO$ or $\MSO$, and let $\tau : \dom(t) \to C$ be a vertex labelling of
some tree $t \in \bbT^\times A$.

    \begin{enuma}
\item
We say that $\tau$~is $L$-definable if, for every $c \in \rng \tau$,
there exists an $L$-formula~$\varphi_c(x)$ such that
\begin{align*}
  \tau(v) = c \quad\iff\quad t \models \varphi_c(v)\,,
  \quad\text{for every } v \in \dom(t)\,.
\end{align*}
The definition for labellings of edges is analogous.

\item We say that the product~$\pi$ of an algebra~$\frakA$ is $L$-definable on~$t$
\emph{in terms of} a vertex labelling $\tau : \dom(t) \to C$ if, for every $a \in A_\xi$
and every $\eta \subseteq \xi$, there exists an $L$-formula~$\psi_{a,\eta}(x,\bar y)$
(where $\bar y$~is an $\eta$-tuple) such that
\begin{align*}
  \pi(t[u,\bar v)) = a \quad\iff\quad
  \langle t,\tau\rangle \models \psi_{a,\eta}(u,\bar v)\,,
\end{align*}
for every factor $[u,\bar v)$ of sort $\eta$. \qedhere
\end{enuma}
\end{Def}
The following result is based on a similar theorem by Colcombet~\cite{Colcombet07b}.
\begin{Thm}\label{Thm: factor products computable from split}
Let\/ $\frakA$~be a finitary\/ $\bbT$-algebra.
For every finite set $C \subseteq A$, there exists a finite (unsorted) set~$S$
such that every $t \in \bbT C$ has a labelling $\sigma : \dom(t) \to S$
such that the product on~$t$ is\/ $\FO$-definable in terms of~$\sigma$.
Furthermore, if\/ $\frakA$~is\/ $\MSO$-definable, then so is~$\sigma$.
\end{Thm}
\begin{proof}
We have shown in Proposition~\ref{Prop: product FO-definable with canonical edge labelling}
that $\frakA$~is $\FO$-definable with respect to the canonical edge labelling~$\lambda$.
Consequently, it is sufficient to find a vertex labelling~$\sigma$ such that
$\lambda$~is $\FO$-definable in terms of~$\sigma$.

By Theorem~\ref{Thm: existence of Ramseyan splits}, there exists an $\MSO$-definable
weak Ramseyan split $\sigma_0 : \dom(t) \to [k]$ for~$\lambda$.
Unfortunately, $\lambda$~does not need to be $\FO$-definable in terms of~$\sigma_0$.
Therefore, we define an extended labelling
\begin{align*}
  \sigma : \dom(t) \to [k] \times A_{\{z\}}
\end{align*}
(for some fixed but arbitrary variable~$z$) as follows.
We denote by $p(v)$~the predecessor of $v \in \dom(t)$ (if it exists).
Fixing an arbitrary element $a_0 \in A_{\{z\}}$, we set
\begin{align*}
  \sigma(v) := \begin{cases}
                 \bigl\langle \sigma_0(v),\,\lambda(p(v),v)\bigr\rangle
                   &\text{if $p(v)$ is defined,} \\
                 \bigl\langle \sigma_0(v),\,a_0\bigr\rangle
                   &\text{otherwise.}
               \end{cases}
\end{align*}
Note that, if $\frakA$~is $\MSO$-definable, then so is $\sigma$.
Hence, it remains to prove that $\lambda$~is $\FO$-definable in
$\langle t,\sigma\rangle$.

We will construct an $\FO$-formula computing $\lambda(u,v)$
by induction on the minimal number~$n$ such that
\begin{align*}
  \sigma_0(w) \leq m\,, \quad\text{for all } u \prec w \prec v\,.
\end{align*}
Let $w_0$~be the minimal vertex $u \prec w_0 \prec v$ with $\sigma_0(w_0) = m$,
let $w_2$~be the maximal one, and let $w_1$~be the minimal vertex $w_0 \prec w_1 \preceq w_2$
with $\sigma_0(w_1) = m$.
We distinguish several cases.
If $v$~is the successor of~$u$, we can read off $\lambda(u,v)$ from~$\sigma(v)$.
If $w_0 = w_2$, we have
\begin{align*}
  \lambda(u,v) = \lambda(u,w_0) \cdot \lambda(w_0,v)\,,
\end{align*}
where both factors can be computed by inductive hypothesis.
Finally, consider the case where $w_0 \prec w_2$.
Then $w_0 \sqsubset_{\sigma_0} w_1 \sqsubseteq_{\sigma_0} w_2$ implies that
\begin{align*}
  \lambda(u,v) &= \lambda(u,w_0) \cdot \lambda(w_0,w_2) \cdot \lambda(w_2,v) \\
               &= \lambda(u,w_0) \cdot \lambda(w_0,w_1) \cdot \lambda(w_2,v)\,.
\end{align*}
All three factors can be computed by inductive hypothesis.
\end{proof}

The problem with the above theorems is that they require access to the canonical
edge-labelling~$\lambda$ and, in order to obtain this labelling, we need to be
able to compute products $\pi([u,v))$ where the factors $[u,v)$ are usually infinite.
Unfortunately, in many applications we only know the products of \emph{finite} factors.
For instance, we cannot combine Theorem~\ref{Thm: regular tree with same labelling limit}
and Proposition~\ref{Prop: product FO-definable with canonical edge labelling}
to conclude that every tree $t \in \bbT A$ over a finitary $\bbT$-algebra~$\frakA$
can be replaced by a regular one with the same product since we do not know whether or not
the regular labelling~$\lambda_0$ constructed in
Theorem~\ref{Thm: regular tree with same labelling limit} is a canonical edge-labelling.

We conclude this section with a counterexample showing that some natural ways to strengthen
Theorem~\ref{Thm: factor products computable from split} do not work.
It turns out that in general, if we want to compute $\pi(t[u,\bar v))$,
we need to know how the factor $[u,\bar v)$ is embedded in the tree.
Just looking at the values $\sigma(u)$ and $\sigma(v_i)$ provided by some labelling~$\sigma$ is
not enough. This is why the $\FO$-formula constructed in
Theorem~\ref{Thm: factor products computable from split}
also inspects the value of~$\sigma$ for vertices on the paths between~$u$ and the~$v_i$.
\begin{Def}
Let $t \in \bbT^\times A$ be a tree.

    \begin{enuma}
\item
For $u,v \in \dom(t)$, we denote by $u \sqcap v$ their infimum in the tree order~$\preceq$.

\item
The \emph{branching pattern} of a tuple~$\bar v$ of vertices is the partial order consisting of
the root of~$t$ and all vertices of the form $v_i \sqcap v_j$ where each edge $u \prec w$
is labelled by the variable~$x$ such that $\suc_x(u) \preceq w$.

\item The \emph{branching type} of a factor $[u,\bar v)$ of~$t$ is the isomorphism type
of the branching pattern of~$\bar v$ in the subtree~$t|_u$.
Alternatively, we can define the branching type as the atomic type of~$\bar v$ in the structure
$\langle t|_u,({\prec_x})_x,{\sqcap},u\rangle$ where
\begin{align*}
  u \prec_x v \quad\defiff\quad \suc_x(u) \preceq v\,.
\end{align*}
The branching type of a tree $t \in \bbT_\xi A$ is the branching type of~$(v_x)_{x \in \xi}$,
where $v_x$ is the vertex labelled by the variable~$x$. \qedhere
\end{enuma}
\end{Def}

\begin{Exam}
We construct a finitary $\bbT$-algebra~$\frakA$ such that, for every $t \in \bbT A$
with factors $[u,\bar v)$ and $[u',\bar v')$,
\begin{align*}
  \pi(t[u,\bar v)) = \pi(t[u',\bar v'))
  \qtextq{implies}
  &[u,\bar v) \text{ and } [u',\bar v') \text{ have the same} \\
  &\text{branching type.}
\end{align*}
For $\xi \in \Xi$, let $A_\xi$~be the set of all branching types of finite trees
$t \in \bbT_\xi\one$ such that
\begin{itemize}
\item every leaf of~$t$ is labelled by a variable and
\item every sort~$\zeta$ of a label in~$t$ is a finite initial segment of~$\omega$.
\end{itemize}
($\one$~is a set with exactly one element of each sort.)
As the arity of all labels in~$t$ is bounded by the sort~$\xi$ of~$t$,
it follows that the set~$A_\xi$ is finite.

Note that, given a tree $s \in \bbT\bbT\Sigma$, we can compute the branching type of $\Flat(s)$
from the branching types of $s(v)$, for $v \in \dom(s)$.
Here the branching type of a tree is defined as the branching type of the factor
$[\emptyseq,\bar v)$ where $\bar v$~are the vertices labelled by a variable.
It follows that there exists a function $\pi : \bbT A \to A$ satisfying
$\pi \circ \bbT\tau = \tau \circ \Flat$, where $\tau : \bbT\Sigma \to A$ is the function
mapping each tree to its branching type.
It follows that $\frakA := \langle A,\pi\rangle$ is a finitary $\bbT$-algebra
with the desired property.

Let $t \in \bbT\Sigma$ be an infinite binary tree and let $\lambda$~be the labelling
mapping a factor $[u,\bar v)$ to its branching type.
We claim that there is no function $\sigma : \dom(t) \to C$ with a finite codomain~$C$
such that the label $\lambda([u,\bar v))$ only depends on the values
$\sigma(u)$ and $\sigma(v_x)$.
We fix a vertex $u \in \dom(t)$ such that the set
\begin{align*}
  C_0 := \set{ \sigma(v) }{ v \succeq u }
\end{align*}
is minimal. Set $c := \sigma(u)$. By choice of~$u$ there are vertices
$v_0,v_1 \in \sigma^{-1}(c)$ with $\suc_0(u) \preceq v_0$ and $\suc_1(u) \preceq v_1$.
Similarly, we can find $v'_0,v'_1 \in \sigma^{-1}(c)$ with
$\suc_0(v_0) \preceq v'_0$ and $\suc_1(v_0) \preceq v'_1$.
\begin{center}
\includegraphics{Expansion-3.mps}
\end{center}
Then $\suc_0(u) \preceq v'_1$ implies that $\lambda([u,v'_0v_1)) \neq \lambda([u,v'_0v'_1))$,
but all four vertices have the same colour~$c$.
\end{Exam}

We conclude this section by briefly mentioning two alternative approaches.
\begin{Rem}
When talking about partition theorems for trees, we also have to mention automata.
Every automaton can be seen as a prescription producing labellings (runs) of trees.
The advantage of automata is that they can be used even if we know very little about
the underlying algebra. In particular, they can be used in cases where we can only
evaluate finite trees. Their disadvantage is that runs are usually not unique and that every
run only contains a limited amount of information. For instance, in general there is no
automaton that allows us to evaluate every factor $\pi(t[u,\bar v))$ of a given tree~$t$,
only factors of a fixed arity.
\end{Rem}

\begin{Rem}
The proof of Theorem~\ref{Thm: existence of Ramseyan splits} is based on
semigroup-theoretic methods, in particular, it makes heavy use of Green's relations.
It looks plausible that, in order to prove a stronger partition theorem
for trees, we have to develop a similar theory of Green's relations for tree algebras.
As it turns out, it is rather straightforward to generalise these relations to the setting of
monoidal categories where all hom-sets are finite.
(We omit the details since the statements and proofs
are virtually identical to those for semigroups.)
Furthermore, every $\bbT^\fin$-algebra~$\frakA$ can be seen as such a category~$\calA$
where the objects are the sorts $\xi \in \Xi$ and the hom-sets $\calA(\xi,\zeta)$
are given by $(A_\xi)^\zeta$.
The question is how to apply the resulting theory to the case at hand.
The main problem with doing so seems to be that, in general,
a finitary $\bbT^\fin$-algebra can have infinitely many $\sfJ$-classes.
\end{Rem}

\section{Expansions and dense submonads}   
\label{Sect:dense}

When looking for a strengthening of the results in the previous section it is always useful
to have an application in mind that can serve as a test case and reality check.
The following problem on expansions seems to be a good candidate for this purpose.

\begin{Def}
Let $\bbT^0 \subseteq \bbT^1 \subseteq \bbT^\times$ be submonads.
We say that a $\bbT^0$-algebra~$\frakA_0$ is a \emph{reduct} of a
$\bbT^1$-algebra $\frakA_1 = \langle A,\pi_1\rangle$ if
$\frakA_0 = \langle A,\pi_0\rangle$ where $\pi_0 : \bbT^0 A \to A$ is the
restriction of $\pi_1 : \bbT^1 A \to A$.
In this case, we call $\frakA_1$~a \emph{$\bbT^1$-expansion} of~$\frakA_0$.
\end{Def}

\paragraph{\normalfont\textbf{Expansion Problem.}}
\textit{Given monads $\bbT^0 \subseteq \bbT^1 \subseteq \bbT^\times$,
which $\bbT^0$-algebras have $\bbT^1$-expansions\?? And for which algebras
are these expansions unique\??}

\medskip
The motivating example of an expansion problem is the result that every finite Wilke
algebra has a unique expansion to an $\omega$-semigroup.
For trees, problems of this kind turn often out to be quite hard and seem to require
advanced techniques from combinatorics. In this article, we develop tools that help answering
such questions, with a focus on the tree monads we have defined above.
For the inclusion $\bbT^\wilke \subseteq \bbT^\thin$,
we will prove in Corollary~\ref{Cor: Twilke has unique Tthin-expansion} below
that expansions (of finitary algebras) always exists.
Unfortunately, the (more interesting) inclusions $\bbT^\reg \subseteq \bbT$ and
$\bbT^\thin \subseteq \bbT$ turn out to be much more complicated and our current techniques
seem to allow us to prove only partial results.

We start our investigation with recalling
some results from~\cite{Blumensath21} that can be used
to prove the uniqueness of expansions, if not their existence.
\begin{Def}
A submonad $\bbT^0 \subseteq \bbT^1$ is \emph{dense} in~$\bbT^1$ over a class~$\calC$
of $\bbT^1$-algebras if, for all algebras $\frakA \in \calC$, subsets $C \subseteq A$,
and trees $s \in \bbT^1 C$, there exists a tree $s^0 \in \bbT^0 C$ with $\pi(s^0) = \pi(s)$.
\end{Def}
\begin{Exam}
The notion of denseness is a generalisation of the fact from the theory of
$\omega$-semigroups that every infinite product has a factorisation of the form~$ab^\omega$.
This translates to the fact that, over the class of all finite $\omega$-semigroups,
the monad for Wilke algebras is dense in the monad for $\omega$-semigroups\?:
for every infinite word~$u$ in a finite $\omega$-semigroup there is an ultimately periodic
word~$u^0$ with the same product.
Details can be found in~\cite{Blumensath21}.
\end{Exam}

We have shown in Lemma~4.13\,(a) of~\cite{Blumensath21} that denseness implies
the uniqueness of expansions (if they exist).
One technical requirement of the proof is that we need to assume that the class in question
is closed under binary products. Such products are defined in the usual way\?:
the product $\frakA \times \frakB$ of two algebras has the universe
$A \times B := (A_\xi \times B_\xi)_{\xi \in \Xi}$ and the product is defined
component-wise, i.e., the product of a tree $t \in \bbT(A \times B)$ is defined
by taking the projections of~$t$ to the two components, multiplying them separately,
and returning the pair of values obtained in this way.
\begin{Prop}\label{Prop: dense expansions are unique}
Let\/ $\bbT^0$ be dense in\/~$\bbT^1$ over some class~$\calC$ that is closed under binary
products. Then every\/ $\bbT^0$-algebra\/~$\frakA$ has at most one\/ $\bbT^1$-expansion
that belongs to~$\calC$.
\end{Prop}
\begin{Rem}
The class of $\MSO$-definable $\bbT^\times$-algebras is closed under binary
products~\cite{Blumensath20}. The same holds for $\MSO$-definable $\bbT$-algebras.
\end{Rem}

The fact that a regular language is uniquely determined by which regular trees it contains
can be generalised to the following theorem (which is a consequence of (the proof of)
Theorem~10.1 of~\cite{Blumensath21}).
\begin{Thm}\label{Thm: regular trees dense}
$\bbT^\reg$~is dense in\/~$\bbT$ over the class of all\/ $\MSO$-definable\/ $\bbT$-algebras,
and\/ $\bbT^{\times\reg}$ is dense in\/~$\bbT^\times$
over the class of all\/ $\MSO$-definable\/ $\bbT^\times$-algebras.
\end{Thm}

It is currently unknown whether this property characterises the class of $\MSO$-definable
algebras.
\begin{Open}
Let\/ $\frakA$~be a finitary\/ $\bbT^\times$-algebra such that, for every $C \subseteq A$
and every $t \in \bbT C$, there is some $t^0 \in \bbT^{\times\reg} C$ with $\pi(t^0) = \pi(t)$.
Is\/ $\frakA$~$\MSO$-definable\??
\end{Open}

For linear trees we can strengthen Theorem~\ref{Thm: regular trees dense} to include existence.
The proof requires some tools from automata theory (for details we refer the
reader to~\cite{Thomas97,Loeding21}).
We only consider automata over trees of sort~$\emptyset$.
When we need an automaton to read a tree $t \in \bbT^\times_\xi\Sigma$,
we will encode it as an element of $\bbT^\times_\emptyset(\Sigma+\xi)$.
Therefore, we will use \emph{(alternating) tree automata} of the form
$\calA = \langle Q,\Sigma + \xi,\Delta,q_0,\Omega\rangle$ where $Q$~is the set of states,
$\Sigma + \xi$ the input alphabet, $\Omega : Q \to \omega$ a priority function,
and $\Delta$~the transition relation.
The latter consists of triples of the form $\langle p,a,(P_z)_{z \in \zeta}\rangle$
for a state $p \in Q$, a letter $a \in (\Sigma + \xi)_\zeta$, and sets of states
$P_z \subseteq Q$.
The behaviour of such an automaton~$\calA$ on a given input tree $t \in \bbT^\times_\xi\Sigma$
is defined via a certain parity game~$\calG$, called the \emph{Automaton-Pathfinder game}
of~$\calA$ on~$t$.
The two players in this game are called \emph{Automaton} and \emph{Pathfinder.}
The positions for the former are the pairs in $\dom(t) \times Q$,
while the positions for the latter are those in $\dom(t) \times \Delta$.
In a position $\langle v,q\rangle$, Automaton chooses a transition
$\langle p,a,\bar P\rangle \in \Delta$ with $p = q$ and $a = t(v)$\?;
Pathfinder replies with some $z$-successor~$u_z$ of~$v$ and some state $r \in P_z$\?;
and the game continues in the position $\langle u_z,r\rangle$.
Automaton wins a play in this game if he either manoevres Pathfinder into a position where
the latter cannot make a move, or if the play is infinite and the corresponding
sequence $(q_n)_{n<\omega}$ of states from the Automaton positions satisfies
the parity condition\?:
\begin{align*}
  \liminf_{n<\omega} \Omega(q_n) \quad\text{is even.}
\end{align*}
We say that the automaton~$\calA$ \emph{accepts} the tree $t \in \bbT^\times_\xi\Sigma$
if Player Automaton has a winning strategy for~$\calG$ when starting in the position
$\langle \emptyseq,q_0\rangle$ that consists of the root~$\emptyseq$ of~$t$ and
the initial state~$q_0$ of~$\calA$.
Finally, we call the automaton \emph{non-deterministic} if, in every transition
$\langle p,a,(P_z)_{z \in \zeta}\rangle \in \Delta$, all the sets $P_z$ are singletons.

It is frequently useful to consider the restriction of an Automaton-Pathfinder game
to some factor $[u,\bar v)$ of the given input tree~$t$.
In such cases we can collect information about the possible strategies of Automaton
on this restriction in a set of \emph{profiles.}
Such a profile is a pair $\langle p,\bar U\rangle$ consisting of a state $p \in Q$
and a tuple $\bar U = (U_z)_{z \in \zeta}$ of sets $U_z \subseteq \rng \Omega \times Q$,
where $\zeta$~is the sort of the factor $[u,\bar v)$.
We say that $\langle p,\bar U\rangle$ is a profile of the factor $[u,\bar v)$ in~$t$
if there exists a strategy for Automaton such that,
when starting the game in position $\langle u,p\rangle$,
Pathfinder has, for each $z \in \zeta$ and every $\langle k,q\rangle \in U_z$,
a strategy to reach the position $\langle v_z,q\rangle$
such that the least priority seen in between is equal to~$k$.

Finally, given a tree $s \in \bbT^\times_\zeta\Sigma$ with an arbitrary sort~$\zeta$,
we say that $\langle p,\bar U\rangle$ is a profile of~$s$ if there exists some tree
$t \in \bbT^\times_\xi\Sigma$ containing a factor $[u,\bar v)$ such that $s = t[u,\bar v)$
and $\langle p,\bar U\rangle$ is a profile of $[u,\bar v)$ in~$t$.
(Note that this fact does not depend on the choice of~$t$.)

By definition, every $\MSO$-definable $\bbT^\times$-algebra~$\frakA$ has a finite set of
generators $C \subseteq A$ such that the preimages $\pi^{-1}(a) \cap \bbT^\times C$ are regular,
that is, recognised by an automaton. For each given sort $\xi \in \Xi$,
we can combine all these automata into a single automaton~$\calA$ that has, for every
element $a \in A_\xi$, some state~$q_a$ such that, when using $q_a$~as the initial state,
the automaton accepts all trees in~$\bbT^\times_\xi C$ with product~$a$.

The\label{automaton over infinite alphabets} fact that this combined automaton~$\calA$ does
not work for all trees in
$\bbT^\times_\xi A$, but only for those in $\bbT^\times_\xi C$, is frequently inconvenient.
The problem is that the set~$A$ of possible labels is infinite.
Given an automaton $\calA = \langle Q,C + \xi,\Delta,q_0,\Omega\rangle$ over the alphabet
$C + \xi$, we can extend it to the infinite alphabet $A + \xi$ as follows.
Since $C$~is a set of generators, there exists some function $\vartheta : A \to \bbT^\times C$
such that $\pi \circ \vartheta = \id$.
As new set of states we use $D \times Q$, where $D := \rng \Omega$ is the set of priorities.
The priority function is $\Omega^+(k,p) := k$ and the initial state
$q_0^+ := \langle 0,q_0\rangle$.
Finally, the extended transition relation~$\Delta_+$ consists of all triples
$\langle \langle l,p\rangle,\,a,\,(U_z)_{z \in \zeta}\rangle$ such that,
in the original automaton, the tree~$\vartheta(a)$ has the profile
$\langle p,(U_z)_{z \in \zeta}\rangle$.
Let $\calA_+ := \langle D \times Q,A + \xi,\Delta_+,q^+_0,\Omega^+\rangle$ be the
automaton with this transition relation.
(Note that $\calA_+$~is not a finite automaton anymore, although it still has only finitely
many states.)
By construction, it follows that $\calA_+$~accepts a tree $t \in \bbT^\times_\xi A$
if, and only if, $\calA$~accepts the tree $\Flat(\bbT^\times\vartheta(t))$.
Since
\begin{align*}
  (\pi \circ \Flat \circ \bbT^\times\vartheta)(t)
  = (\pi \circ \bbT^\times\pi \circ \bbT^\times\vartheta)(t)
  = (\pi \circ \bbT^\times(\pi \circ \vartheta))(t)
  =  \pi(t)\,,
\end{align*}
it follows that $\calA_+$~recognises the language $\pi^{-1}(a)$ (with no restriction
on the labels allowed in the trees).

For linear trees, we are able to improve upon Theorem~\ref{Thm: regular trees dense}
by also establishing uniqueness. We start with a technical lemma which is based on an argument
originally due to Boja\'nczyk and Klin
(the first published version of which appears in~\cite{Blumensath20}).
\begin{Lem}\label{Lem: regular relation}
Let $\Sigma$~be a finite sorted set, $m < \omega$, and
let $R \subseteq \bbT \Sigma \times \bbT \Sigma$ be a (sorted) binary relation between trees.
For every tree $U \in \bbT_\xi R$, there exists a tree $U^0 \in \bbT^\reg_\xi R$
such that, for each $i < 2$, the trees
\begin{align*}
  \Flat(\bbT p_i(U))
  \qtextq{and}
  \Flat(\bbT p_i(U^0))
\end{align*}
have identical $\MSO$-theories of quantifier-rank~$m$,
where $p_0,p_1 : \bbT \Sigma \times \bbT \Sigma \to \bbT \Sigma$ are the two projections.
\end{Lem}
\begin{proof}
To construct~$U^0$ we use a variant of the Automaton-Pathfinder Game from above.
Given two non-deterministic tree automata $\calA$~and~$\calB$, we define a game
$\calG_R(\calA,\calB)$ where the first player wins if, and only if, there exist two trees
$S,T \in \bbT\bbT\Sigma$ with the same domain such that
\begin{itemize}
\item $\langle S(v), T(v)\rangle \in R$, for all vertices $v$,
\item $\calA$~accepts $\Flat(S)$,
\item $\calB$~accepts $\Flat(T)$.
\end{itemize}
The difference to the usual Automaton-Pathfinder Game is that we
simulate two automata at the same time and that, instead of playing single letters,
we play larger trees in each step.
The game has two players \emph{Automaton} and \emph{Pathfinder.}
Each round starts in a position of the form $\langle p,q\rangle$,
where $p$~is a state of~$\calA$ and $q$~one of~$\calB$.
In the first round of the game, $p$~and~$q$ are the initial states of the respective automata.
Given such a position $\langle p,q\rangle$, Automaton chooses
\begin{itemize}
\item a pair of trees $\langle s,t\rangle \in R_\xi$, for some $\xi \in \Xi$,
\item a profile~$\langle p,\bar V\rangle$ for~$\calA$ on~$s$ that starts in the state~$p$, and
\item a profile~$\langle q,\bar W\rangle$ for~$\calB$ on~$t$ that starts in the state~$q$.
\end{itemize}
Pathfinder responds by selecting a variable $x \in \xi$
and pairs $\langle k,p'\rangle \in V_x$ and $\langle l,q'\rangle \in W_x$.
The \emph{outcome} of this round is the pair
$\langle \langle p,k,p'\rangle,\, \langle q,l,q'\rangle\rangle$
and the game will continue in the position $\langle p',q'\rangle$.

If at some point in the game one of the players cannot make his choice,
that player loses the game.
Otherwise, the players produce an infinite sequence
$\langle \delta_0,\varepsilon_0\rangle,\langle \delta_1,\varepsilon_1\rangle,\dots$ of outcomes.
Let $k_i$~be the priority in~$\delta_i$ and $l_i$~the priority in~$\varepsilon_i$.
Player Automaton wins the game if each of the sequences $k_0,k_1,\dots$ and $l_0,l_1,\dots$
satisfies the parity condition. Otherwise, Pathfinder wins.

Clearly, if there are two trees $S,T \in \bbT\bbT\Sigma$ with the same domain
such that
\begin{itemize}
\item $\langle S(v), T(v)\rangle \in R$, for all vertices $v$,
\item $\calA$~accepts $\Flat(S)$, and
\item $\calB$~accepts $\Flat(T)$,
\end{itemize}
then Automaton has the following winning strategy in $\calG_R(\calA,\calB)$.
He fixes two winning strategies $\varrho$~and~$\sigma$ for the games on, respectively,
$\Flat(S)$ and $\Flat(T)$.
During the game he descends through the trees $S$~and~$T$.
When the game reaches a vertex~$v$, Automaton chooses the trees $S(v)$~and~$T(v)$ and
the profiles on the corresponding factors of $\Flat(S)$ and $\Flat(T)$
that are associated with $\varrho$~and~$\sigma$, respectively.

Conversely, if Automaton has a winning strategy in the game, we can use it to construct
\begin{itemize}
\item two trees $S,T \in \bbT\bbT\Sigma$ such that
  $\langle S(v), T(v)\rangle \in R$, for all~$v$, and
\item winning strategies for Automaton in the games for $\Flat(S)$ and $\Flat(T)$, respectively.
\end{itemize}

Having defined the game $\calG_R(\calA,\calB)$, we prove the statement of the lemma as follows.
Fix $U \in \bbT R$, let $\theta_i$~be the $\MSO$-theory of quantifier-rank~$m$ of the tree
\begin{align*}
  \Flat(\bbT p_i(U))\,,
\end{align*}
and let $\calA_i$~be an automaton recognising the class of all trees with theory~$\theta_i$.
It follows that Player Automaton has a winning strategy in the game $\calG_R(\calA_0,\calA_1)$.
As the winning condition of this game is regular, we can apply the
B\"uchi-Landweber Theorem~\cite{BuchiLandweber69}, which tells us that Automaton even has a
finite-memory winning strategy.
Since the choice of $S(v)$~and~$T(v)$ by Automaton in the game only depends
on the current position $\langle p,q\rangle$ and on the contents of the memory,
it follows that the resulting trees $S$~and~$T$ are regular, i.e., they belong to
$\bbT^\reg\bbT\Sigma$.
The tree $U^0 \in \bbT^\reg R$ with labels
\begin{align*}
  U^0(v) := \langle S(v),T(v)\rangle
\end{align*}
has the desired properties.
\end{proof}

\begin{Thm}\label{Thm: unique MSO-definable expansion of Tree algebras}
Every\/ $\MSO$-definable\/ $\bbT^\reg$-algebra can be expanded to a unique\/
$\MSO$-definable\/ $\bbT$-algebra.
\end{Thm}
\begin{proof}
Uniqueness follows by Theorem~\ref{Thm: regular trees dense}
and Proposition~\ref{Prop: dense expansions are unique}.
For existence, consider an $\MSO$-definable $\bbT^\reg$-algebra~$\frakA$
and let $C \subseteq A$ be a finite set of generators.
For each $a \in A$, we fix an $\MSO$-formula~$\varphi_a$ defining the set
$\pi^{-1}(a) \cap \bbT^\reg C$.
Intuitively, we obtain the desired expansion $\frakA_+ = \langle A,\pi_+\rangle$
by taking for~$\pi_+$ the function on all trees defined by these formulae.
To make this work we have to \textsc{(i)}~find a way to use the formulae~$\varphi_a$ on
trees containing labels from $A \setminus C$,
\textsc{(ii)}~show that the resulting function is well-defined,
\textsc{(iii)}~show that it extends~$\pi$, and
\textsc{(iv)}~show that it satisfies the axioms of a $\bbT$-algebra.

\textsc{(i)}
We define $\pi_+$~as follows.
As $C$~generates~$\frakA$, there exists a function
$\sigma : A \to \bbT^\reg C$ such that $\pi(\sigma(a)) = a$, for all $a \in A$.
We choose~$\sigma$ such that $\sigma(c) = \sing(c)$, for $c \in C$.
Let $\hat\sigma := \Flat \circ \bbT\sigma : \bbT A \to \bbT C$ be its extension to~$\bbT A$.
We set
\begin{align*}
  \pi_+ := \pi_0 \circ \hat\sigma\,,
\end{align*}
where
\begin{align*}
  \pi_0(t) := a \quad\defiff\quad t \models \varphi_a\,,
  \quad\text{for } t \in \bbT C\,.
\end{align*}

\textsc{(ii)}
To see that $\pi_+$~is well-defined, we have to check that, for every tree $t \in \bbT_\xi C$,
there is exactly one element $a \in A_\xi$ with $t \models \varphi_a$.
For a contradiction, suppose otherwise. Then we can find a tree $t \in \bbT_\xi C$ such that
\begin{align*}
  t \models
    \Lor_{a \neq b} (\varphi_a \land \varphi_b) \lor \neg\Land_{a \in A_\xi} \varphi_a\,.
\end{align*}
Since every non-empty $\MSO$-definable tree language contains a regular tree, it follows that
we can choose $t \in \bbT^\reg C$. By choice of the formulae~$\varphi_a$ this means
that $\pi(t)$ has either no value, or more than one. A~contradiction.

\textsc{(iii)}
$\pi_+$~extends~$\pi$ since, for $t \in \bbT^\reg A$, we have
\begin{align*}
  \pi_+(t)
  = \pi_0(\hat\sigma(t))
  = \pi(\hat\sigma(t))
  = \pi(t)\,,
\end{align*}
where the second step follows since $\hat\sigma(t) \in \bbT^\reg C$, for $t \in \bbT^\reg C$.

\textsc{(iv)}
It remains to check the axioms of a $\bbT$-algebra.
First, note that
\begin{alignat*}{-1}
  \pi_+ \circ \hat\sigma
  &= \pi_0 \circ \hat\sigma \circ \hat\sigma \\
  &= \pi_0 \circ \Flat \circ \bbT\sigma \circ \Flat \circ \bbT\sigma \\
  &= \pi_0 \circ \Flat \circ \Flat \circ \bbT(\bbT\sigma \circ \sigma)
     &&\qquad\text{[$\Flat$ nat. trans.]}\\
  &= \pi_0 \circ \Flat \circ \Flat \circ \bbT(\bbT\sing \circ \sigma) \\
  &= \pi_0 \circ \Flat \circ \bbT\sing \circ \Flat \circ \bbT\sigma
     &&\qquad\text{[$\Flat$ nat. trans.]}\\
  &= \pi_0 \circ \Flat \circ \bbT\sigma
     &&\qquad\text{[$\bbT$ monad]}\\
  &= \pi_0 \circ \hat\sigma \\
  &= \pi_+\,.
\end{alignat*}

For the unit law, it therefore follows that
\begin{align*}
  \pi_+ \circ \sing
  = \pi_+ \circ \hat\sigma \circ \sing
  = \pi_+ \circ \sigma
  = \pi \circ \sigma
  = \id\,.
\end{align*}

It remains to check the associative law.
For a set of sorts $\Delta \subseteq \Xi$ and a sorted set~$A$, we denote by~$A|_\Delta$ the
sorted set obtained from~$A$ by removing all elements whose sort does not belong to~$\Delta$.
Below we will establish the following two claims.
\begin{enuma}
\item The associative law holds for all trees $t \in \bbT\bbT^\reg A$.
\item The associative law holds for all trees $t \in \bbT(\bbT C)|_\Delta$ with finite~$\Delta$
  satisfying $C \subseteq A|_\Delta$.
\end{enuma}
Then we can prove the general case as follows.
Given $t \in \bbT\bbT A$, we consider the tree $s := \bbT\hat\sigma(t)$.
Each tree $r \in \bbT C$ can be written as $r = p(\bar u,\bar x)$ for a finite tree~$p$,
variables~$\bar x$, and infinite trees~$\bar u$ that do not contain variables.
This implies that each tree $r \in \bbT C$ can be written as
$r = \Flat(R)$, for some $R \in \bbT^\fin(\bbT C)|_\Delta$, where $\Delta \subseteq \Xi$
is the finite set consisting of the sort~$\emptyset$ and the sorts of the elements of~$C$.
(Note that this argument does not work for~$\bbT^\times C$.)
Consequently, there exist a finite set $\Delta \subseteq \Xi$ and a tree
$S \in \bbT\bbT^\fin(\bbT C)|_\Delta$ with $s = \bbT\Flat(S)$.
By the two claims above, it follows that
\begin{alignat*}{-1}
  (\pi_+ \circ \Flat)(t)
  &= (\pi_+ \circ \hat\sigma \circ \Flat)(t)
     &&\qquad\text{[$\pi_+ \circ \hat\sigma = \pi_+$]} \\
  &= (\pi_+ \circ \Flat \circ \bbT\sigma \circ \Flat)(t) \\
  &= (\pi_+ \circ \Flat \circ \Flat \circ \bbT\bbT\sigma)(t)
     &&\qquad\text{[$\Flat$ nat. trans.]}\\
  &= (\pi_+ \circ \Flat \circ \bbT\Flat \circ \bbT\bbT\sigma)(t)
     &&\qquad\text{[$\bbT$ monad]} \\
  &= (\pi_+ \circ \Flat \circ \bbT\hat\sigma)(t) \\
  &= (\pi_+ \circ \Flat)(s) \\
  &= (\pi_+ \circ \Flat \circ \bbT\Flat)(S) \\
  &= (\pi_+ \circ \Flat \circ \Flat)(S)
     &&\qquad\text{[$\bbT$ monad]} \\
  &= (\pi_+ \circ \bbT\pi_+ \circ \Flat)(S)
     &&\qquad\text{[by (b)]}\\
  &= (\pi_+ \circ \Flat \circ \bbT\bbT\pi_+)(S)
     &&\qquad\text{[$\Flat$ nat. trans.]}\\
  &= (\pi_+ \circ \bbT\pi_+ \circ \bbT\bbT\pi_+)(S)
     &&\qquad\text{[by (a)]}\\
  &= (\pi_+ \circ \bbT\pi_+ \circ \bbT\Flat)(S)
     &&\qquad\text{[by (b)]}\\
  &= (\pi_+ \circ \bbT\pi_+)(s) \\
  &= (\pi_+ \circ \bbT\pi_+ \circ \bbT\hat\sigma)(t) \\
  &= (\pi_+ \circ \bbT\pi_+)(t)\,.
     &&\qquad\text{[$\pi_+ \circ \hat\sigma = \pi_+$]}
\end{alignat*}
Hence, it remains to prove the two claims.

\begin{enuma}
\item
Fix a tree $t \in \bbT\bbT^\reg A$.
We consider the relation $R \subseteq \bbT C \times \bbT C$ given by
\begin{align*}
  \langle s,s'\rangle \in R
  \quad\defiff\quad
  s,s' \in \bbT^\reg C \text{ and } \pi(s) = \pi(s')\,,
\end{align*}
and the tree $U \in \bbT R$ with labels
\begin{align*}
  U(v) := \bigl\langle \hat\sigma(t(v)),\, \sigma(\pi_+(t(v)))\bigr\rangle\,,
  \quad\text{for } v \in \dom(t)\,.
\end{align*}
Let $p_0,p_1 : \bbT C \times \bbT C \to \bbT C$ be the two projections. Then
\begin{alignat*}{-1}
  a :={}& (\pi_+ \circ \Flat \circ \bbT p_0)(U) \\
     ={}& (\pi_+ \circ \Flat \circ \bbT \hat\sigma)(t) \\
     ={}& (\pi_+ \circ \Flat \circ \bbT(\Flat \circ \bbT\sigma))(t) \\
     ={}& (\pi_+ \circ \Flat \circ \Flat \circ \bbT\bbT\sigma)(t)
          &&\qquad\text{[$\bbT$ monad]} \\
     ={}& (\pi_+ \circ \Flat \circ \bbT\sigma \circ \Flat)(t)
          &&\qquad\text{[$\Flat$ nat. trans.]} \\
     ={}& (\pi_+ \circ \hat\sigma \circ \Flat)(t) \\
     ={}& \pi_+(\Flat(t))\,,
          &&\qquad\text{[$\pi_+ \circ \hat\sigma = \pi_+$]} \displaybreak[0]\\
  b :={}& (\pi_+ \circ \Flat \circ \bbT p_1)(U) \\
     ={}& (\pi_+ \circ \Flat \circ \bbT(\sigma \circ \pi_+))(t) \\
     ={}& (\pi_+ \circ \hat\sigma \circ \bbT\pi_+)(t) \\
     ={}& \pi_+(\bbT\pi_+(t))\,.
          &&\qquad\text{[$\pi_+ \circ \hat\sigma = \pi_+$]}
\end{alignat*}
Hence, we have to prove that $a = b$.
By Lemma~\ref{Lem: regular relation}, there exists a regular tree $U^0 \in \bbT^\reg R$ with
\begin{align*}
  \pi_0(\Flat(\bbT p_0(U^0)))
  &= \pi_0(\Flat(\bbT p_0(U))) = a\,, \\
  \pi_0(\Flat(\bbT p_1(U^0)))
  &= \pi_0(\Flat(\bbT p_1(U))) = b\,.
\end{align*}
Setting $s_i := \bbT p_i(U^0) \in \bbT^\reg\bbT^\reg C$,
it follows by associativity of~$\pi$ and the fact that we have
$\langle s_0(v),s_1(v)\rangle \in R$, for all~$v$, that
\begin{align*}
  a = \pi(\Flat(s_0))
    = \pi(\bbT\pi(s_0))
    = \pi(\bbT\pi(s_1))
    = \pi(\Flat(s_1))
    = b\,.
\end{align*}

\item
Fix $t \in \bbT(\bbT C)|_\Delta$ where $\Delta \subseteq \Xi$ is a finite set
such that $C \subseteq A|_\Delta$.
Recall that $\varphi_a$~is a formula defining the set $\pi^{-1}(a) \cap \bbT^\reg C$.
Since the $\MSO$-definable tree languages are closed under inverse homomorphisms (see,
e.g., Lemma~A.1 of~\cite{Blumensath20}), we can construct an $\MSO$-formula~$\varphi^\Delta_a$
such that
\begin{align*}
  s \models \varphi^\Delta_a \quad\iff\quad \hat\sigma(s) \models \varphi_a\,,
  \quad\text{for } s \in \bbT(A|_\Delta)\,.
\end{align*}
For $a \in A$, let $\psi^\Delta_a$~be a formula stating that
\begin{itemize}
\item[] ``For every factorisation of the given tree such that each factor has a sort in~$\Delta$,
  there exists a labelling of the factors by elements of~$A|_\Delta$ such that
  \begin{itemize}
  \item each factor with label $b \in A|_\Delta$ satisfies the formula~$\varphi_b$, and
  \item the tree consisting of the guessed labels satisfies~$\varphi^\Delta_a$.''
  \end{itemize}
\end{itemize}
As every factor (in~$\bbT C$) of a regular tree in~$\bbT^\reg C$ is regular,
it follows by~(a) that
\begin{align*}
  s \models \psi^\Delta_a \liff \varphi_a\,, \quad\text{for every } s \in \bbT^\reg C\,.
\end{align*}
(We can choose $\bbT^\reg\sing(s)$ for the factorisation guessed by~$\psi^\Delta_a$.)
It follows that the same is true for every $s \in \bbT C$.
Consequently,
\begin{align*}
  \Flat(t) \models \varphi_a \qtextq{implies} \bbT\pi_+(t) \models \varphi^\Delta_a\,,
  \quad\text{for all } a\,.
\end{align*}
Furthermore, by choice of~$\varphi^\Delta_a$, we have
\begin{align*}
  \bbT\pi_+(t) \models \varphi^\Delta_a
  &\quad\iff\quad (\hat\sigma \circ \bbT\pi_+)(t) \models \varphi_a \\
  &\quad\iff\quad (\pi_0 \circ \hat\sigma \circ \bbT\pi_+)(t) = a \\
  &\quad\iff\quad (\pi_+ \circ \bbT\pi_+)(t) = a\,.
\end{align*}
Consequently, it follows that
\begin{align*}
    (\pi_+ \circ \Flat)(t) = (\pi_+ \circ \bbT\pi_+)(t)\,. \tag*{\qed}
\end{align*}
\upqed
\end{enuma}
\renewcommand{\qed}{}
\end{proof}

Note that our uniqueness result in Proposition~\ref{Prop: dense expansions are unique}
only concerns expansions in the given class~$\calC$.
It is possible that there exist additional expansions outside of~$\calC$.
\begin{Exam}\label{Ex:BK-algebra}
In~\cite{BojanczykKlin18} Boja\'nczyk and Klin have presented an example of a finitary
$\bbT^\times$-algebra that is not $\MSO$-definable.
This algebra can be used to find an $\MSO$-definable $\bbT^\reg$-algebra
with several $\bbT$-expansions, one of them $\MSO$-definable.
(By the preceding theorem, there can only be one of the latter.)
The construction of this $\bbT^\reg$-algebra $\frakA^0 = \langle A,\pi\rangle$
and two of its $\bbT$-expansions $\frakA^\reg = \langle A,\pi_\reg\rangle$
($\MSO$-definable) and $\frakA^{\mathrm{non}} = \langle A,\pi_{\mathrm{non}}\rangle$
(not $\MSO$-definable) is as follows.

For~$\frakA^{\mathrm{non}}$, we take (a simplified version) of the algebra
from~\cite{BojanczykKlin18}.
Let $\Sigma := \{a,b\}$ where both elements have arity~$2$ and set
$\Delta_\xi := \bbT^\fin_\xi\Sigma$.
As $\Sigma$~contains only binary elements, every leaf of a tree $t \in \Delta_\xi$
must be labelled by a variable. Hence, $t$~has at most~$\abs{\xi}$ leaves and, therefore,
at most $\abs{\xi}-1$ internal vertices. This implies that $\Delta_\xi$~is a finite set.

We call a tree $t \in \bbT_\emptyset\Sigma$ \emph{antiregular} if no two subtrees of~$t$
are isomorphic. It is \emph{densely antiregular} if every subtree of~$t$ has an antiregular
subtree.

The domains of all three algebras are
\begin{align*}
  A_\xi := \Delta_\xi \cupdot \{\bot,{*}\}\,,
  \quad\text{for } \xi \in \Xi\,,
\end{align*}
which we order such that $\bot$~is the least element and all other elements are incomparable.
Intuitively, the elements~$\Delta_\xi$ encode finite trees (by themselves),
while the element~$*$ encodes a densely antiregular tree and
$\bot$~encodes an infinite tree that is not densely antiregular.

Let us call a tree $t \in \bbT_\emptyset A$ \emph{good} if
every subtree~$s$ of~$t$ contains some vertex~$v$ such that $s(v) = {*}$,
or such that $s|_v \in \bbT\Delta$ and $\Flat(s|_v)$~is densely antiregular.

For $t \in \bbT_\xi A$, we define the product $\pi_{\mathrm{non}}(t)$ of~$\frakA^{\mathrm{non}}$
by the following case distinction.
(We tacitly assume in each case that the conditions for the previous cases are not satisfied.)
\begin{itemize}
\item $\pi_{\mathrm{non}}(t) = \bot$ if $t$~contains the label~$\bot$.
\item $\pi_{\mathrm{non}}(t) = \Flat(t)$ if $t \in \bbT^\fin_\xi\Delta$.
\item $\pi_{\mathrm{non}}(t) = {*}$ if every subtree without variables is good.
\item $\pi_{\mathrm{non}}(t) = \bot$ if $t$~has a subtree without variables that is not good.
\end{itemize}
The product $\pi_\reg(t)$ of~$\frakA^\reg$ is defined as follows.
\begin{itemize}
\item $\pi_\reg(t) = \bot$ if $t$~contains the label~$\bot$.
\item $\pi_\reg(t) = \Flat(t)$ if $t \in \bbT^\fin_\xi\Delta$.
\item $\pi_\reg(t) = {*}$ otherwise.
\end{itemize}
Note that this product is $\MSO$-definable and that the restrictions of
$\pi_\reg$~and~$\pi_{\mathrm{non}}$ to~$\bbT^\reg A$ coincide.
Thus $\frakA^\reg$~and~$\frakA^{\mathrm{non}}$ are $\bbT$-expansions of the same
$\bbT^\reg$-algebra.
\end{Exam}

To conclude this section, let me mention one of the main open problems concerning
the relation between regular trees and arbitrary ones (for details, see~\cite{Blumensath21}).
\begin{Open}
Does there exist a system of equations (in the algebra\/~$\bbT^\times\Sigma$) modulo which
every tree is equivalent to a regular one\?? If so, does it have an explicit description\??
\end{Open}
Having such an equational characterisation would be invaluable for applications,
where proofs frequently require a reduction to regular trees.
For instance, there exists an equational characterisation of bisimulation-invariant
languages of \emph{regular} trees~\cite{BojanczykId09}, but so far nobody was able
to generalise it to languages of arbitrary trees.

\section{Evaluations}   
\label{Sect:evaluations}

The notion of denseness seems to be only useful if we already know that the algebra in
question has an expansion. To actually prove existence we need different techniques.
We start with the following simple idea.
Given a submonad $\bbT^0 \subseteq \bbT$ and a $\bbT^0$-algebra~$\frakA$,
we try to compute the product of a tree $t \in \bbT A$
inductively bottom-up using the given $\bbT^0$-product.
That is, we factorise~$t$ into pieces that belong to~$\bbT^0$, evaluate them,
and then recursively evaluate the remaining tree. If we can show that,
\begin{itemize}
\item after a finite number of such steps, the tree~$t$ is reduced to a single vertex and
\item that the final result does not depend on the choice of factorisation used in
  each step,
\end{itemize}
it follows that we can uniquely evaluate every tree in $\bbT A$
using only the $\bbT^0$-product. In particular, the product of~$\frakA$ can be
uniquely extended to the set of all trees.

A well-known use of this technique is given by Simon's Factorisation Tree Theorem
(see, e.g.,~\cite{Colcombet21}).
Such a factorisation tree is just a hierarchical decomposition of a given semigroup-product
using binary products and products of idempotents only.
A~second example of this approach was used in~\cite{CartonColcombetPuppis18} to prove, in our
terminology, that a certain inclusion between monads of countable linear orders is dense.
The aim of the current section is to make this idea of using an inductive approach
work for trees. Below we will use recursive factorisations to settle the
expansion problem for thin trees and we will derive partial results for general ones.

The definition below is a bit more general than the above intuitive description.
We will need the added generality for the more powerful decompositions used further below.
Suppose that we are given a $\bbT^0$-algebra which we want to expand to a $\bbT^1$-algebra,
for some $\bbT^0 \subseteq \bbT^1 \subseteq \bbT^\times$,
and suppose that we have already found some sorted set $S \supseteq \bbT^0 A$ such that we can
extend the product $\pi : \bbT^0 A \to A$ to $\rho : S \to A$.
To extend~$\rho$ further to a function $\bbT^1 A \to A$, consider a tree $t \in \bbT^1 A$.
We choose a factorisation~$T$ of~$t$ where we have already inductively assigned some value
$\val(T(v))$ to each factor.
If the reduced tree $\bbT^1\val(T)$ belongs to~$S$, then we can set
$\val(t) := \rho(\bbT^1\val(T))$.

In the following formal definition, $\bbE_\alpha(\rho,\bbT^1)$ is the set of all recursive
factorisations with $\alpha$~levels of recursion,
$\term_\alpha : \bbE_\alpha(\rho,\bbT^1) \to \bbT^1 A$ maps each such factorisation
to the tree it factorises, and $\val_\alpha : \bbE_\alpha(\rho,\bbT^1) \to A$ maps each
factorisation to the value obtained by recursively evaluating every factor.
\begin{Def}
Let $\bbT^0 \subseteq \bbT^1 \subseteq \bbT^\times$ be submonads, $\frakA$~a $\bbT^0$-algebra,
and $\rho : S \to A$ a function with domain $S \supseteq \bbT^0 A$
with $\rho \restriction \bbT^0 A = \pi$.

    \begin{enuma}
\item
For each ordinal~$\alpha$, we inductively define a sorted set $\bbE_\alpha(\rho,\bbT^1)$
of \emph{$\rho$-eval\-u\-ations} and two functions
\begin{align*}
  \term_\alpha : \bbE_\alpha(\rho,\bbT^1) \to \bbT^1 A
  \qtextq{and}
  \val_\alpha : \bbE_\alpha(\rho,\bbT^1) \to A
\end{align*}
by
\begin{align*}
  \bbE_0(\rho,\bbT^1)          &:= A\,, \\
  \bbE_{\alpha+1}(\rho,\bbT^1) &:= \bbE_\alpha(\rho,\bbT^1) +
    \bigset{ \gamma \in \bbT^1\bbE_\alpha(\rho,\bbT^1 \cap \bbT) }
           { \bbT^1\val_\alpha(\gamma) \in \dom(\rho) }\,, \\
  \bbE_\delta(\rho,\bbT^1)     &:= \bigcup_{\alpha < \delta} \bbE_\alpha(\rho,\bbT^1)\,,
    \qquad\text{for limit ordinals } \delta\,,
\end{align*}
and
\begin{alignat*}{-1}
  \term_0                &:= \sing\,,
  &\qquad\val_0          &:= \id\,, \\
  \term_{\alpha+1}       &:= \term_\alpha + \Flat \circ \bbT^1\term_\alpha\,,
  &\qquad\val_{\alpha+1} &:= \val_\alpha + \rho \circ \bbT^1\val_\alpha\,, \\
  \term_\delta           &:= \bigcup_{\alpha < \delta} \term_\alpha\,.
  &\qquad\val_\delta     &:= \bigcup_{\alpha < \delta} \val_\alpha\,.
\end{alignat*}
Finally, we set\footnote{Note that these classes are sets since the unions stabilise after
at most~$\omega_1$ steps.}%
\begin{align*}
  \bbE(\rho,\bbT^1) := \bigcup_\alpha \bbE_\alpha(\rho,\bbT^1)\,,\quad
  \term := \bigcup_\alpha \term_\alpha\,,
  \qtextq{and}
  \val := \bigcup_\alpha \val_\alpha\,.
\end{align*}

\item We call $\term(\gamma)$ the \emph{underlying term} of $\gamma \in \bbE(\rho,\bbT^1)$
and $\val(\gamma)$ its \emph{value.} If $t = \term(\gamma)$, we say that
$\gamma$~is a \emph{$\rho$-evaluation} of~$t$.

\item
We say that the algebra~$\frakA$ \emph{has $\rho$-evaluations for~$\bbT^1$} if
$\term : \bbE(\rho,\bbT^1) \to \bbT^1 A$ is surjective,
and we say that it has \emph{essentially unique $\rho$-evaluations} if furthermore
\begin{align*}
  \term(\gamma) = \term(\gamma')
  \qtextq{implies}
  \val(\gamma) = \val(\gamma')\,.
\end{align*}

\item In the special case where $\rho = \pi$, we also write
$\bbE(\frakA,\bbT^1) := \bbE(\pi,\bbT^1)$ and we call the elements of $\bbE(\frakA,\bbT^1)$
\emph{simple $\bbT^0$-evaluations.} \qedhere
\end{enuma}
\end{Def}
\begin{Rem}
Note that in the recursive definition of $\bbE_{\alpha+1}(\rho,\bbT^1)$,
we only use decompositions into \emph{linear} factors.
The reason why we allow $\bbT^1 \subseteq \bbT^\times$ is to be able to have
evaluations of non-linear trees.
\end{Rem}
\begin{Exam}
Let $\frakA$~be a $\bbT^\reg$-algebra, $a \in A_{\{x,y\}}$, $b \in A_{\{x\}}$,
$c \in A_\emptyset$ elements, and $t$~the tree consisting of an infinite branch
labelled~$a$ attached to which are trees of the form $s_n := b^n(c)$, for every $n < \omega$
(see Figure~\ref{Fig: a simple evaluation}).
\begin{figure}
\centering
\includegraphics{Expansion-4.mps}
\caption{The trees $t$~and~$t'$.}\label{Fig: a simple evaluation}
\end{figure}
We construct a simple $\bbT^\reg$-evaluation of~$t$ as follows.
In the first step, we evaluate each of the subtrees $s_n \in \bbT^\reg A$.
The resulting tree has the form
\begin{align*}
  t' := a(d_0,a(d_1,\dots))\,,
  \quad\text{where } d_n := \pi(s_n)\,,
\end{align*}
i.e., $t'$~consists of an infinite branch labelled~$a$ to each vertex of which we have attached
a leaf with label~$d_n$.
Since $A_{\{x\}}$ forms a finite semigroup and $d_{n+1} = b(d_n)$,
the sequence $d_0,d_1,\dots$ is ultimately periodic.
Fixing indices $k < l$ with $d_k = d_l$, we can write $t' = uv^\omega$ where
\begin{align*}
  u &:= a(d_0,a(d_1,\cdots a(d_{k-1},x)\cdots))\,, \\
  v &:= a(d_k,a(d_{k+1},\cdots a(d_{l-1},x)\cdots))\,.
\end{align*}
Consequently, $t'$~is regular and we can take it as the second and final level of our evaluation.
\end{Exam}

As a first application of evaluations let us note that the existence of simple evaluations
implies the denseness of the corresponding monads.
\begin{Lem}\label{Lem: evaluations imply denseness}
Let $\bbT^0 \subseteq \bbT^1 \subseteq \bbT^\times$ and
let $\calC$~be a class of~$\bbT^1$-algebras such that every $\frakA \in \calC$
has simple $\bbT^0$-evaluations for~$\bbT^1$.
Then $\bbT^0$~is dense in~$\bbT^1$ over~$\calC$.
\end{Lem}
\begin{proof}
We make use of a characterisation of denseness from Lemma~4.21 in~\cite{BlumensathLN2}
according to which $\bbT^0$~is dense in~$\bbT^1$ over~$\calC$ if, and only if,
for all $\frakA \in \calC$,
every subalgebra of the $\bbT^0$-reduct of~$\frakA$ is also a subalgebra of~$\frakA$.
Hence, fix $\frakA \in \calC$ and let $C \subseteq A$ be a set inducing a subalgebra of the
$\bbT^0$-reduct of~$\frakA$. We have to show that $C$~is closed under the product of~$\frakA$.

Let $t \in \bbT^1C$.
If $t \in \bbT^0 C$, we have $\pi(t) \in C$ since $C$~is closed under
$\pi \restriction \bbT^0 A$. Hence, suppose otherwise.
By assumption, $t$~has a simple $\bbT^0$-evaluation~$\gamma$. Let $\alpha$~be the ordinal
such that $\gamma \in \bbE_{\alpha+1}(\frakA,\bbT^1) \setminus \bbE_\alpha(\frakA,\bbT^1)$.
By induction on~$\alpha$, we prove that
\begin{align*}
  \pi(t) = \val_{\alpha+1}(\gamma)
  \qtextq{and}
  \val_{\alpha+1}(\gamma) \in C\,.
\end{align*}

For the first claim, associativity of~$\pi$ implies that
\begin{align*}
  \pi(t) &= \pi(\term_{\alpha+1}(\gamma)) \\
         &= \pi(\Flat(\bbT^1\term_\alpha(\gamma))) \\
         &= \pi(\bbT^1\pi(\bbT^1\term_\alpha(\gamma))) \\
         &= \pi(\bbT^1\val_\alpha(\gamma)) \\
         &= \val_{\alpha+1}(\gamma)\,.
\end{align*}

For the second claim, note that
every subevaluation~$\gamma(v)$ is an evaluation of some factor of~$t$.
Hence, we can use the inductive hypothesis to prove that
$\bbT^1\val_\alpha(\gamma) \in \bbT^1C$.
Since $C$~is closed under $\pi \restriction \bbT^0 A$, it follows that
\begin{align*}
  \val_{\alpha+1}(\gamma) = \pi(\bbT^1\val_\alpha(\gamma)) \in C\,.
\end{align*}
\upqed
\end{proof}

The only reason why we allow evaluations $\gamma \in \bbE_{\alpha+1}(\rho,\bbT^1)$
with components~$\gamma(v)$ belonging to $\bbE_\beta(\rho,\bbT^1)$, for $\beta < \alpha$,
is to define $\bbE_\alpha(\rho,\bbT^1)$ for all ordinals~$\alpha$.
In all of our applications below, the case $\alpha < \omega$ will actually be sufficient.
In this case, we can always assume that
$\gamma(v) \in \bbE_\alpha(\rho,\bbT^1) \setminus \bbE_{\alpha-1}(\rho,\bbT^1)$ for all~$v$.
(If not, we can replace $\gamma(v)$ by $\sing(\gamma(v))$.)
\begin{Def}
We say that an evaluation $\gamma \in \bbE_\alpha(\rho,\bbT^1)$ has
\emph{uniform depth~$\alpha$} if $\alpha = 0$ or $\gamma \in \bbT^\times\bbT^{\alpha-1}(A)$.
\end{Def}

As another example, let us show how to encode an evaluation of uniform depth~$n$ by a function
$\tau : \dom_0(t) \to [n]$ similar to a Ramseyan split.
By definition, such an evaluation is nothing but a nested factorisation
$\gamma \in \bbT^\times\bbT\cdots\bbT A$.
We can encode such a factorisation by assigning to every vertex $w \in \dom_0(t)$ the depth at
which it appears in this nesting (see Figure~\ref{Fig:evaluation and split} for an example).
\begin{figure}
\centering
\includegraphics{Expansion-5.mps}
\caption{An evaluation (on the right) encoded by a function (on the left).
For simplicity, we have omitted the labels from~$A$ and drawn just the $\tau$-labelling.}%
\label{Fig:evaluation and split}
\end{figure}
\begin{Lem}\label{Lem: evaluation from a split}
Let $0 < n < \omega$ and let $t \in \bbT^\times A$.
The construction in Figure~\ref{Fig:evaluation and split} induces
a bijection between all simple $\bbT^\times$-evaluations
$\gamma \in \bbE_n(\frakA,\bbT^\times)$ of~$t$ of uniform depth~$n$
and all functions $\tau : \dom_0(t) \to [n]$ with $\tau(\emptyseq) = n-1$.
\end{Lem}
\begin{proof}
Given an evaluation~$\gamma$ of~$t$ of uniform depth~$n$,
we construct the corresponding function~$\tau$ by induction on~$n$.
If $n = 1$, we set $\tau(w) := 0$, for all~$w$.
Otherwise, let $\tau_v : \dom_0(\term_{n-1}(\gamma(v))) \to [n-1]$, for $v \in \dom_0(\gamma)$,
be the functions obtained from the inductive hypothesis for
$\gamma(v) \in \bbE_{n-1}(\frakA,\bbT)$.
Note that $t = \term_n(\gamma) = \Flat(\bbT^\times\term_{n-1}(\gamma))$.
Hence, every vertex $w \in \dom_0(t)$ corresponds to a pair $\langle v,u\rangle$ of vertices
with $v \in \dom_0(\gamma)$ and $u \in \dom_0(\term_{n-1}(\gamma(v)))$.
We set
\begin{align*}
  \tau(w) := \begin{cases}
               n-1       &\text{if } u \text{ is the root of } \term_{n-1}(\gamma(v))\,, \\
               \tau_v(u) &\text{otherwise}\,.
             \end{cases}
\end{align*}

Conversely, let $\tau : \dom_0(t) \to [n]$ be a function with $\tau(\emptyseq) = n-1$.
If $n = 1$, we set $\gamma := t$.
Otherwise, let $T$~be the factorisation of~$t$ induced by the set
\begin{align*}
  H := \set{ w \in \dom_0(t) }{ \tau(w) = n-1 }\,,
\end{align*}
that is, $H$~is the set of vertices corresponding to the roots of the factors~$T(v)$.
For each $v \in \dom_0(T)$, we can use the inductive hypothesis to find an evaluation
$\gamma_v \in \bbE_{n-1}(\frakA,\bbT^\times)$ of $T(v)$.
(The corresponding function~$\tau'$ is obtained from the restriction of~$\tau$ to~$T(v)$
by setting the value at the root to $n-2$.)
Since $T(v) \in \bbT A$ is linear, we have $\gamma_v \in \bbE_{n-1}(\frakA,\bbT)$.
Consequently, we can set
\begin{align*}
  \gamma(v) := \gamma_v\,.
\end{align*}

It is straightforward to check that the two functions defined above are inverse to each other.
\end{proof}

Let us next explain how to use $\rho$-evaluations to construct $\bbT^1$-expansions.
The proof makes use of the following glueing operation for evaluations.
\begin{Lem}\label{Lem: gluing simple evaluations}
Let $\bbT^0 \subseteq \bbT^1 \subseteq \bbT$ be monads,
$\gamma \in \bbT^1\bbE(\rho,\bbT^1)$ a tree of evaluations,
and let $\beta$~be a $\rho$-evaluation of the tree $t := \bbT^1\val(\gamma)$.
There exists a $\rho$-evaluation $\beta|\gamma$ of the tree
$(\Flat\circ\bbT^1\term)(\gamma)$ such that $\val(\beta|\gamma) = \val(\beta)$.
\end{Lem}
\begin{proof}
We define $\beta|\gamma$ by the following induction on the ordinal~$\alpha$ with
$\beta \in \bbE_\alpha(\rho,\bbT^1)$.
If $\alpha = 0$, then $\beta = a \in A$ and $\gamma = \sing(\gamma_0)$, for some
$\gamma_0 \in \bbE(\rho,\bbT^1)$ with $\val(\gamma_0) = a$.
Setting $\beta|\gamma := \gamma_0$ we obtain
$\val(\beta|\gamma) = \val(\gamma_0) = a = \val(\beta)$.

For the inductive step,
suppose that $\beta \in \bbE_{\alpha+1}(\rho,\bbT^1) \setminus \bbE_\alpha(\rho,\bbT^1)$.
Then $\beta \in \bbT^1\bbE_\alpha(\rho,\bbT^1)$.
Let $\gamma' \in \bbT^1\bbT^1\bbE(\rho,\bbT^1)$ be the tree with the same domain as~$\beta$
where $\gamma'(v)$, for $v \in \dom_0(\beta)$, is (isomorphic to) the restriction of~$\gamma$ to
$\dom_0(\term(\beta(v))) \subseteq \dom(\gamma)$ (plus vertices for the variables).
We can use the inductive hypothesis to obtain evaluations
$\beta(v)|\gamma'(v) \in \bbE(\rho,\bbT^1)$, for $v \in \dom_0(\beta)$.
We choose for $\beta|\gamma$ the tree with the same domain as~$\beta$ and
\begin{align*}
  (\beta|\gamma)(v) := \beta(v)|\gamma'(v)\,,
  \quad\text{for } v \in \dom_0(\beta)\,.
\end{align*}
By inductive hypothesis it then follows that
\begin{alignat*}{-1}
  \term((\beta|\gamma)(v)) &= \term(\beta(v)|\gamma'(v))
                           &&= (\Flat \circ \bbT^1\term)(\gamma'(v))\,, \\
  \val((\beta|\gamma)(v))  &= \val(\beta(v)|\gamma'(v))
                           &&= \val(\beta(v))\,.
\end{alignat*}
Consequently, we have
\begin{alignat*}{-1}
  \term(\beta|\gamma)
  &= (\Flat\circ\bbT^1\term)(\beta|\gamma) \\
  &= (\Flat \circ \bbT^1(\Flat\circ\bbT^1\term))(\gamma') \\
  &= (\Flat \circ \Flat \circ \bbT^1\bbT^1\term)(\gamma')
     &&\qquad\text{[$\bbT^1$ monad]} \\
  &= (\Flat \circ \bbT^1\term \circ \Flat)(\gamma')
     &&\qquad\text{[$\Flat$ nat. trans.]} \\
  &= (\Flat\circ\bbT^1\term)(\gamma)\,, \\
  \val(\beta|\gamma)
  &= (\rho \circ \bbT^1\val)(\beta|\gamma) \\
  &= (\rho \circ \bbT^1\val)(\beta)
   = \val(\beta)\,.
\end{alignat*}
Furthermore, note that $\beta|\gamma$ really is an evaluation since
\begin{align*}
  \bbT^1\val(\beta|\gamma) = \bbT^1\val(\beta) \in \dom(\rho)\,.
\end{align*}

Finally, if $\beta \in \bbE_\delta(\rho,\bbT^1)$, for some limit ordinal~$\delta$,
then there is some $\alpha < \delta$ with $\beta \in \bbE_\alpha(\rho,\bbT^1)$.
Hence, the claim follows by inductive hypothesis.
\end{proof}

\begin{Thm}\label{Thm: simple evaluations imply expansion}
Let $\bbT^0 \subseteq \bbT^1 \subseteq \bbT$ be submonads.
Every $\bbT^0$-algebra~$\frakA$ with essentially unique $\rho$-evaluations
has a $\bbT^1$-expansion $\langle A,\pi_+\rangle$ such that
\begin{align*}
  \pi_+ \circ \term = \val\,.
\end{align*}
\end{Thm}
\begin{proof}
For $t \in \bbT^1 A$, we define
\begin{align*}
  \pi_+(t) := \val(\gamma)\,, \quad\text{for some } \gamma \in \term^{-1}(t)\,.
\end{align*}
As $\rho$-evaluations are essentially unique, it does not matter which
evaluation~$\gamma$ we choose. Hence, this function is well-defined.
We claim that $\frakA_+ := \langle A,\pi_+\rangle$ is the desired $\bbT^1$-expansion of~$\frakA$.

The equation $\pi_+ \circ \term = \val$ follows immediately from the definition of~$\pi_+$.
Since every tree $t \in \bbT^0 A \subseteq \bbE_1(\rho,\bbT^1)$ is its own evaluation,
we further have
\begin{align*}
  \pi_+(t) = \val_1(t) = \rho(t) = \pi(t)\,, \quad\text{for } t \in \bbT^0 A\,.
\end{align*}
Consequently, $\pi_+$~is an extension of~$\pi$ and it remains to check
the axioms of a $\bbT^1$-algebra.

For the unit law, we have $\pi_+(\sing(a)) = \pi(\sing(a)) = a$ since $\sing(a) \in \bbT^0 A$.
For associativity, let $t \in \bbT^1\bbT^1 A$.
Then there exists a tree of $\rho$-evaluations $\gamma \in \bbT^1\bbE(\rho,\bbT^1)$ such that
$t = \bbT^1\term(\gamma)$.
Furthermore, we can fix an evaluation $\beta \in \bbE(\rho,\bbT^1)$ of the tree
$\bbT^1\val(\gamma)$.
Let $\beta|\gamma$~be the $\rho$-evaluation from Lemma~\ref{Lem: gluing simple evaluations}.
Then
\begin{align*}
  (\pi_+ \circ \Flat)(t)
  &= (\pi_+ \circ \Flat \circ \bbT^1\term)(\gamma) \\
  &= (\pi_+ \circ \term)(\beta|\gamma) \\
  &= \val(\beta|\gamma) \\
  &= \val(\beta) \\
  &= (\pi_+ \circ \term)(\beta) \\
  &= (\pi_+ \circ \bbT^1\val)(\gamma) \\
  &= (\pi_+ \circ \bbT^1\pi_+ \circ \bbT^1\term)(\gamma)
   = (\pi_+ \circ \bbT^1\pi_+)(t)\,.
\end{align*}
\upqed
\end{proof}

The theorem tells us how to use evaluations to construct expansions.
Let us see next where the limits of this method are.
\begin{Prop}\label{Prop: val = pi o term}
Let $\frakA = \langle A,\pi\rangle$ be a $\bbT^0$-algebra with a $\bbT^1$-expansion
$\frakA_+ = \langle A,\pi_+\rangle$, and
suppose that $\rho : S \to A$ is the restriction of~$\pi_+$ to some set $S \supseteq \bbT^0A$.
Then
\begin{align*}
  \val(\gamma) = (\pi_+ \circ \term)(\gamma)\,,
  \quad\text{for every $\rho$-evaluation } \gamma\,.
\end{align*}
\end{Prop}
\begin{proof}
We prove that $\val_\alpha = \pi_+\circ\term_\alpha$ by induction on~$\alpha$.
For $\alpha = 0$, we have
\begin{align*}
  \val_0(\gamma) = \gamma = \pi_+(\sing(\gamma)) = \pi_+(\term_0(\gamma))\,,
  \quad\text{for } \gamma \in \bbE_0(\rho,\bbT^1) = A\,.
\end{align*}
For the successor step, suppose that the equation holds for~$\alpha$ and consider some
$\gamma \in \bbE_{\alpha+1}(\rho,\bbT^1) \setminus \bbE_\alpha(\rho,\bbT^1)$. Then
\begin{align*}
  \val_{\alpha+1}(\gamma)
  &= \rho(\bbT^1\val_\alpha(\gamma)) \\
  &= \rho(\bbT^1(\pi_+ \circ \term_\alpha)(\gamma)) \\
  &= \pi_+(\bbT^1\pi_+(\bbT^1\term_\alpha(\gamma))) \\
  &= \pi_+(\Flat(\bbT^1\term_\alpha(\gamma))) \\
  &= \pi_+(\term_{\alpha+1}(\gamma))\,.
\end{align*}
Finally, for a limit ordinal~$\alpha$, the claim follows immediately from
the inductive hypothesis.
\end{proof}
\begin{Cor}\label{Cor: expansions and evaluations}
Let\/ $\bbT^0 \subseteq \bbT^1 \subseteq \bbT$ and let\/ $\frakA$~be a\/ $\bbT^0$-algebra.
    \begin{enuma}
\item If\/ $\frakA$ has several different\/ $\bbT^1$-expansions,
  there exist trees $t \in \bbT^1 A$ without a simple\/ $\bbT^0$-evaluation.
\item If\/ $\frakA$ has simple\/ $\bbT^0$-evaluations for\/~$\bbT^1$ that
  are not essentially unique, it has no\/ $\bbT^1$-expansion.
\end{enuma}
\end{Cor}

\begin{Exam}
Before using these results to study thin trees, let us quickly recall
the results of~\cite{CartonColcombetPuppis18} about countable linear orders.
We denote by~$\bbC A$ the (unsorted) set of all countable $A$-labelled linear orders
and by $\bbC^\reg A \subseteq \bbC A$ the subset of all \emph{regular} linear orders.
By definition, a linear order is regular if it can be denoted by a finite term using the
following operations\?: \textsc{(i)}~constants for singletons, \textsc{(ii)}~binary
ordered sums, \textsc{(iii)}~multiplication by $\omega$ and $\omega^\op$ ($\omega$~with the
reverse ordering), and \textsc{(iv)}~dense shuffles.
In~\cite{CartonColcombetPuppis18} (Propositions 3.8 and 3.9) it is shown that
every finite $\bbC^\reg$-algebra has essentially unique $\bbC^\reg$-evaluations for~$\bbC$.
This fact can be used to prove the following results (for the proofs,
see~\cite{CartonColcombetPuppis18,Bojanczyk20,BlumensathLN2}).
\begin{itemize}
\item $\bbC^\reg \subseteq \bbC$ is dense over the class of all finite $\bbC$-algebras
  (Theorem~3.14 of~\cite{Bojanczyk20}).
\item Every finite $\bbC^\reg$-algebra has a unique $\bbC$-expansion
  (Theorem~3.11 of~\cite{CartonColcombetPuppis18}).
\item Every finite $\bbC$-algebra is $\MSO$-definable
  (Theorem~5.1 of~\cite{CartonColcombetPuppis18}).
\item A language $K \subseteq \bbC\Sigma$ of countable linear orders is $\MSO$-definable
  if, and only if, it is recognised by some finite $\bbC$-algebra
  (Theorems 4.1~and~5.1 of~\cite{CartonColcombetPuppis18}).
\item Every $\MSO$-definable language $K \subseteq \bbC\Sigma$ is recognised by a finite
  $\bbC$-algebra (Theorem~3.12 of~\cite{Bojanczyk20}).
  \qedhere
\end{itemize}
\end{Exam}

\subsection{Thin trees}   
\label{Sect:thin trees}

As an application of simple evaluations we consider thin trees, where we can use the
Theorem of Ramsey and other tools from semigroup theory.
Note that, with every $\bbT^\fin$-algebra~$\frakA$, we can associate the semigroup
with universe~$A_{\{z\}}$ (for some fixed variable~$z$ whose choice does not matter)
and where the product is defined by
\begin{align*}
  a \cdot b := a(b(z))\,, \quad\text{for } a,b \in A_{\{z\}}\,.
\end{align*}
If $\frakA$~is a $\bbT^\wilke$-algebra, this semigroup can be expanded to a Wilke algebra
$\langle A_{\{z\}},A_\emptyset\rangle$ by setting
\begin{align*}
  a \cdot c := a(c)
  \qtextq{and}
  a^\omega := \pi(t_a)\,,
  \quad\text{for } a \in A_{\{z\}} \text{ and } c \in A_\emptyset\,,
\end{align*}
where $t_a$~is an infinite path each vertex of which is labelled by~$a$.
Finally, if $\frakA$~is a $\bbT^\thin$-algebra, we obtain an $\omega$-semigroup with
infinite product
\begin{align*}
  \pi(a_0,a_1,\dots) := a_0(a_1(\cdots))\,,
  \quad\text{for } a_i \in A_{\{z\}}\,.
\end{align*}

We start by generalising the fact that every finite Wilke algebra has a unique expansion
to an $\omega$-semigroup to the monads $\bbT^\wilke \subseteq \bbT^\thin$.
\begin{Prop}\label{Prop: unique Twilke-evaluations}
Every finitary $\bbT^\wilke$-algebra has essentially unique simple $\bbT^\wilke$-eval\-u\-ations
for~$\bbT^\thin$.
\end{Prop}
\begin{proof}
Let $\frakA$~be a finitary $\bbT^\wilke$-algebra and $t \in \bbT^\thin A$ a thin tree.
We construct the desired simple evaluation of~$t$ by induction on the Cantor-Bendixson
rank~$\alpha$ of~$t$.
(Recall that the Cantor-Bendixson rank of a tree~$t$ is the least ordinal~$\alpha$ such that
$\partial^{\alpha+1}(t)$ is empty, where $\partial(t)$~denotes the tree obtained from~$t$
by removing every subtree with only finitely many infinite branches,
and $\partial^\alpha$~is the $\alpha$-th iteration of~$\partial$.
One can show that such an ordinal~$\alpha$ exists if, and only if, the given tree~$t$ is thin.)

By inductive hypothesis, every subtree~$t|_v$ of rank less than~$\alpha$ has a
simple evaluation $\gamma_v \in \bbE(\frakA,\bbT^\thin)$.
Let $s$~be tree obtained from~$t$ by replacing every such subtree~$t|_v$ by $\val(\gamma_v)$.
It is sufficient to find a simple evaluation of~$s$. Then we can use the glueing operation
from Lemma~\ref{Lem: gluing simple evaluations} to construct the desired evaluation of~$t$.

By construction, $s$~has only finitely many infinite branches.
We distinguish three cases.

\textsc{(i)}
If $s$~is finite, it is its own evaluation.

\textsc{(ii)}
Next, suppose that $s$~has a single infinite branch.
By the Theorem of Ramsey, we can find a factorisation $s = p_0p_1p_2\dots$
such that $\pi(p_i) = \pi(p_j)$, for all $i,j > 0$.
As each factor~$p_i$ is finite, we obtain simple evaluations~$\beta_i$ of~$p_i$ by~\textsc{(i).}
The path $\rho := \pi(p_0),\pi(p_1),\pi(p_2),\dots$ is of the form $ae^\omega$ for
$a := \pi(p_0)$ and $e := \pi(p_1)$. In particular, it is regular.
Let $\beta_*$~be the path $\beta_0,\beta_1,\beta_2,\dots$.
Then
\begin{align*}
  \bbT^\thin\val(\beta_*) = aeee\cdots \in \bbT^\reg A
  \qtextq{and}
  \term(\beta_*) = p_0p_1p_2\cdots = s\,.
\end{align*}
Hence, $\beta_*$~is the desired simple evaluation of~$s$.

\textsc{(iii)} Finally, suppose that $s$~has at least two infinite branches.
Then we can factorise~$s$ into a finite prefix and finitely many trees with a single infinite
branch. By~\textsc{(i)}~and~\textsc{(ii),} each of these factors has a simple evaluation.
Let~$\beta$~be the finite tree consisting of these evaluations.
Then $\beta$~is a simple evaluation of~$s$.
\end{proof}
Using Lemma~\ref{Lem: evaluations imply denseness} we obtain the following corollary.
\begin{Cor}
$\bbT^\wilke \subseteq \bbT^\thin$ is dense over the class of all finitary $\bbT^\thin$-algebras.
\end{Cor}
\begin{Cor}\label{Cor: Twilke has unique Tthin-expansion}
Every finitary $\bbT^\wilke$-algebra has a unique $\bbT^\thin$-expansion.
\end{Cor}

It follows that the step from a $\bbT^\wilke$-algebra to a $\bbT^\thin$-expansion is fairly well
understood. The inclusion $\bbT^\fin \subseteq \bbT^\wilke$ is slightly more complicated
since expansions are no longer unique.
\begin{Def}
Let $\frakA$~be a $\bbT^\wilke$-algebra. The corresponding \emph{$\omega$-power operation}
${-}^\omega : A_{\{z\}} \to A_\emptyset$ is defined by
\begin{align*}
  a^\omega := \pi(aaa\dots)\qquad
  \text{(an infinite path labelled~$a$)}\,.
\end{align*}
\upqed
\end{Def}
\begin{Prop}\label{Prop: fin to wilke expansions}
Let\/ $\frakA = \langle A,\pi\rangle$ be a finitary\/ $\bbT^\fin$-algebra.
The function mapping a\/ $\bbT^\wilke$-expansions of\/~$\frakA$ to the
corresponding $\omega$-power operation provides
a bijection between all\/ $\bbT^\wilke$-expansions of\/~$\frakA$ and
all functions ${-}^\omega : A_{\{z\}} \to A_\emptyset$ satisfying the equations
\begin{align*}
  (ab)^\omega = a(ba)^\omega
  \qtextq{and}
  (a^n)^\omega = a^\omega,
  \quad\text{for all } a,b \in A_{\{z\}}\,.
\end{align*}
\end{Prop}
\begin{proof}
Clearly the $\omega$-power operation $a \mapsto a^\omega$ associated with a
$\bbT^\wilke$-expansion $\frakA_+ = \langle A,\pi_+\rangle$ satisfies the two axioms above
since the product~$\pi_+$ is associative.
It therefore remains to show that the correspondence is bijective.

Note that every tree $t \in \bbT^\wilke A$ is the unravelling of a finite graph
all of which strongly connected components are either singletons or induced cycles
(cycles without any additional edges).

For injectivity, suppose that there are two expansions
$\frakA_0 = \langle A,\pi_0\rangle$ and $\frakA_1 = \langle A,\pi_1\rangle$ of~$\frakA$
with the same associated $\omega$-power.
Let $t \in \bbT^\wilke A$ be the unravelling of a graph~$\frakG$ with
$n$~strongly connected components. We prove that
\begin{align*}
  \pi_0(t) = \pi_1(t)
\end{align*}
by induction on~$n$.

Since both products agree on finite trees and we can write every tree $s \in \bbT^\wilke_\xi A$
as $s = p(\bar u,\bar x)$ where $\bar x$~is an enumeration of~$\xi$, $p \in \bbT^\fin A$ is
finite, and each $u_i \in \bbT^\wilke_\emptyset A$ is a tree without variables,
it is sufficient to prove the claim for $t \in \bbT^\wilke_\emptyset A$.

Let $C$~be the strongly connected component of~$\frakG$ containing the root of~$t$.
First, suppose that $C$~is a singleton that is not a cycle.
Then $t = a(s_0,\dots,s_{m-1})$, for some $a \in A$ and $s_i \in \bbT^\wilke A$.
Hence, it follows by inductive hypothesis that
\begin{align*}
  \pi_0(t) = a\bigl(\pi_0(s_0),\dots,\pi_0(s_{m-1})\bigr)
           = a\bigl(\pi_1(s_0),\dots,\pi_1(s_{m-1})\bigr)
           = \pi_1(t)\,.
\end{align*}

Next, suppose that $C$~is a cycle.
For every vertex $v \in \dom_0(t) \setminus C$, it follows by inductive hypothesis that
\begin{align*}
  \pi_0(t|_v) = \pi_1(t|_v)\,.
\end{align*}
Replacing these subtrees by their product (and merging the obtained leaves into
their predecessor) we may assume that~$\frakG$ has a single strongly connected component~$C$.
Since $t$~is thin, this component forms a cycle and there exists a finite path~$p$
such that $t = p^\omega$. Consequently,
\begin{align*}
  \pi_0(t) = \pi(p)^\omega = \pi_1(t)\,.
\end{align*}

For surjectivity, suppose that ${-}^\omega : A_{\{z\}} \to A_\emptyset$ is an $\omega$-power
operation. To construct an expansion $\frakA^+ = \langle A,\pi_+\rangle$ of~$\frakA$,
we define a function~$\hat\pi$ mapping every finite graph~$\frakG$ whose unravelling
belongs to $\bbT^\wilke A$ to some element $\hat\pi(\frakG) \in A$.
Then we set $\pi_+(t) := \hat\pi(\frakG)$ for some~$\frakG$ with unravelling~$t$.

We define $\hat\pi(\frakG)$ by induction on the number of connected components of~$\frakG$.
Let $C$~be the strongly connected component of~$\frakG$ containing the root.
For every vertex $v \notin C$ that is not a variable, we can compute $\hat\pi(\frakG|_v)$
by inductive hypothesis, where $\frakG|_v$~denotes the subgraph of~$\frakG$ consisting of
all vertices reachable from~$v$.
Let $\frakG'$~be the graph obtained from~$\frakG$ by replacing every such subgraph~$\frakG|_v$
by its product $\hat\pi(\frakG|_v)$ and then merging the resulting leaves into their predecessor.
If $\frakG'$~is a tree with a root with label~$a$ attached to which are variables~$\bar x$,
we set
\begin{align*}
  \hat\pi(\frakG) := a(\bar x)\,.
\end{align*}
Otherwise, $\frakG'$~consists of a cycle. (Since the unravelling of~$\frakG'$ is a linear
tree, it follows that $\frakG'$~cannot contain any variables.)
Let $p \in \bbT^\fin A$ be the finite path obtained
from~$\frakG'$ by removing the edge leading to the root and replacing it by a variable.
We set
\begin{align*}
  \hat\pi(\frakG) := \pi(p)^\omega\,.
\end{align*}

It remains to check that that the value of~$\hat\pi(\frakG)$ only depends on the unravelling
of~$\frakG$, that the resulting function~$\pi_+$ satisfies the axioms of a $\bbT^\wilke$-algebra,
and that the associated $\omega$-operation coincides with the given one.
We start with the latter.
Let $a \in A_{\{z\}}$ and let $\frakG$~be a graph with unravelling $aaa\ldots$.
Then $\frakG$~is a path of length $m<\omega$ leading to a cycle of length $n < \omega$.
It follows that
\begin{align*}
  \hat\pi(\frakG) = a^m \cdot \pi(a^n)^\omega = a^m \cdot (a^n)^\omega = a^\omega,
\end{align*}
as desired.

Next let us show that $\hat\pi(\frakG) = \hat\pi(\frakH)$, for graphs $\frakG$~and~$\frakH$
with the same unravelling.
Given a graph~$\frakG$ and a vertex~$v$, we define an equivalence relation~$\sim_\frakG$
on its set of vertices by
\begin{align*}
  u \sim_\frakG v \quad\defiff\quad
  \text{the unravellings of } \frakG|_u \text{ and } \frakG|_v \text{ are isomorphic.}
\end{align*}
Since two graphs $\frakG$~and~$\frakH$ have the same unravelling if, and only if,
their quotients by, respectively, $\sim_\frakG$~and~$\sim_\frakH$ are isomorphic,
it is sufficient to show that
\begin{align*}
  \hat\pi(\frakG) = \hat\pi(\frakG/{\sim_\frakG})\,,
  \quad\text{for every graph } \frakG\,.
\end{align*}

We prove the claim by induction on the number of strongly connected components of~$\frakG$.
Let $C$~be the strongly connected component of~$\frakG$ containing the root.
We distinguish several cases.

First, suppose that $C$~is a union of $\sim_\frakG$-classes.
Then the image $C/{\sim_\frakG}$ of~$C$ under the quotient map forms a strongly connected
component of~$\frakG/{\sim_\frakG}$.

If $C = \{v\}$ is a singleton that is not a cycle, we argue as follows.
Let $u_0,\dots,u_{n-1}$ be the non-variable successors of~$v$ and let $y_0,\dots,y_{m-1}$
be the variables attached to~$v$.
By inductive hypothesis, we have
\begin{align*}
  \hat\pi(\frakG|_{u_i}) = \hat\pi((\frakG/{\sim_\frakG})|_{u_i})\,.
\end{align*}
Let $a$~be the label of~$v$ and set $c_i := \hat\pi(\frakG|_{u_i})$, for $i < n$.
It follows that
\begin{align*}
  \hat\pi(\frakG) = a(c_0,\dots,c_{n-1},y_0,\dots,y_{m-1}) = \hat\pi(\frakG/{\sim_\frakG})\,.
\end{align*}

If $C$~(and therefore~$C/{\sim_\frakG}$) form cycles, the product $\hat\pi(\frakG/{\sim_\frakG})$
is computed by \textsc{(i)}~evaluating all products $\hat\pi((\frakG/{\sim_\frakG})|_v)$
for $v \notin C/{\sim_\frakG}$\?; \textsc{(ii)}~merging the resulting values into the label
of the predecessor of~$v$\?; and \textsc{(iii)}~computing~$p^\omega$, where $p$~is the path
obtained by~\textsc{(ii).}
By inductive hypothesis it follows that, when performing the same procedure for~$\frakG$,
we obtain a path of the form~$p^m$, for some $0 < m < \omega$. Hence,
\begin{align*}
  \hat\pi(\frakG) = (p^m)^\omega = p^\omega = \hat\pi(\frakG/{\sim_\frakG})\,.
\end{align*}

It remains to consider the case where there are vertices outside of~$C$
$\sim_\frakG$-equivalent to some vertex of~$C$.
Then the strongly connected component~$D$ of~$\frakG/{\sim_\frakG}$ containing the root
forms a cycle.

Suppose that $C = \{v\}$ is a singleton that is not a cycle.
Let $u_0,\dots,u_{n-1}$ be the successors of~$v$.
(Note that none of them can be variables since there is some vertex $v' \sim_\frakG v$,
but each variable occurs only once in~$\frakG$.)
Note that there can be only one successor~$u_i$ whose $\sim_\frakG$-class belongs to~$D$ since,
otherwise, the unravelling of~$\frakG$ would not be thin.
We choose the ordering of $u_0,\dots,u_{n-1}$ such that this vertex is~$u_0$.
Let $a$~be the label of~$v$ and set $c_i := \hat\pi(\frakG|_{u_i})$.
By inductive hypothesis, we have
\begin{align*}
  \hat\pi((\frakG/{\sim_\frakG})|_{u_i}) = c_i\,.
\end{align*}
Let $p$~be the path such that
\begin{align*}
  \hat\pi(\frakG/{\sim_\frakG}) = p^\omega\,.
\end{align*}
Then the first label of~$p$ is $b := a(z,c_1,\dots,c_{n-1})$.
Hence, $p = bq$, for some path~$q$, and it follows by inductive hypothesis that
\begin{align*}
  c_0 = (qb)^\omega.
\end{align*}
Consequently, we have
\begin{align*}
  \hat\pi(\frakG)
  &= a(c_0,\dots,c_{n-1}) \\
  &= a\bigl((qb)^\omega,c_1,\dots,c_{n-1}\bigr) \\
  &= b(qb)^\omega \\
  &= (bq)^\omega \\
  &= \hat\pi(\frakG/{\sim_\frakG})\,.
\end{align*}

Finally, suppose that $C$~is a cycle. We claim that this case cannot occur.
For a contradiction, suppose otherwise. Then there is some vertex $v \in C$
with two different successors $u,u'$ from that we can reach vertices
$w$~and~$w'$, respectively, with $w \sim_\frakG w'$ and such that
the corresponding $\sim_\frakG$-class belongs to~$D$.
Since $D$~is strongly connected,
it follows that we can reach from $w$~and~$w'$ vertices $\hat v$~and~$\hat v'$ with
$\hat v \sim_\frakG v \sim_\frakG \hat v'$.
This implies that the unravelling of~$\frakG$ is not thin.
A~contradiction.

It remains to check the axioms for a $\bbT^\wilke$-algebra.
It follows directly by definition that
\begin{align*}
  \pi_+(\sing(a)) = a\,.
\end{align*}

For associativity, fix $t \in \bbT^\wilke\bbT^\wilke A$ and let $\frakG$~be
a finite graph with unravelling~$t$.
For each vertex~$v$ of~$\frakG$, we fix a finite graph~$\frakH_v$ with unravelling~$t(v)$.
Then $\Flat(t)$ is the unravelling of the graph obtained from the disjoint union
of all~$\frakH_v$ by adding edges according to~$\frakG$.
Let us call this graph~$\frakK$.
We will prove that
\begin{align*}
  \hat\pi(\frakK) = \hat\pi(\frakG')\,,
\end{align*}
where $\frakG'$~is the graph obtained from~$\frakG$ by replacing each label by the
corresponding product $\hat\pi(\frakH_v)$.
Then it follows that
\begin{align*}
  \pi_+(\Flat(t)) = \pi_+(\bbT^\wilke\pi_+(t))\,,
\end{align*}
as desired.

We proceed by induction on the number of vertices of~$\frakK$.
Let $C$~be the strongly connected component of~$\frakG$ containing the root.
For every vertex $v \in \dom_0(\frakG) \setminus C$, it follows by inductive hypothesis that
\begin{align*}
  \hat\pi(\frakK|_{\mu(v)}) = \hat\pi(\frakG'|_v)\,,
\end{align*}
where $\mu : \dom_0(\frakG) \to \dom_0(\frakK)$ is the function
mapping each vertex~$v$ of~$\frakG$ to the root of~$\frakH_v$.
Note that the products $\hat\pi(\frakK)$ and $\hat\pi(\frakG')$
can be computed by \textsc{(i)}~replacing all the subgraphs $\frakK|_{\mu(v)}$ and $\frakG'|_v$,
for $v \in \dom_0(\frakG) \setminus C$, by their respective products,
\textsc{(ii)}~merging the resulting leaves into their parents, and then
\textsc{(iii)}~computing the products of the remaining graphs.
(For the graph~$\frakG'$, this statement follows immediately from the definition of~$\hat\pi$\?;
for~$\frakK$, it follows by a straightforward induction on the size of~$\frakK$.)
We may therefore assume that $\frakG$~consists of a single strongly connected component~$C$
(plus possibly some variables in case $C$~is a singleton).
If $C = \{v\}$ is a singleton which is not a cycle, we have $\frakK = \frakH_v$ and
\begin{align*}
  \hat\pi(\frakK) = \hat\pi(\frakH_v) = \hat\pi(\frakG')\,.
\end{align*}
Hence, suppose that $C$~is a cycle.
Each graph~$\frakH_v$ consists of a finite path~$p_v$ leading to a variable to which are
possibly attached additional graphs without variables.
By inductive hypothesis, associativity holds for these subgraphs.
Again, replacing each such subgraph by its product, we may assume that $\frakH_v$~is equal
to~$p_v$.
Consequently, $\frakK$~is a single path consisting of the concatenation of all~$p_v$,
while $\frakG'$~is the path labelled by the products $\pi(p_v)$.
The product of these two paths is the same.
\end{proof}

Combining this result with Corollary~\ref{Cor: Twilke has unique Tthin-expansion} we
obtain the corresponding statement for $\bbT^\thin$-expansions.
\begin{Cor}\label{Cor: fin to thin expansions}
Let\/ $\frakA = \langle A,\pi\rangle$ be a finitary\/ $\bbT^\fin$-algebra.
There exists a bijection between all\/ $\bbT^\thin$-expansions of\/~$\frakA$ and
all functions ${-}^\omega : A_{\{z\}} \to A_\emptyset$ satisfying the axioms of a Wilke algebra.
\end{Cor}
\begin{Cor}
Every\/ $\bbT^\thin$-algebra\/~$\frakA$ is uniquely determined by
\textup{\textsc{(i)}}~its\/ $\bbT^\fin$-reduct and
\textup{\textsc{(ii)}}~the associated $\omega$-semigroup.
\end{Cor}

\subsection{Evaluations with merging}   

When we try to go beyond~$\bbT^\thin$ our machinery breaks down since we cannot use the
results for semigroups anymore.
The following counterexample shows that a na\"ive generalisation of our definitions does not
work.
\begin{Lem}
There exists an $\MSO$-definable $\bbT^\reg$-algebra~$\frakA$ and a tree $t \in \bbT A$ that
has no simple $\bbT^\reg$-evaluation.
\end{Lem}
\begin{proof}
Let $\frakA$~be the $\bbT^\reg$-reduct of the Boja\'nczyk-Klin algebra from the example
on page~\pageref{Ex:BK-algebra}.
Then the claim follows immediately from Corollary~\ref{Cor: expansions and evaluations}\,(a).
Nevertheless we give an explicit proof to see what exactly is going wrong.
Set $\Delta := \bbT^\fin\{a,b\}$ and recall that $\Delta \subseteq A$.
We will prove by induction on~$\alpha$ that
\begin{align*}
  t \notin \rng \term_\alpha\,,
  \qquad&\text{for all } t \in \bbT\Delta \text{ where every subtree has vertices of} \\
  &\text{arbitrarily high arity.}
\end{align*}

For a contradiction, suppose otherwise.
Let $\alpha$~be the minimal ordinal such that there is some simple evaluation
$\gamma \in \bbE_\alpha(\frakA,\bbT)$ where every subtree of $\term(\gamma)$
has vertices of arbitrarily high arity.
If $\alpha = 0$, then $\term(\gamma) = \sing(a)$ in contradiction to our choice of~$\gamma$.
Hence, $\alpha = \beta + 1$, for some~$\beta$.
Fix $v \in \dom(\gamma)$.
Note that every subtree~$s$ of $\term_\beta(\gamma(v))$ has a simple evaluation in
$\bbE_\beta(\frakA,\bbT)$ which, by inductive hypothesis, means that $s$~has a subtree
where the arity of the vertices is bounded.
We claim that this implies that $t_v := \term(\gamma(v))$ is finite.
Suppose otherwise. Since $t_v$~has only finitely many variables,
it has some infinite subtree~$s$ without variables.
But $s$~is also a subtree of $\term(\gamma)$.
By choice of~$\gamma$ this implies that the arities of the vertices of~$s$ are unbounded.
A~contradiction.

Hence, we have $\term(\gamma(v)) \in \bbT^\fin\Delta$, which implies that
\begin{align*}
  \val(\gamma(v)) = \pi(\term(\gamma(v))) = \term(\gamma(v))\,.
\end{align*}
Furthermore, $\term(\gamma(v))$ being finite its arity is at least as high as the maximal
arity of a vertex in $\dom(\gamma(v))$.
It follows that, for every $n < \omega$, there is some $v \in \dom(\gamma)$ such that
$\val(\gamma(v))$ has arity at least~$n$.
But $\gamma \in \bbE_{\alpha+1}(\frakA,\bbT)$ implies that
$\bbT\val(\gamma) \in \bbT^\reg A$.
In particular, $\bbT\val(\gamma)$ uses only finitely many different labels.
This implies that their arity is bounded. A~contradiction.
\end{proof}

A closer look at the above proof reveals two possible reasons making simple evaluations
impossible. Firstly, our counterexample uses a tree with infinitely many different labels.
It still might be possible that trees with only finitely many different labels always have
simple evaluations.
Secondly, we made essential use of the fact that every factor of an infinite binary tree
has a subtree that is itself an infinite binary tree.
To be able to use factorisations of trees into pieces that are significantly simpler,
we will probably have to allow more general factors, which then necessarily have infinitely
many variables. Unfortunately, it is hard to combine these two modifications since factors with
infinitely many variables usually give rise to infinitely many different elements
of the algebra. What seems to be missing is some technique that, given a tree with infinitely
many different labels, allows us to bound their arity by merging different variables
(e.g., replacing $a(x,y,z)$ by, say, $a(x,x,z)$).

This observation leads to the following attempt to allow for evaluations where variables
are merged. To make our definitions precise we need a bit of terminology.
First, as we want to identify variables, we need to work in~$\bbT^\times$ instead of~$\bbT$.
We also need a set of labels telling us which variables to identify.
\begin{Def}
    \begin{enuma}
\item For a tree $t \in \bbT^\times_\zeta A$ and a function $\sigma : \zeta \to \xi$,
we denote by ${}^\sigma t \in \bbT^\times_{\xi_0} A$
the tree obtained from~$t$ by replacing every variable~$z$ by~$\sigma(z)$
(where $\xi_0 \subseteq \xi$ is the range of~$\sigma$).

If $\bbT^0 \subseteq \bbT^\times$ is closed under the operation ${}^\sigma{-}$
we can extend this operation to $\bbT^0$-algebras~$\frakA$ by setting
\begin{align*}
  {}^\sigma a := \pi({}^\sigma\sing(a))\,,
  \quad\text{for } a \in A_\zeta\,.
\end{align*}

\item
For a sort $\xi \in \Xi$, we set $\Gamma(\xi) := (\Gamma_\zeta(\xi))_{\zeta \in \Xi}$ where
\begin{align*}
  \Gamma_\zeta(\xi) := \set{ \sigma }{ \sigma : \zeta \to \xi }\,.
\end{align*}
\upqed
\end{enuma}
\end{Def}

Given a tree $t \in \bbT^\times A$ we can choose some sort $\xi \in \Xi$ and
functions $\sigma_v \in \Gamma(\xi)$, for every $v \in \dom(t)$, and then
replace every label $t(v)$ by~${}^{\sigma_v} t(v)$.
The problem is that the resulting tree is not well-formed since the sorts do not match anymore.
For instance, in the tree $a(b,c)$ with $a = a(z_0,z_1)$ we can replace $z_0$~and~$z_1$
by the same variable~$x$. This produces the label $a' := a(x,x)$ of arity~$\{x\}$.
Consequently, we need to produce a tree where the corresponding vertex has a single
successor instead of two. Given the tree $a(b,c)$ the only obvious choices for such
a tree would be $a'(b)$ or $a'(c)$. This idea can be generalised as follows.
\begin{Def}
Let $\frakA$~be a $\bbT^0$-algebra where $\bbT^0 \subseteq \bbT^\times$ is closed under the
operations~${}^\sigma{-}$ and let $p : \Gamma(\xi) \times A \to \Gamma(\xi)$ and
$q : \Gamma(\xi) \times A \to A$ be the two projections.

    \begin{enuma}
\item A \emph{condensation} of a tree $t \in \bbT^\times A$ is a tree
$s \in \bbT^\times(\Gamma(\xi) \times A)$ such that
\begin{align*}
  t = \bbT^\times q(s)\,.
\end{align*}

\item
Let $s$~be a condensation of~$t$ and set $\sigma_v := p(s(v))$, for $v \in \dom_0(s)$.
A \emph{choice function} for~$s$ is a family $\mu = (\mu_v)_{v \in \dom_0(s)}$ of functions
$\mu_v : \rng \sigma_v \to \dom \sigma_v$ such that $\sigma_v \circ \mu_v = \id$.

\item
Let $s$~be a condensation of~$t$ and $\mu = (\mu_v)_v$ a choice function for~$s$.
We define the tree $s \mVert \mu \in \bbT^\times A$ as follows. For every $v \in \dom_0(s)$,
\begin{itemize}
\item we delete from~$s$ all subtrees~$s|_{\suc_x(v)}$ with $x \notin \rng \mu_v$,
\item for $x \in \rng \mu_v$, we change the $x$-successor of~$v$ to a $\sigma_v(x)$-successor,
  where $\sigma_v := p(s(v))$, and
\item we replace every label $s(v) = \langle\sigma,a\rangle$ by~${}^\sigma a$
  (which we consider to be an element of sort $\rng \sigma$).
  If $s(v)$~is a variable, we leave it unchanged.
  \qedhere
\end{itemize}
\end{enuma}
\end{Def}
\begin{Exam}
Let $t \in \bbT_\emptyset A$ be an infinite tree where all labels on the same level are equal,
and let $s \in \bbT(\Gamma(\{x\}) \times A)$ be the tree with the same domain as~$t$
such that
\begin{align*}
  s(v) = \langle\sigma_n,a_n\rangle\,, \quad\text{for every vertex $v$ with } \abs{v} = n\,,
\end{align*}
where $a_n = t(v) \in A_{\zeta_n}$ is the label on level~$n$ of~$t$ and
$\sigma_n : \zeta_n \to \{x\}$ is the function mapping all variables of~$a_n$ to
the same variable~$x$.
For every choice function~$\mu$ for~$s$, we obtain a path
\begin{align*}
  s \mVert \mu = b_0b_1\cdots
  \qtextq{with}
  b_n(x) := a_n(x,\dots,x)\,.
\end{align*}
\upqed
\end{Exam}

\begin{Rem}
Note that, in a tree of the form $s \mVert \mu$, every vertex has a sort
which is a subset of~$\xi$. In particular, the number of sorts used is finite.
\end{Rem}

We can produce well-formed trees $s \mVert \mu$ using a choice function~$\mu$.
But which one do we take\??
The easiest case is if all choice functions produce the same result (cf.~\cite{Puppis10}),
then it does not matter.
(A~more general construction will be presented further below.)
\begin{Def}
Let $\bbT^0 \subseteq \bbT^1 \subseteq \bbT^\times$ be submonads such that
$\bbT^0$~is closed under all operations~${}^\sigma{-}$, and let $\frakA$~be a $\bbT^0$-algebra.

    \begin{enuma}
\item
A condensation $s \in \bbT^\times(\Gamma(\xi) \times A)$~is \emph{uniform} if
\begin{align*}
  s \mVert \mu = s \mVert \mu'\,,
  \quad\text{for all choice functions } \mu,\mu'\,.
\end{align*}

\item A \emph{uniform $\bbT^0$-condensation} of $t \in \bbT^1 A$ is a uniform condensation
$s \in \bbT^1(\Gamma(\xi) \times A)$ of~$t$ such that
\begin{align*}
  s \mVert \mu \in \bbT^0 A\,, \quad\text{for some/all choice functions } \mu\,.
\end{align*}

\item Let $\tau$~be a partial function mapping each tree~$t$ to some
uniform $\bbT^0$-condensation of~$t$ (if such a condensation exists).
We set
\begin{align*}
  \pi^\rmu_\tau(t) := \pi(\tau(t) \mVert \mu)\,,
  \quad\text{where } \mu \text{ is an arbitrary choice function}\,.
\end{align*}
If $\tau(t)$~is undefined, we let $\pi^\rmu_\tau(t)$~be undefined as well.
We call $\pi^\rmu_\tau$-evaluations \emph{$\bbT^0$-evaluations with uniform merging,}
and we denote the corresponding set by
\begin{align*}
  \bbE^{\rmu,\tau}_\alpha(\frakA,\bbT^1) := \bbE_\alpha(\pi^\rmu_\tau,\bbT^1)\,.
\end{align*}
\upqed
\end{enuma}
\end{Def}
\begin{Rem}
Since $\bbT^0$-evaluations with uniform merging are $\pi^\rmu_\tau$-evaluations,
Theorem~\ref{Thm: simple evaluations imply expansion} and Proposition~\ref{Prop: val = pi o term}
apply to them. Hence, we can use such more general evaluations to study $\bbT^1$-expansions.
A~similar statement holds for evaluations with consistent merging, which we will define below.
\end{Rem}

\begin{Exam}
The tree~$s$ from the preceding example is a uniform condensation of~$t$
with $s \mVert \mu \in \bbT^\thin A$.
For technical reasons, it is not a uniform $\bbT^\thin$-condensation of~$t$\?:
the set~$\bbT^\thin A$ is not closed under the operations ${-}^\sigma$.
Instead, $s$~is a uniform $\bbT^0$-condensation where $\bbT^0$~is the monad
`generated' by $\bbT^{\times\thin}$, that is,
the set~$\bbT^0 X$ is the closure of $\bbT^{\times\thin} X$ under $\Flat$.
\end{Exam}

\begin{Exam}
$\bbT^{\times\reg}$-evaluations with uniform merging were introduced in~\cite{Puppis10} where
they were used to derive decidability results for trees.
To do so Puppis considers trees $t \in \bbT^\times X$ such that (in our terminology),
for every $\MSO$-definable $\bbT^{\times\reg}$-algebra~$\frakA$ and every function
$\beta : X \to A$, the image~$\bbT^\times\beta(t)$ has an evaluation
$\gamma \in \bbE^{\rmu,\tau}_n(\frakA,\bbT^\times)$, for some number
$n < \omega$ independent of~$\beta$ and~$\frakA$.
The function~$\tau$ chooses condensations based on the runs of an automaton recognising
the product of~$\frakA$. (The details can be found in~\cite{Puppis10}.
A~similar construction is used at the beginning of the proof of
Lemma~\ref{Lem: split for definable algebras} below.)
Let us call such trees \emph{reducible.}

By induction on~$n$, we can transform every reducible tree~$t$ into a regular tree~$t_0$
with the same value as~$t$. Puppis considers reducible trees~$t$ where this transformation
is computable using a particular algorithm. (We again omit the details.)
Let us call such trees~\emph{effectively reducible.}
\cite{Puppis10}~contains the following results.
\begin{itemize}
\item Every deterministic tree in the Caucal hierarchy is effectively reducible.
\item The class of effectively reducible trees is closed under a number of natural operations.
\item Every effectively reducible tree has a decidable $\MSO$-theory.
  \qedhere
\end{itemize}
\end{Exam}

The key technical result of~\cite{Puppis10} is the following recipe of how
to evaluate products in an $\MSO$-definable $\bbT^\times$-algebra using evaluations
with uniform merging (cf.~Theorem~5 of~\cite{Puppis10}).
Intuitively, the following proposition states that every $\MSO$-definable property
of trees is also $\MSO$-definable in every given uniform $\bbT^\times$-condensation of them.
\begin{Prop}[Puppis]
Let\/ $\frakA$~be an\/ $\MSO$-definable\/ $\bbT^\times$-algebra and $\xi \in \Xi$ a sort.
There exists an\/ $\MSO$-definable\/ $\bbT^\times$-algebra\/~$\frakB$, a morphism
$\rho : \bbT^\times A \to \frakB$,
and\/ $\MSO$-formulae~$\varphi_a$, for $a \in A_\xi$, such that,
given a tree $T \in \bbT^\times_\xi\bbT^\times A$ and a uniform\/ $\bbT^\times$-condensation~$s$
of\/~$\bbT^\times\rho(T)$, we have
\begin{align*}
  \pi(\Flat(T)) = a \quad\iff\quad s \mVert \mu \models \varphi_a\,,
  \quad&\text{for all } a \in A_\xi \text{ and all choice} \\
  &\text{functions } \mu\,.
\end{align*}
\end{Prop}
\noindent
The proof uses similar techniques as that of Lemma~\ref{Lem: split for definable algebras} below.

Since evaluations with uniform merging generalise simple evaluations,
they allow us to decompose more trees. Unfortunately, there are still trees
left without an evaluation. We can generalise our evaluations even further
by not requiring that all choice functions lead to the same tree, but
only to one that is `equivalent'.
\begin{Def}
Let $\bbT^0 \subseteq \bbT^1 \subseteq \bbT^\times$ be submonads such that
$\bbT^0$~is closed under all operations~${}^\sigma{-}$. Let $\frakA = \langle A,\pi\rangle$
be a $\bbT^0$-algebra, and $t \in \bbT^1 A$.

    \begin{enuma}
\item A condensation $s \in \bbT^\times(\Gamma(\xi) \times A)$ is
\emph{$\pi$-consistent} if
\begin{alignat*}{-1}
  &s|_v \mVert \mu \in \bbT^0 A\,,
  &&\quad\text{for every choice function~$\mu$ of~$s|_v$ and each} \\
  &&&\quad\text{vertex } v \in \dom(s)\,, \\
  &\pi(s|_v \mVert \mu) = \pi(s|_v \mVert \mu')\,,
  &&\quad\text{for all choice functions~$\mu,\mu'$ of~$s|_v$ and each} \\
  &&&\quad\text{vertex } v \in \dom_0(s)\,.
\end{alignat*}

\item A $\pi$-consistent condensation $s \in \bbT^1(\Gamma(\xi) \times A)$ of~$t$
is also called a \emph{consistent $\bbT^0$-condensation} of~$t$.

\item Let $\tau$~be a partial function mapping each tree~$t$ to some
consistent $\bbT^0$-condensation of~$t$ (if such a condensation exists).
We set
\begin{align*}
  \pi^\rmc_\tau(t) := \pi(\tau(t) \mVert \mu)\,,
  \quad\text{where } \mu \text{ is an arbitrary choice function}\,.
\end{align*}
If $\tau(t)$~is undefined, we let $\pi^\rmc_\tau(t)$~be undefined as well.
We call $\pi^\rmc_\tau$-evaluations \emph{$\bbT^0$-evaluations with consistent merging,}
and we denote the corresponding set by
\begin{align*}
  \bbE^{\rmc,\tau}_\alpha(\frakA,\bbT^1) := \bbE_\alpha(\pi^\rmc_\tau,\bbT^1)\,.
\end{align*}
\upqed
\end{enuma}
\end{Def}

Clearly, consistent merging generalises uniform merging. While
$\bbT^{\times\reg}$-eval\-u\-ations with uniform merging seem to exist only in special cases,
our hope is that $\bbT^{\times\reg}$-evaluations with consistent merging always exist
(at least for $\MSO$-definable algebras).
At the moment we are only able to obtain partial results.
To do so we need a bit of terminology.
First, it is convenient to work with arbitrary directed graphs instead of just trees.
\begin{Def}
Let $A$~be a sorted set.

    \begin{enuma}
\item An \emph{$A$-labelled rooted graph}~$g$ is a countable directed graph
with a distinguished vertex~$r$, the \emph{root} of~$g$, where every
vertex is labelled by some element of~$A$ and where every edge is labelled by some variable
in such a way that, if a vertex~$v$ is labelled by $a \in A_\xi$, then $v$~has exactly
one out-going edge labelled~$x$, for every $x \in \xi$.
As usual, we treat an $A$-labelled rooted graph as a function $g : \dom(g) \to A$.
We denote by $\bbR_\xi A$ the set of all $(A + \xi)$-labelled rooted graphs
(where, as usual, the variables are considered to be elements of sort~$\emptyset$),
and we set $\bbR A := (\bbR_\xi A)_\xi$.

\item Given a graph $g \in \bbR\bbR A$, we denote by $\Flat(g)$ the graph obtained from
the disjoint union $\sum_{v \in \dom(g)} g(v)$ of all component graphs by
replacing, in every component $g(v)$, every $x$-labelled edge $u \to u'$ to
a vertex labelled by some variable~$z$ to an $x$-labelled edge $u \to w$ where $w$~is
the root of the the component~$g(v')$ for the $z$-successor~$v'$ of~$v$.
Then we delete all vertices labelled by a variable.
(Again, if $g(v)$~is just a variable~$z$, we consider it a $1$-vertex graph with label~$z$
and we do not delete this vertex.)

\item We denote by $\bbR^\thin A \subseteq \bbR A$ the set of all graphs whose unravelling
is a thin tree. \qedhere
\end{enuma}
\end{Def}

We start with a technical lemma which is based on the following variant of a condensation.
In a usual condensation we can only redirect edges to another successor of the same vertex.
Below we will need a variant where we can also redirect edges to vertices that are farther
away. We specify which destinations are allowed in such a redirection via
a labelling~$\sigma$ of the tree by numbers.
\begin{Def}
Let $t \in \bbT^\times_\xi A$ be a tree and $\sigma : \dom(t) \to [N]$ a function.

    \begin{enuma}
\item The \emph{$k$-th $\sigma$-parent}~$p^k_\sigma(v)$ of $v \in \dom(t)$ is the maximal
vertex~$u$ satisfying
\begin{align*}
  u \prec v \qtextq{and} \sigma(u) \geq k\,.
\end{align*}
If no such vertex exists, we set $p^k_\sigma(v) := \bot$ and we say that $v$~does not have
a $k$-th $\sigma$-parent.
For $k = \sigma(v)$, we omit the superscript and write just
$p_\sigma(v) := p^{\sigma(v)}_\sigma(v)$.
We define a relation $\approx_\sigma$ on~$\dom(t)$ by
\begin{align*}
  u \approx_\sigma v \quad\defiff\quad
  \sigma(u) = \sigma(v) \qtextq{and} p_\sigma(u) = p_\sigma(v)\,.
\end{align*}

\item We say that a tree $t' \in \bbT^\times A$ is a \emph{$\sigma$-rewiring} of~$t$ if
it is the unravelling of some graph $g \in \bbR_\xi A$ satisfying the following conditions.
\begin{itemize}
\item $\dom(g) \subseteq \dom(t)$ is prefix-closed and non-empty.
\item If $v$~is the $x$-successor of~$u$ in~$t$ and $u,v \in \dom(g)$, then
  $v$~is also the $x$-successor of~$u$ in~$g$.
\item If $v$~is the $x$-successor of~$u$ in~$g$
  and $v'$~is its $x$-successor in~$t$, then $v \approx_\sigma v'$.
\end{itemize}
We say that $g$~is the graph \emph{inducing} the $\sigma$-rewiring~$t'$. \qedhere
\end{enuma}
\end{Def}

\begin{Exam}
Let $t$~be the tree on the left and $\sigma$~the depicted labelling of~$t$.
Then the (unravelling of the) graph~$g$ on the right is a $\sigma$-rewiring of~$t$.
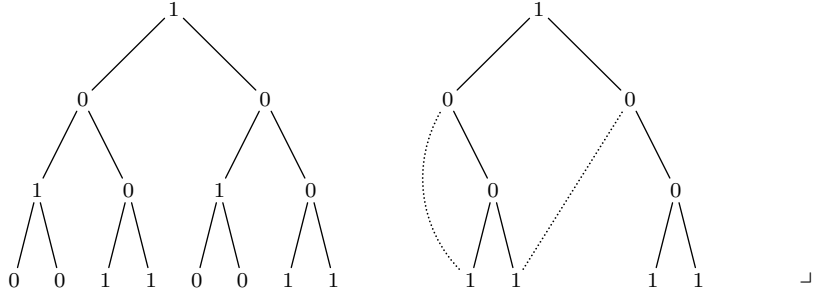
\begin{center}
\includegraphics{Expansion-6.mps}
\end{center}
\upqed
\end{Exam}
\begin{Lem}\label{Lem: every path contains p_sigma}
Let $t'$~be a $\sigma$-rewiring of~$t$ and let $\rho$~be a path in~$t'$.
    \begin{enuma}
\item If $\rho$~starts at the root and it contains a vertex~$v$, it also contains~$p_\sigma(v)$
  (if $p_\sigma(v) \neq \bot$).
\item Suppose that $\rho$~starts at~$u$, ends in~$v$, and that
  \mathindent=1em%
  \begin{align*}
    \bot \neq p_\sigma(x) \npreceq p_\sigma(u) \neq \bot\,,
    \quad\text{for all vertices } x \text{ of } \rho \text{ different from } u\,.
  \end{align*}
  Then $u \preceq v$.
\item There is no pair of vertices $x \preceq y$ such that $x$~is after~$y$ along the
  path~$\rho$.
\end{enuma}
\end{Lem}
\begin{proof}
In the following we will always use the notation
$\preceq$~and~$p_\sigma(v)$ with respect to the tree~$t$.
We use the notation $u \to v$ to indicate that, in~$t'$, there is an edge from~$u$
to~$v$.

    \begin{enuma}
\item We prove the claim by induction on the length of~$\rho$.
Let $u \to v$ be the last edge of the path~$\rho$.
By inductive hypothesis, $\rho$~contains the vertex~$p_\sigma(u)$.
Repeating this argument, it follows that it also contains all iterates $(p_\sigma)^i(u)$
(that are defined). It is therefore sufficient to show that $(p_\sigma)^i(u) = p_\sigma(v)$,
for some~$i$. We distinguish two cases.

If $u \prec v$, the claim is immediate.
Hence, suppose that $u \to v$ is a redirected edge.
Then there exists a successor~$v'$ of~$u$ (in~$t$) such that $v' \approx_\sigma v$.
It follows that there is some~$i$ with $(p_\sigma)^i(u) = p_\sigma(v') = p_\sigma(v)$.

\item
We proceed by induction on the length of~$\rho$.
If $v = u$, the claim is trivial. For the inductive step, consider the last edge $w \to v$
of~$\rho$. By inductive hypothesis, we know that $u \preceq w$.
If the edge is one of the original edges of~$t$, we have $u \preceq w \prec v$.
Hence, suppose that the edge is a redirected one. Let $v'$~be the successor of~$w$ (in~$t$)
with $v' \approx_\sigma v$.
If $u \preceq p_\sigma(v')$, we have $u \preceq p_\sigma(v') = p_\sigma(v) \prec v$,
as desired.
Hence, suppose otherwise. Then $p_\sigma(v') \prec u \prec v'$ implies that
$\sigma(u) < \sigma(v')$, and it follows that $p_\sigma(v) = p_\sigma(v') \preceq p_\sigma(u)$.
A~contradiction.

\item
Without loss of generality we may assume that the path~$\rho$ starts at the root.
In order to avoid case distinctions, we add a new root~$r$ to~$t$
that is mapped by~$\sigma$ to the maximal possible value.
We also add~$r$ to the beginning of~$\rho$, so that $\rho$~starts at~$r$.
Then $p_\sigma(v) \neq \bot$, for all vertices~$v$ (except for the new root).
For a contradiction, suppose that there are vertices $x \preceq y$
such that $x$~is after~$y$ along the path~$\rho$.
We choose~$x$ $\preceq$-minimal and, given~$x$, we choose~$y$ $\preceq$-minimal.
Note that, since~$\rho$ contains the root~$r$ only once, it follows that $x \neq r$
and, hence, $y \neq r$.
\end{enuma}

It follows by~(a) that the vertex~$p_\sigma(y)$ lies on~$\rho$ before~$y$ and, hence,
also before~$x$. By minimality of~$y$, it therefore follows that $p_\sigma(y) \prec x$.
Let $u_0 \to v_0,\dots,u_m \to v_m$ be an enumeration (in order) of all edges of~$\rho$
that lie between $y$~and~$x$ and that satisfy $p_\sigma(v_i) \preceq p_\sigma(x)$.
Let $v'_i$~be the successor of~$u_i$ (in~$t$) such that $v'_i \approx_\sigma v_i$.
(If the edge $u_i \to v_i$ is not redirected, we have $v'_i = v_i$.)
Note that the part of~$\rho$ between $v_i$~and~$u_{i+1}$ cannot contain a vertex~$z$
with $p_\sigma(z) \preceq p_\sigma(v_i)$ since this would imply
$p_\sigma(z) \preceq p_\sigma(x)$, meaning $z$~would be one of the~$v_j$.
Hence, we can apply~(b) to the part of~$\rho$ between $v_i$~and~$u_{i+1}$ and
we obtain $v_i \preceq u_{i+1} \prec v'_{i+1}$.
Since $p_\sigma(x) \prec x$, it follows by minimality of~$x$ that $v_i \npreceq p_\sigma(x)$.
Hence, $p_\sigma(v'_{i+1}) = p_\sigma(v_{i+1}) \preceq p_\sigma(x)$ implies that
$v_i \npreceq p_\sigma(v'_{i+1})$.
Therefore, we have $p_\sigma(v'_{i+1}) \prec v_i \prec v'_{i+1}$ and
$\sigma(v_i) < \sigma(v'_{i+1}) = \sigma(v_{i+1})$.
Similarly, applying~(b) to the part of~$\rho$ between $y$~and~$v_0$, we obtain
$p_\sigma(v'_0) = p_\sigma(v_0) \preceq p_\sigma(x) \prec y \preceq u_0 \prec v'_0$,
which means that $\sigma(y) < \sigma(v'_0) = \sigma(v_0)$.
Furthermore, $p_\sigma(y) \prec x \preceq y$ implies that $\sigma(x) \leq \sigma(y)$.
Finally, we have $v_m = x$ since minimality of~$x$ implies that the edge of~$\rho$ leading
to~$x$ is a redirected one and we trivially have $p_\sigma(x) \preceq p_\sigma(x)$.
Altogether, we obtain
\begin{align*}
  \sigma(x) = \sigma(v_m) \geq \sigma(v_0) > \sigma(y) \geq \sigma(x)\,.
\end{align*}
A~contradiction.
\end{proof}

After these preparations we can state and prove our key technical lemma.
\begin{Lem}\label{Lem: split for definable algebras}
Let $\frakA$~be an $\MSO$-definable $\bbT^\times$-algebra and $\xi \in \Xi$ a sort.
There exists a constant $N < \omega$ with the following property.
For every tree $t \in \bbT^\times_\xi A$, we can find a function $\sigma : \dom(t) \to [N]$
such that
\begin{align*}
  \pi(t') = \pi(t)\,, \quad\text{for every $\sigma$-rewiring $t'$ of } t\,,
\end{align*}
and $\sigma$~maps the root of~$t$ to $N-1$.
\end{Lem}
\begin{proof}
Fix a tree $t \in \bbT^\times_\xi A$,
let $\calA = \langle Q,A + \xi,\Delta,q_0,\Omega\rangle$ be the automaton
checking that the product of a given tree in $\bbT^\times_\xi A$ is equal to $\pi(t)$
(extended to the infinite alphabet $A + \xi$ as explained in Section~\ref{Sect:dense}
(page~\pageref{automaton over infinite alphabets})),
and let $\calG$~be the corresponding Automaton-Pathfinder game for~$\calA$ on the input tree~$t$.
Without loss of generality, we may assume that $\calA$~is a non-deterministic automaton.
Since $\calG$~is a parity game, there exists a positional winning strategy~$\tau$ for Automaton.

Given a vertex $v \in \dom(t)$, we denote by $\mu(v)$ the state~$q$ such that the
unique play of~$\calG$ conforming to~$\tau$ that reaches the vertex~$v$ does so in state~$q$.
Fix some bijective function $h : Q \to [N]$, for $N < \omega$, such that
\begin{align*}
  \Omega(p) < \Omega(q) \qtextq{implies} h(p) > h(q)\,,
\end{align*}
and set
\begin{align*}
  \sigma(v) :=
    \begin{cases}
      N-1       &\text{if } v \text{ is the root,} \\
      h(\mu(v)) &\text{otherwise.}
    \end{cases}
\end{align*}
We claim that $\sigma$ is the desired function.

Let $t'$~be a $\sigma$-rewiring and let $\calG'$~be the Automaton-Pathfinder game on~$t'$.
We have to show that $\pi(t') = \pi(t)$.
Note that the Automaton-Pathfinder game~$\calG'$ for~$\calA$ on~$t'$ can be obtained from
the game~$\calG$ for~$t$ by removing some positions and redirecting some of the edges.
Consequently, the strategy~$\tau$ for~$t$ induces a strategy~$\tau'$ for the game on~$t'$.
For a path~$\nu$ (in~$\calG$ or~$\calG'$),
we denote by $\Omega(\nu)$ the least priority seen along~$\nu$.
We start by proving the following claim.

\begin{Claim}
Let $\langle u_1,q_1\rangle$ be a position of Automaton in~$\calG'$ that is reachable by some
play confoming to~$\tau'$, and let $\nu,\nu'$ be two plays of~$\calG'$ confoming
to~$\tau'$ that start at $\langle u_1,q_1\rangle$ and that end in two positions
$\langle u_2,q_2\rangle$ and $\langle u_2,q'_2\rangle$, respectively, with
the same vertex~$u_2$.
Suppose that
\begin{align*}
  u_1 \prec u_2
  \qtextq{and}
  \sigma(u_1) \geq \sigma(x)\,, \quad\text{for all } u_1 \preceq x \preceq u_2\,.
\end{align*}
Then we have
\begin{align*}
  q_2 = q'_2 \qtextq{and} \Omega(\nu) = \Omega(\nu')\,.
\end{align*}
\end{Claim}

We prove the claim by an induction on the following two numbers (in order of
decreasing priority)\?:
\begin{itemize}
\item first on the total number of edges in $\nu$~or~$\nu'$
  that are not edges of $\calG$,
\item then on the sum of the lengths of $\nu$~and~$\nu'$.
\end{itemize}
We distinguish three cases.

\textsc{(i)}
Suppose that there is some vertex $u_1 \prec u_3 \prec u_2$ such that
\begin{align*}
  \sigma(u_3) \geq \sigma(x)\,, \quad\text{for all } u_3 \preceq x \preceq u_2\,,
\end{align*}
$\nu$~contains the position $\langle u_3,q_3\rangle$ and
$\nu'$~the position $\langle u_3,q'_3\rangle$.
Then we can split the respective plays at these positions.
Let $\nu_0,\nu_1,\nu'_0,\nu'_1$ be
the corresponding parts. Applying the inductive hypothesis twice, it follows that
\begin{align*}
  \Omega(\nu_0) = \Omega(\nu'_0)\,, \quad
  q_3 = q'_3\,, \quad
  \Omega(\nu_1) = \Omega(\nu'_1)\,,
  \qtextq{and}
  q_2 = q'_2\,.
\end{align*}
This implies that $\Omega(\nu) = \Omega(\nu')$.

\textsc{(ii)}
Suppose that the second but last positions of $\nu$~and~$\nu'$
contain the same vertex~$u_3$ with $u_1 \prec u_3 \prec u_2$.
Let $\langle u_3,q_3\rangle$ and $\langle u_3,q'_3\rangle$ be these positions
and let $\nu_0$~and~$\nu'_0$ be the corresponding prefixes of the two plays.
By inductive hypothesis, we have $q_3 = q'_3$ and
$\Omega(\nu_0) = \Omega(\nu'_0)$.
As the plays $\nu$~and~$\nu'$ both conform to the same strategy~$\tau'$,
it follows that $q_2 = q'_2$ and $\Omega(\nu) = \Omega(\nu')$.

\textsc{(iii)}
Finally, suppose that cases \textsc{(i)}~and~\textsc{(ii)} do not hold.
Let $\langle v,s\rangle$ and $\langle v',s'\rangle$ be the second but last positions
of Automaton in the plays $\nu$~and~$\nu'$, respectively
(see Figure~\ref{Fig: plays in G and G'}\,(a)).
\begin{figure}
\centering
\includegraphics{Expansion-7.mps}
\caption{(a)~Left\?: the plays $\nu$~and~$\nu'$.
Dashed lines represent plays in~$\calG'$, solid ones plays in~$\calG$.
(b)~Right\?: the plays $\rho_i$ and $\nu'_i$, projected to the game~$\calG$.}%
\label{Fig: plays in G and G'}
\end{figure}
Since we are not in case~\textsc{(ii),} we have either $v \neq v'$ or $v = v' \nprec u_2$.
In both cases, at least one of the edges $v \to u_2$ and $v' \to u_2$ is an edge of~$t'$
but not one of~$t$.
Let $w$~and~$w'$ be the successors of, respectively, $v$~and~$v'$ in~$t$ that correspond
to these two edges and let $q_3$~and~$q'_3$ be the states at $w$~and~$w'$.
Let $\rho$~be the play that conforms to~$\tau$, starts at $\langle u_1,q_1\rangle$, ends in
$\langle w,q_3\rangle$, and that uses only edges that are present in~$t$.
Similarly, let $\rho'$~be the corresponding play to $\langle w',q'_3\rangle$.

Note that, by definition of a $\sigma$-rewiring, we have
$w \approx_\sigma u_2 \approx_\sigma w'$. This implies that
\begin{align*}
  \sigma(w) = \sigma(u_2) = \sigma(w')
  \qtextq{and}
  p_\sigma(w) = p_\sigma(u_2) = p_\sigma(w')\,.
\end{align*}
Since we are not in case~\textsc{(i),} it further follows by
Lemma~\ref{Lem: every path contains p_sigma} that $p_\sigma(u_2) = u_1$.

Let $\nu_0$,~$\nu'_0$, $\rho_0$, and~$\rho'_0$ be the prefixes of the plays
$\nu$,~$\nu'$, $\rho$, and~$\rho'$ ending in the position
$\langle v,s\rangle$ and $\langle v',s'\rangle$, respectively.
Since $\rho_0$~and~$\rho'_0$ do not contain redirected edges,
$p_\sigma(w) = u_1 = p_\sigma(w')$, and $u_1 \preceq v \prec w$ and $u_1 \preceq v' \prec w'$,
we can use the inductive hypothesis to show that
\begin{align*}
  \Omega(\rho_0) = \Omega(\nu_0)
  \qtextq{and}
  \Omega(\rho'_0) = \Omega(\nu'_0)\,.
\end{align*}
Furthermore, $\sigma(w) = \sigma(u_2) = \sigma(w')$ implies that
$q_3 = \mu(w) = \mu(u_2) = \mu(w') = q'_3$. Let $k := \Omega(q_3)$ be the priority of this state.
Since
\begin{align*}
  \Omega(\rho) = \min {\{k, \Omega(\rho_0)\}}\,,
\end{align*}
and similarly for the other plays, it follows that
\begin{align*}
  \Omega(\rho) = \Omega(\nu)
  \qtextq{and}
  \Omega(\rho') = \Omega(\nu')\,.
\end{align*}

Hence, it remains to show that $\Omega(\rho) = \Omega(\rho')$.
We claim that
\begin{align*}
  \Omega(\rho) = \Omega(\mu(u_1))
  \qtextq{and}
  \Omega(\rho') = \Omega(\mu(u_1))\,.
\end{align*}
By symmetry, it is sufficient to prove the first equation.
Since $w \approx_\sigma u_2$, we have $p_\sigma(w) = p_\sigma(u_2)$.
By definition of~$p_\sigma$, we have
\begin{align*}
  \sigma(x) \leq \sigma(w) \leq \sigma(p_\sigma(w))\,,
  \quad\text{for all } p_\sigma(u_2) = p_\sigma(w) \preceq x \preceq w\,.
\end{align*}
Since, as we have shown above, $p_\sigma(u_2) = u_1$, it follows that
\begin{align*}
  \sigma(x) \leq \sigma(u_1)\,, \quad\text{for all } u_1 \preceq x \preceq w\,.
\end{align*}
Consequently, we have $\Omega(\rho) = \Omega(\mu(u_1))$.

\smallskip
Having proved the above claim we can now show that $\pi(t') = \pi(t)$,
i.e., that $\calA$~accepts the tree~$t'$.
We claim that the strategy~$\tau'$ defined above is winning.
To do so it is sufficient to show that, every play~$\nu'$ in~$\calG'$ conforming to~$\tau'$
induces a play~$\nu$ in~$\calG$ conforming to~$\tau$ such that the minimal priorities seen
infinitely often along $\nu$~and~$\nu'$ are the same.

Let $k$~be the maximal number such that $\nu'$~contains infinitely many positions
$\langle v,q\rangle$ of Automaton with $\sigma(v) = k$.
Then there exists an infinite sequence $\langle v_i,q_i\rangle_{i<\omega}$
of positions of~$\nu'$ with $\sigma(v_i) = k$ and
\begin{align*}
  v_0 \sqsubset_\sigma v_1 \sqsubset_\sigma \cdots\,.
\end{align*}
We choose this sequence maximal (with respect to inclusion), which implies
that $v_i = p_\sigma(v_{i+1})$.
Let $\nu_0 = \nu'_*\nu'_0\nu'_1\dots$ be the factorisation of
the play~$\nu_0$ where $\nu'_*$~is the prefix ending in~$\langle v_0,q_0\rangle$
and $\nu'_i$~is the part between the positions $\langle v_i,q_i\rangle$
and $\langle v_{i+1},q_{i+1}\rangle$.

For every $i < \omega$, let $\rho_i$~be the partial play of~$\calG'$
from $\langle v_i,q_i\rangle$ to $\langle v_{i+1},q'_{i+1}\rangle$, for some state~$q'_{i+1}$,
that conforms to~$\tau'$ and that uses only edges belonging to~$\calG$,
see Figure~\ref{Fig: plays in G and G'}\,(b).
(To see that such a play exists note that a position of the form
$\langle v_{i+1},q'_{i+1}\rangle$ is reachable by some play conforming to~$\tau'$
and this play contains the vertex $p_\sigma(v_{i+1}) = v_i$ by
Lemma~\ref{Lem: every path contains p_sigma}.
Furthermore, it follows by the above claim that the state at this vertex is equal to~$q_i$.)
By the above claim, we have $q'_i = q_i$ and $\Omega(\rho_i) = \Omega(\nu'_i)$, for all~$i$.
Consequently, the composition $\nu := \nu'_*\rho_0\rho_1\rho_2\ldots$
forms a play in~$\calG$ conforming to~$\tau$ and
the least priority seen infinitely often in~$\nu$ is the same as in~$\nu'$.
\end{proof}

\begin{Thm}\label{Thm: definable algebras have condensations}
Let $\frakA$~be an $\MSO$-definable $\bbT^\times$-algebra.
Every tree $t \in \bbT^\times A$ has a consistent $\bbT^\times$-condensation~$s$ such that
\begin{align*}
  \pi(t) = \pi(s \mVert \mu)\,,
  \quad\text{for all choice functions } \mu\,.
\end{align*}
\end{Thm}
\begin{proof}
Let $\sigma : \dom(t) \to [N]$ be the function from
Lemma~\ref{Lem: split for definable algebras}.
Fix a set of variables~$\xi$ with $\abs{\xi} = N$ and let $\mu : [N] \to \xi$ be a bijection.
We define the desired condensation~$s$ by
\begin{align*}
  s(v) := \langle\tau_v,t(v)\rangle
  \qtextq{with}
  \tau_v(x) := \mu(\sigma(u_x))\,,
\end{align*}
where $u_x$~is the $x$-successor of~$v$.
Then every tree of the form $s \mVert \mu$ is a $\sigma$-rewiring of~$t$.
By choice of~$\sigma$ this implies that $\pi(s \mVert \mu) = \pi(t)$.
\end{proof}
In particular, it follows that every $\MSO$-definable $\bbT^\times$-algebra has
$\bbT^\times$-eval\-u\-ations with consistent merging. Note that this statement is not as trivial
as it sounds since trees can contain labels of arbitrarily high arity,
while every $\bbT^\times$-con\-dens\-a\-tion produces a tree where these arities are bounded.
In particular, the statement is false for $\bbT^\times$-evaluations with uniform merging.

Our hope is that a more elaborate version of the construction from the proof
of Lemma~\ref{Lem: split for definable algebras} can be used to construct a
$\bbT^{\times\reg}$-condensation instead of a $\bbT^\times$-one,
or at least that we can iterate such a construction to obtain a $\bbT^{\times\reg}$-evaluation.
\begin{Conj}
Let\/ $\frakA$~be an\/ $\MSO$-definable\/ $\bbT^{\times\reg}$-algebra.
Then every tree $t \in \bbT A$ has a\/ $\bbT^{\times\reg}$-evaluation with consistent merging.
\end{Conj}

If we relax our notion of an evaluation and of consistent merging a bit, we even
obtain something similar to `$\bbT^{\times\thin}$-evaluations'. (Note that, formally,
this notion does not exist, since $\bbT^{\times\thin}$~does not form a monad.)
\begin{Thm}\label{Thm: reduction to Txthin}
Let $\frakA$~be an $\MSO$-definable $\bbT^\times$-algebra and $C \subseteq A$ a subset.
For every $t \in \bbT^\times_\xi C$, there is some $s \in (\bbT^{\times\thin})^n C$ with
$\pi(\Flat^{n-1}(s)) = \pi(t)$, where the exponent $n < \omega$ only depends on
$\frakA$~and~$\xi$.
\end{Thm}
\begin{proof}
Let $\sigma : \dom(t) \to [n]$ be the function from Lemma~\ref{Lem: split for definable algebras}
and let
\begin{align*}
  D := \set{ {}^\tau c }{ c \in C_\zeta,\ \tau : \zeta \to \eta \text{ surjective, }
                          \eta,\zeta \in \Xi }
\end{align*}
be the closure of~$C$ under variable substitutions. By induction on $i \leq n$, we construct
graphs $s_i \in (\bbT^\times)^{n-i}(\bbR^{\times\thin})^i D$ satisfying the following conditions.
\begin{itemize}
\item The graph $\Flat^{n-1}(s_i) \in \bbR D$ induces a $\sigma$-rewiring of~$t$.
  In particular, $\Flat^{n-1}(s_i)$ is obtained from a subgraph of~$t$ by redirecting
  all edges leaving this subgraph.
\item Let $s'_i$~be the tree obtained from~$s_i$ by deleting all redirected edges
  and let $t'_i$~be the tree obtained from~$t$ by deleting all subtrees reachable
  by an edge that has been redirected in~$s_i$. (Note that $s'_i$~and~$t'_i$ are not
  elements of $(\bbT^\times)^nD$ and $\bbT^\times C$, respectively, since some
  vertices are missing some of their successors.)
  Then $s'_i$~is the recursive factorisation of~$t'_i$ that corresponds
  via the construction of Lemma~\ref{Lem: evaluation from a split} and
  Figure~\ref{Fig:evaluation and split}
  to the restriction of~$\sigma$ to $\dom(t'_i) \subseteq \dom(t)$.
\end{itemize}
Then we can produce the desired tree~$s$ as follows.
As each $d \in D$ is of the form $d = \pi(r)$, for some $r \in \bbT^\fin C$,
there exists a function $\vartheta : D \to \bbT^\fin C$ with
$\pi \circ \vartheta = \id$.
We chose for $s \in (\bbT^{\times\thin})^{n+1} C$ the unravelling (at all $n+1$ levels)
of the graph $\bbR^n\vartheta(s_n) \in (\bbR^{\times\thin})^{n+1} C$.
Since the unravelling~$s'_n$ of $\Flat^{n-1}(s_n)$ is a $\sigma$-rewiring of~$t$ it follows that
\begin{align*}
  \pi(\Flat^n(s)) = \pi(s'_n) = \pi(t)\,,
\end{align*}
as desired.

It remains to construct $s_0,\dots,s_n$.
For~$s_0$, we choose the $\bbT^\times$-evaluation of~$t$ induced by~$\sigma$ as in
Lemma~\ref{Lem: evaluation from a split}.
For the inductive step, suppose that we have already defined~$s_i$.
For every component $s_i(u_0)\cdots(u_{n-i-2})$ there exists a canonical injection
\begin{align*}
  \dom_0(s_i(u_0)\cdots(u_{n-i-2})) \to \dom_0(\Flat^{n-1}(s_i))
\end{align*}
mapping every $u_{n-i-1} \in \dom_0(s_i(u_0)\cdots(u_{n-i-2}))$
to the vertex of the flattening corresponding to the root of the graph
$\Flat^{i-1}(s_i(u_0)\cdots(u_{n-i-2})(u_{n-i-1}))$.
Since $\dom(\Flat^{n-1}(s_i))$ is contained in $\dom(t)$,
the restriction of~$\sigma$ to the range of this injection induces a function
\begin{align*}
  \sigma_i : \dom_0(s_i(u_0)\cdots(u_{n-i-2})) \to [n]\,.
\end{align*}
By inductive hypothesis, we have $\sigma_i(v) \geq i$, for all~$v$.
Let $s_{i+1}$~be obtained from~$s_i$ by replacing every component
$s_i(u_0)\cdots(u_{n-i-2}) \in \bbT^\times(\bbR^{\times\thin})^i A$
by a path $r \in (\bbR^{\times\thin})^{i+1} A$ which we construct as follows.
We define a branch $v_0,v_1,\dots$ of the tree $s_i(u_0)\cdots(u_{n-i-2})$
starting at the root~$v_0$. This branch will form the domain of~$r$.
For the inductive step, suppose that we have already defined the vertex~$v_j$ and
the labels $r(v_k)$, for all $k < j$.
Suppose that
\begin{align*}
  s_i(u_0)\cdots(u_{n-i-2})(v_j) = p(\bar x,\bar y)\,,
  \quad\text{for } p \in (\bbR^{\times\thin})^iA\,,
\end{align*}
where $\bar x$~are the variables such that the corresponding successors~$w$ of~$v_j$
are labelled by non-variables (which implies that $\sigma_i(w) = i$),
while $\bar y$~are the remaining ones.
Let $\bar y'$~be the variables labelling the successors corresponding to~$\bar y$.

If $\bar x$~is empty, we set
\begin{align*}
  r(v_j) := p(\bar y')
\end{align*}
and the construction terminates.
Otherwise, we pick one variable $x' \in \bar x$, we choose for $v_{j+1}$ the $x'$-successor
of~$v_j$, and we set
\begin{align*}
  r(v_j) := p(x'\!\ldots x',\bar y')\,.
\end{align*}

The flattening $\Flat^{n-1}(s_{i+1})$ of the resulting graph~$s_{i+1}$
induces the desired $\sigma$-rewiring of~$t$.
\end{proof}
As already mentioned above, we cannot conclude that $\bbT^{\times\thin}$~is dense
in~$\bbT^\times$ since the former does not form a monad.
But we obtain the following, weaker statements.
\begin{Cor}\label{Cor: T0 dense in Tx}
Let $\bbT^0 \subseteq \bbT^\times$ be the closure of~$\bbT^{\times\thin}$ under $\Flat$.
Then $\bbT^0$~is dense in~$\bbT^\times$ over the class of all $\MSO$-definable
$\bbT^\times$-algebras.
\end{Cor}
\begin{Cor}
Let\/ $\frakA$~be an\/ $\MSO$-definable\/ $\bbT^\times$-algebra.
A set $C \subseteq A$ induces a subalgebra of\/~$\frakA$ if, and only if,
\begin{align*}
  \pi(t) \in C\,, \quad\text{for all } t \in \bbT^{\times\thin} C\,.
\end{align*}
\end{Cor}
\begin{Cor}
The product of every\/ $\MSO$-definable\/ $\bbT^\times$-algebra\/~$\frakA$ is uniquely determined
by its restriction to the set\/ $\bbT^{\times\thin}A$.
\end{Cor}
\begin{proof}
Suppose that there are two $\MSO$-definable $\bbT^\times$-algebras
$\frakA_0 = \langle A,\pi_0\rangle$ and $\frakA_1 = \langle A,\pi_1\rangle$ with
the same universe~$A$ and whose products have the same restriction to $\bbT^{\times\thin}A$.
Fix a tree $t \in \bbT^\times A$.
We have to show that $\pi_0(t) = \pi_1(t)$.

Let $\delta : A \to A \times A$ be the diagonal map, let
$\Delta := \rng \delta \subseteq A \times A$ be its range,
and set $s := \bbT^\times\delta(t) \in \bbT^\times\Delta$.
As the product $\frakA_0 \times \frakA_1$ is also $\MSO$-definable,
we can use Theorem~\ref{Thm: reduction to Txthin} to find a tree
$r \in (\bbT^{\times\thin})^n\Delta$ with $\pi(\Flat^{n-1}(r)) = \pi(s)$.
Since $\pi_0$~and~$\pi_1$ agree on all trees in $\bbT^{\times\thin}A$, we have
\begin{align*}
  \pi(u) \in \Delta\,, \quad\text{for all } u \in \bbT^{\times\thin}\Delta\,.
\end{align*}
Consequently, we can evaluate
\begin{align*}
  \pi(\Flat^{n-1}(r)) = \pi(\bbT^{\times\thin}\pi(\bbT^{\times\thin}\bbT^{\times\thin}\pi(\cdots
    (\bbT^{\times\thin})^{n-1}\pi(r)\cdots)))
\end{align*}
by recursion using only products of trees in~$\bbT^{\times\thin}\Delta$.
In particular, we have $\pi(\Flat^{n-1}(r)) \in \Delta$ and it follows that
\begin{align*}
  \langle\pi_0(t),\pi_1(t)\rangle = \pi(s) = \pi(\Flat^{n-1}(r)) \in \Delta
  \qtextq{implies}
  \pi_0(t) = \pi_1(t)\,.
\end{align*}
\upqed
\end{proof}
\begin{Rem}
Using a variant of Proposition~\ref{Prop: unique Twilke-evaluations} for non-linear trees,
we can strenghten Theorem~\ref{Thm: reduction to Txthin} to obtain a tree
$s \in (\bbT^{\times\wilke})^n C$.
Since the closure of~$\bbT^{\times\wilke}$ under $\Flat$ coincides with~$\bbT^{\times\reg}$,
the argument in Corollary~\ref{Cor: T0 dense in Tx} then provides an alternative
proof of Theorem~\ref{Thm: regular trees dense}.
\end{Rem}

\section{Consistent labellings}   
\label{Sect:labellings}

As we have seen in the previous section, we can construct expansions with the help of evaluations
if the two monads in question are sufficiently well-behaved. What do we do if they are not\??
Let us turn to a second idea of how to prove that a $\bbT^0$-algebra~$\frakA$ has a
$\bbT$-expansion\?:
when we want to define the product of $t \in \bbT A$, we first annotate~$t$ with additional
information that makes it easier to determine the value of the product.
For instance, for each vertex~$v$, we can guess the value~$\pi(t|_v)$ of
the corresponding subtree and then check that these guesses are correct.
\begin{Def}
Let $\bbT^\fin \subseteq \bbT^0 \subseteq \bbT$ be a submonad,
$\frakA = \langle A,\pi\rangle$ a $\bbT^0$-algebra, and $t \in \bbT_\xi A$.

    \begin{enuma}
\item A \emph{labelling} of~$t$ is a function $\lambda : \dom_0(t) \to A$ (not necessarily
arity-preserving) such that, for every vertex~$v$,
\begin{align*}
  \lambda(v) \in A_\zeta \quad\iff\quad
  \zeta \text{ is the set of variables appearing in } t|_v\,.
\end{align*}

\item A labelling $\lambda : \dom_0(t) \to A$ is \emph{weakly $\bbT^0$-consistent} if,
for every factor $[u,\bar v)$ with $t[u,\bar v) \in \bbT^0 A$,
\begin{align*}
  \lambda(u) = \pi\bigl(t[u,\bar v)(\lambda(v_0),\dots,\lambda(v_{n-1})\bigr)\,.
\end{align*}
\upqed
\end{enuma}
\end{Def}
\begin{Exam}
For every $\bbT$-algebra~$\frakA$ and every tree $t \in \bbT A$, we can define a labelling by
\begin{align*}
  \lambda(v) := \pi(t|_v)\,.
\end{align*}
This labelling is obviously weakly $\bbT$-consistent and, hence, weakly $\bbT^0$-consistent
for every $\bbT^0 \subseteq \bbT$.
In particular, if a $\bbT^0$-algebra has a $\bbT$-expansion, then every tree has at least
one weakly $\bbT^0$-consistent labelling.
\end{Exam}

Weak consistency is based on factors with finitely many variables.
In many situations this is not sufficient and we have to use the following stronger
version of consistency where we also allow factors with infinitely many variables.
\begin{Def}
Let $\bbT^0 \subseteq \bbT$ be a submonad, $\frakA = \langle A,\pi\rangle$ a $\bbT^0$-algebra,
and $t \in \bbT_\xi A$.

    \begin{enuma}
\item Given a factor $[u,\bar v)$ of~$t$, possibly with infinitely many holes~$\bar v$,
we denote by $t[u,\bar v)(a_0,a_1,\dots)$ the tree obtained from $t[u,\bar v)$
by replacing each leaf labelled by a variable~$x_i$ by the tree $\sing(a_i)$.

\item A labelling $\lambda : \dom_0(t) \to A$ is \emph{strongly $\bbT^0$-consistent} if,
for every factor $[u,\bar v)$, possibly with infinitely many holes~$\bar v$,
with $t[u,\bar v)(\lambda(v_0),\lambda(v_1),\dots) \in \bbT^0 A$, we have
\begin{align*}
  \lambda(u) = \pi\bigl(t[u,\bar v)(\lambda(v_0),\lambda(v_1),\dots)\bigr)\,.
\end{align*}
\upqed
\end{enuma}
\end{Def}

We start with the following easy observation.
\begin{Lem}
Let\/ $\frakA$~be a finitary\/ $\bbT^\fin$-algebra.
Every tree $t \in \bbT A$ has a\/ strongly $\bbT^\fin$-consistent labelling.
\end{Lem}
\begin{proof}
We call a labelling~$\lambda$ of some tree~$t$ \emph{locally consistent} if
\begin{align*}
  \lambda(v) = t(v)\bigl(\lambda(u_0),\dots,\lambda(u_{n-1})\bigr)\,,
\end{align*}
for every vertex~$v$ with successors $u_0,\dots,u_{n-1}$.
Fix an increasing sequence $P_0 \subset P_1 \subset \dots \subset \dom(t)$
of finite prefixes of~$t$ with $\bigcup_i P_i = \dom(t)$,
and let $\Lambda_i$~be the set of all locally consistent labellings of~$P_i$, for $i < \omega$.
Then $\Lambda := \bigcup_i \Lambda_i$ ordered by~$\subset$ forms a finitely-branching tree.
By K\H onig's Lemma, there exists an infinite branch $\lambda_0 \subset \lambda_1 \subset\dots$.
Let $\lambda$~be its limit. Then $\lambda$~is locally consistent.

It therefore, remains to prove that every locally consistent labelling of~$t$
is strongly $\bbT^\fin$-consistent.
Consider a finite factor~$H$ of~$t$ with root~$v$ and leaves $u_0,\dots,u_{m-1}$.
By induction on $\abs{H}$ it follows that
\begin{align*}
  \lambda(v) = \pi\bigl((t \restriction H)(\lambda(u_0),\dots,\lambda(u_{m-1})\bigr)\,.
\end{align*}
\upqed
\end{proof}

Next, let us show how to use consistent labellings to characterise possible $\bbT$-expansions
of a given $\bbT^0$-algebra. We need the following additional property.
\begin{Def}
Let $\bbT^0 \subseteq \bbT$.

    \begin{enuma}
\item A \emph{weak labelling scheme} for a $\bbT^0$-algebra~$\frakA$ is a function~$\sigma$
assigning to each tree $t \in \bbT A$ a weakly $\bbT^0$-consistent labelling $\sigma(t)$ of~$t$.
Similarly, a \emph{strong labelling scheme}~$\sigma$ assigns
to each tree $t \in \bbT A$ a strongly $\bbT^0$-consistent labelling~$\sigma(t)$.

\item A labelling scheme~$\sigma$ for~$\frakA$ is \emph{associative} if,
for every tree $T \in \bbT\bbT A$, we have
\begin{align*}
  \sigma(t) = \sigma(\Flat(T)) \circ \mu\,,
\end{align*}
where $\mu : \dom_0(T) \to \dom_0(\Flat(T))$ maps each vertex
$v \in \dom_0(T)$ to the vertex of~$\Flat(T)$ corresponding to the root of~$T(v)$,
and $t \in \bbT A$ is the tree such that
\begin{align*}
  t(v) := \sigma(T(v))(\emptyseq)\,,
  \quad\text{for } v \in \dom_0(T)\,.
\end{align*}
\upqed
\end{enuma}
\end{Def}

\begin{Exam}
There are algebras with several associative strong labelling schemes.
Let $\frakA$~be the $\bbT^\thin$-algebra with domains $A_\xi := [n]$,
for some fixed number $n < \omega$, where the product is just the maximum
\begin{align*}
  \pi(t) := \max {\set{ t(v) }{ v \in \dom_0(t) }}\,.
\end{align*}
For every $k < n$, we obtain an associative labelling scheme~$\sigma_k$ defined by
\begin{align*}
  \sigma_k(t)(v) := \begin{cases}
                      \pi(t|_v)    &\text{if } t|_v \in \bbT^\thin A\,, \\
                      \max {\{k\} \cup \set{ t(u) }{ u \succeq v }}\,,
                                   &\text{otherwise}\,.
                    \end{cases}
\end{align*}
\upqed
\end{Exam}

There is a tight connection between $\bbT$-expan\-sions and associative labelling schemes
(weak or strong, it does not matter).
\begin{Prop}\label{Prop: expansions and labelling schemes}
Let\/ $\bbT^0 \subseteq \bbT$ and let\/ $\frakA$~be a\/ $\bbT^0$-algebra.
    \begin{enuma}
\item Every associative weak labelling scheme for\/~$\frakA$ is strong.
\item There exists a bijective correspondence between associative labelling
  schemes~$\sigma$ and\/ $\bbT$-expansions of\/~$\frakA$.
\end{enuma}
\end{Prop}
\begin{proof}
We define two mutually-inverse functions mapping
\textsc{(i)}~every associative weak labelling scheme to a $\bbT$-expansion of~$\frakA$
and \textsc{(ii)}~every such expansion to an associative strong labelling scheme.

\textsc{(i)}
Given a weak scheme $\sigma$ we define the corresponding function~$\pi_+$ by
\begin{align*}
  \pi_+(t) := \sigma(t)(\emptyseq)\,,
  \quad\text{for } t \in \bbT A\,.
\end{align*}
Then $\pi_+$~extends~$\pi$ since weak $\bbT^0$-consistency of~$\sigma$ implies that
\begin{align*}
  \pi(t) = \sigma(t)(\emptyseq) = \pi_+(t)\,,
  \quad\text{for } t \in \bbT^0 A\,.
\end{align*}
Hence, it remains to show that~$\pi_+$ is associative. Fix $T \in \bbT\bbT A$.
By the definition of associativity of~$\sigma$, we have
\begin{align*}
  \sigma(\bbT\pi_+(T)) = \sigma(\Flat(T)) \circ \mu\,,
\end{align*}
which in particular implies that
\begin{align*}
  \pi_+(\bbT\pi_+(T))
  &= \sigma(\bbT\pi_+(T))(\emptyseq) \\
  &= \sigma(\Flat(T))(\emptyseq)
   = \pi_+(\Flat(T))\,.
\end{align*}

\textsc{(ii)}
Conversely, given a product $\pi_+ : \bbT A \to A$ we define a scheme~$\sigma$ by
\begin{align*}
  \sigma(t)(v) := \pi_+(t|_v)\,,
  \quad\text{for } t \in \bbT A \text{ and } v \in \dom_0(t)\,.
\end{align*}
To show that this function~$\sigma$ is a strong labelling scheme,
fix a factor~$[u,\bar v)$ (possibly with infinitely many holes~$\bar v$)
of some tree $t \in \bbT A$.
Let $T \in \bbT\bbT A$ be the tree with $\dom(T) = \dom(t[u,\bar v))$
(plus some leaves for the variables below some of the~$v_i$) and labelling
\begin{align*}
  T(w) :=
    \begin{cases}
      \sing(t(w)) &\text{for } w \in [u,\bar v)\,, \\
      t|_{v_i}    &\text{for } w = v_i\,.
    \end{cases}
\end{align*}
Then it follows by associativity of~$\pi_+$ that
\begin{align*}
  \sigma(t)(u)
  &= \pi_+(t|_u) \\
  &= \pi_+(\Flat(T)) \\
  &= \pi_+(\bbT\pi_+(T)) \\
  &= \pi_+\bigl(t[u,\bar v)(\pi_+(t|_{v_0}),\pi_+(T|_{v_1}),\dots)\bigr) \\
  &= \pi_+\bigl(t[u,\bar v)(\sigma(t)(v_0),\sigma(t)(v_1),\dots)\bigr)\,,
\end{align*}
as desired.
To show that $\sigma$~is associative,
let $T \in \bbT\bbT A$, $v \in \dom_0(T)$, and let $t$~be the tree from the
definition of associativity. Then
\begin{align*}
  t(x) = \sigma(T(x))(\emptyseq) = \pi_+(T(x)|_\emptyseq) = \pi_+(T(x))\,,
  \quad\text{for } x \in \dom_0(t)\,,
\end{align*}
implies that
\begin{align*}
  \sigma(t)(v)
  &= \pi_+(t|_v) \\
  &= \pi_+(\bbT\pi_+(T|_v)) \\
  &= \pi_+(\Flat(T|_v)) \\
  &= \pi_+(\Flat(T)|_{\mu(v)})
   = \sigma(\Flat(T))(\mu(v))\,.
\end{align*}

It remains to show that the mappings $\sigma \mapsto \pi_+$ and $\pi_+ \mapsto \sigma$
are inverse to each other.
Translating a product~$\pi_+$ to a scheme and back, we obtain the product
\begin{align*}
  \pi'_+(t) = \pi_+(t|_\emptyseq) = \pi_+(t)\,.
\end{align*}
Conversely, translating a scheme~$\sigma$ to a product and back, we obtain the scheme
\begin{align*}
  \sigma'(t)(v) = \sigma(t|_v)(\emptyseq)\,.
\end{align*}
To see that this value is equal to $\sigma(t)(v)$, consider the tree $T \in \bbT\bbT A$
consiting of a root labelled by the tree $t[\emptyseq,v)$ to which we attach the
tree~$\sing(t|_v)$. Let $w$~be the successor of the root of~$T$
and let $t'$~be tree with domain $\dom(t') = \dom(T)$ and labels
\begin{align*}
  t'(u) := \sigma(T(u))(\emptyseq)\,, \quad\text{for } u \in \dom_0(T)\,,
\end{align*}
as in the definition of associativity. By associativity of~$\sigma$, it follows that
\begin{align*}
  \sigma(t|_v)(\emptyseq)
  &= \sigma(T(w))(\emptyseq) \\
  &= t'(w) \\
  &= \sigma(\Flat(T))(\mu(w)) \\
  &= \sigma(t)(\mu(w)) \\
  &= \sigma(t)(v)\,.
\end{align*}
\upqed
\end{proof}

In particular, if labellings are unique, so is the expansion.
In fact, the following proposition shows that we do not need to assume associativity of
the labelling scheme here.
\begin{Prop}\label{Prop: unique labelling implies unique expansion}
Let\/ $\bbT^\fin \subseteq \bbT^0 \subseteq \bbT$ and let\/ $\frakA$~be a\/ $\bbT^0$-algebra
satisfying at least one of the following two conditions.
\begin{enumr}
\item Every tree $t \in \bbT A$ has a unique\/ weakly $\bbT^0$-consistent labelling.
\item Every tree $t \in \bbT A$ has a unique\/ strongly $\bbT^0$-consistent labelling.
\end{enumr}
Then\/ $\frakA$~has a unique\/ $\bbT$-expansion.
\end{Prop}
\begin{proof}
Let $\sigma$~be the unique labelling scheme (weak or strong).
By Proposition~\ref{Prop: expansions and labelling schemes},
it is sufficient to prove that $\sigma$~is associative.
Hence, fix a tree $T \in \bbT\bbT A$ and let $t$~and~$\mu$ be as in the definition of
associativity. We claim that the labelling
\begin{align*}
  \lambda := \sigma(\Flat(T)) \circ \mu
\end{align*}
is a strongly $\bbT^0$-consistent labelling of~$t$. Then uniqueness of labellings implies that
\begin{align*}
  \sigma(t) = \lambda = \sigma(\Flat(T)) \circ \mu\,,
\end{align*}
as desired.

For the proof, fix a factor $[u,\bar v)$ of~$t$ (possibly with infinitely many holes~$\bar v$).
We have to show that
\begin{align*}
  \lambda(u) = \pi\bigl(t[u,\bar v)(\lambda(v_0),\lambda(v_1),\dots)\bigr)\,.
\end{align*}
Since every tree $T(w)$ (with $w$~corresponding to some vertex in~$[u,\bar v)$)
only contains finitely many variables, we can replace in~$T(w)$ some subtrees~$T(w)|_{w'}$
(without variables) by the corresponding constant~$\sigma(T(w))(w')$.
Let $P(w)$~be a finite tree obtained in this way from~$T(w)$.
Using the consistency of $\sigma(P(w))$ and $\sigma(T(w))$, we can show
by induction on~$w'$ (starting at the leaves) that
\begin{align*}
  \sigma(P(w))(w') = \sigma(T(w))(w')\,, \quad\text{for all } w' \in \dom_0(P(w))\,.
\end{align*}
Consequently, consistency implies that
\begin{align*}
  \pi(P(w)) = \sigma(P(w))(\emptyseq) = \sigma(T(w))(\emptyseq) = t(w)\,.
\end{align*}
We regard the family~$(P(w))_w$ as a tree $P \in \bbT\bbT A$ with domain
\begin{align*}
  \dom(P) = [u,\bar v) \cup \{v_0,v_1,\dots\} \cup \bigcup_i L_i\,,
\end{align*}
where $L_i$~is a set of leaves attached to~$v_i$ corresponding to the variables
appearing in the subtree~$t|_{v_i}$, and the vertices~$v_i$ are labelled by
\begin{align*}
  P(v_i) := \lambda(v_i)\,.
\end{align*}
Then the domain of~$P$ is that of a tree in~$\bbT^0$
(if we ignore the fact that the root is~$u$ and not~$\emptyseq$).
Hence, we have $P \in \bbT^0\bbT^\fin A \subseteq \bbT^0\bbT^0 A$.
This implies by consistency that $\Flat(P) \in \bbT^0 A$ and
\begin{align*}
  \sigma(\Flat(P))(\emptyseq)
  &= \pi(\Flat(P)) \\
  &= \pi(\bbT^0\pi(P))
   = \pi\bigl(t[u,\bar v)(\lambda(v_0),\lambda(v_1),\dots)\bigr))\,,
\end{align*}
where the last step follows by the fact that
\begin{align*}
  \pi(P(w)) = \begin{cases}
                t(w)      &\text{if } w \in [u,\bar v)\,, \\
                \lambda(v_i) &\text{if } w = v_i\,.
              \end{cases}
\end{align*}
Hence, consistency of~$\sigma$ implies that
\begin{align*}
  \pi\bigl(t[u,\bar v)(\lambda(v_0),\lambda(v_1),\dots)\bigr)
  &= \sigma(\Flat(P))(\emptyseq) \\
  &= \sigma(\Flat(T)|_{\mu(u)})(\emptyseq) \\
  &= \sigma(\Flat(T))(\mu(u))
   = \lambda(u)\,,
\end{align*}
where the second step follows by consistency of~$\sigma$ and the third one by
uniqueness of the labellings.
(The function $w \mapsto \sigma(\Flat(T))(\mu(u)w)$ is a consistent labelling
of~$\Flat(T)|_{\mu(u)}$.)
\end{proof}

As an application let us show how to use consistent labellings to prove
that an algebra is definable.
\begin{Prop}\label{Prop: thin-algebras definable}
Every finitary\/ $\bbT^\thin$-algebra is\/ $\MSO$-definable.
\end{Prop}
\begin{proof}
Let $\frakA$~be a finitary $\bbT^\thin$-algebra, fix a finite set $C \subseteq A$
of generators, a sort~$\xi$, and an element $a \in A_\xi$.
Set $B := \bigcup_{\zeta \subseteq \xi} A_\zeta$.
We construct a formula that, given a tree $t \in \bbT^\thin_\xi C$, guesses
the labelling $\lambda : \dom(t) \to B$ induced by the product $\lambda(v) := \pi(t|_v)$ and
then verifies the correctness of its guess by checking for each vertex~$v$ that
\begin{itemize}
\item $\lambda(v) = t(v)(\lambda(u_0),\dots,\lambda(u_{n-1}))$, where $u_0,\dots,u_{n-1}$
  are the successors of~$v$,
\item $\lambda(v) = \pi(s_\beta)$, for every branch~$\beta$ starting at~$v$,
  where $s_\beta$~is the path obtained from~$t|_v$ by replacing every subtree attached at
  a vertex~$u$ not belonging to~$\beta$ by the tree $\sing(\lambda(u))$.
\end{itemize}
The latter condition can be expressed in $\MSO$ since this logic
can evaluate products in finite $\omega$-semigroups and
the $\omega$-semigroup we are working with has the domains
$\bigcup_{\zeta \subseteq \xi} A_{\zeta + \{x\}}$ and~$B$, both of which are finite.
By induction on the Cantor-Bendixson rank of~$t$ it follows that
the above checks ensure that the guessed labelling coincides with the intended one.
\end{proof}
\begin{Cor}\label{Cor: unambiguous algebras definable}
Let\/ $\frakA$~be a finitary\/ $\bbT$-algebra where every tree $t \in \bbT A$ has
exactly one weakly\/ $\bbT^\thin$-consistent labelling.
Then\/ $\frakA$~is\/ $\MSO$-definable.
\end{Cor}
\begin{proof}
Fix a finite set $C \subseteq A$ of generators of~$\frakA$.
We claim that $D := C \cup A_\emptyset$ is a set of generators of the
$\bbT^\thin$-reduct~$\frakA^0$ of~$\frakA$.
For the proof, fix $a \in A_\xi$. By assumption, there is some tree $t \in \bbT_\xi C$ with
$\pi(t) = a$. Let $s$~be the tree obtained from~$t$ by replacing every subtree without
a variable by its product. Then $s \in \bbT^\thin D$ and $\pi(s) = \pi(t) = a$.

Consequently, $\frakA^0$~is finitary and we can use
Proposition~\ref{Prop: thin-algebras definable} to show that $\frakA^0$~is $\MSO$-definable.
In particular, there exists a finite set $C \subseteq A$ generating~$A$ via
the product of~$\frakA^0$. It follows that $C$~is also a set of generators for~$\frakA$.
We claim that we can evaluate products of trees in $\bbT C$ in~$\MSO$.

Given a tree $t \in \bbT C$, we guess a labelling~$\lambda$ of~$t$
and check that it is weakly $\bbT^\thin$-consistent.
As the canonical labelling given by $v \mapsto \pi(t|_v)$ is weakly consistent,
it then follows by uniqueness of such labellings that $\lambda(\emptyseq) = \pi(t)$.
To see that consistency of~$\lambda$ can be expressed in~$\MSO$ note that,
by $\MSO$-definability of~$\frakA^0$, we can evaluate products of the form
\begin{align*}
  \pi\bigl(t[u,\bar v)(\lambda(v_0),\dots,\lambda(v_{n-1})\bigr)
  \qtextq{with}
  t[u,\bar v) \in \bbT^\thin C\,.
\end{align*}
Furthermore, note that the class of all thin trees is $\MSO$-definable.
(We only have to say that there is no embedding of the infinite binary tree.)
\end{proof}

\section{Unambiguous algebras}   
\label{Sect:unambiguous}

In Corollary~\ref{Cor: fin to thin expansions}, we have obtained a complete classification
of all $\bbT^\thin$-expansions of a $\bbT^\fin$-algebra.
In the present section, we use consistent labellings to study the inclusion
$\bbT^\thin \subseteq \bbT$. First let us remark that it is not dense.
\begin{Lem}\label{Lem: non-unique definable T-expansions}
There exists a\/ $\bbT^\thin$-algebra\/~$\frakA$ with two different\/ $\MSO$-definable\/
$\bbT$-expansions.
\end{Lem}
\begin{proof}
Let $A$~be the set with two elements $0_\xi,1_\xi$ for every sort~$\xi$.
We consider two different products $\pi_0,\pi_1 : \bbT A \to A$ on this set.
The first one is just the minimum operation\?:
\begin{align*}
  \pi_0(t) := \min {\set{ t(v) }{ v \in \dom(t) }}\,, \quad\text{for } t \in \bbT A\,.
\end{align*}
The second one is given by
\begin{align*}
  \pi_1(t) := \begin{cases}
                1 &\text{if } t \in \bbT^\thin C\,, \\
                0 &\text{if } t \in \bbT A \setminus \bbT^\thin C\,, \\
              \end{cases}
\end{align*}
where $C \subseteq A$ is the subset consisting of the elements~$1_\xi$, $\xi \in \Xi$.
Note that both functions coincide when restricted to thin trees.
Since there exists an $\MSO$-formula expressing that a given tree is thin,
both products are $\MSO$-definable. Furthermore, $\pi_0$~is clearly associative.
To show that so is~$\pi_1$, fix a tree $t \in \bbT\bbT A$. We distinguish three cases.
\begin{itemize}
\item If there is some $v \in \dom(t)$ with $t(v) \in \bbT A \setminus \bbT^\thin C$, we have
  $\Flat(t) \notin \bbT^\thin C$ and $\pi_1(\bbT\pi_1(t)) = 0 = \pi_1(\Flat(t))$.
\item If $t \in \bbT^\thin\bbT^\thin C$, then $\Flat(t) \in \bbT^\thin C$
  and $\pi_1(\bbT\pi_1(t)) = 1 = \pi_1(\Flat(t))$.
\item Finally, suppose that $t \in \bbT\bbT^\thin C \setminus \bbT^\thin\bbT^\thin C$.
  Then we have $\Flat(t) \in \bbT C \setminus \bbT^\thin C$ and
  $\pi_1(\bbT\pi_1(t)) = 0 = \pi_1(\Flat(t))$.
  \qedhere
\end{itemize}
\end{proof}

Consistent labellings have been used in~\cite{BilkowskiSkrzypczak13}
to study unambiguous tree languages. Let us give a brief overview over these results.
The central notion is the following one.
\begin{Def}
Let $\bbT^0 \subseteq \bbT$.
A $\bbT^0$-algebra~$\frakA$ is \emph{unambiguous} if every tree $t \in \bbT A$
has at most one strongly $\bbT^0$-consistent labelling.
\end{Def}
\begin{Rems}
    \begin{enuma}
\item For $\bbT^0 = \bbT^\thin$ these algebras were introduced
in~\cite{BilkowskiSkrzypczak13} under the name \emph{prophetic thin algebras.}

\item The fact that a given tree has a unique strongly $\bbT^\thin$-consistent labelling
is expressible in $\MSO$. \qedhere
\end{enuma}
\end{Rems}

First, note that there exist $\bbT^\thin$-algebras which are not unambiguous.
\begin{Exam}
Let $\frakA$~be the $\bbT^\thin$-algebra generated by the elements $0,1$ (of~arity~$0$),
$b_0,b_1,c_0,c_1$ (of~arity~$1$), and $a$~(of~arity~$2$) subject to the following equations.
\begin{alignat*}{-1}
  b_i(j) &= j\,, \qquad
  &a(x,i) &= b_i(x)\,, \\
  b_i(b_j(x)) &= b_{\max\{i,j\}}(x)\,, \qquad
  &a(i,x) &= c_i(x)\,, \\
  c_i(j) &= i\,, \qquad
  &b_i^\omega &= 1-i\,, \\
  c_i(b_j(x)) &= c_i(x)\,, \qquad
  &c_i^\omega &= i\,, \\
  c_i(c_j(x)) &= c_i(x)\,,
\end{alignat*}
for $i,j \in \{0,1\}$.
This algebra is not unambiguous since the (unique) tree $t \in \bbT_\emptyset\{a\}$
has several consistent labellings, including
\begin{alignat*}{-1}
  \lambda(w) := \abs{w}_1 \bmod 2
  \qtextq{and}
  \mu(w) := (\abs{w}_1 + 1) \bmod 2\,,
\end{alignat*}
where $\abs{w}_1$ denotes the number of letters~$1$ in $w \in \{0,1\}^*$.
\end{Exam}

The connection between unambiguous $\bbT^\thin$-algebras and unambiguous tree languages
is given by the following theorem.
\begin{Def}
    \begin{enuma}
\item A tree automaton is \emph{unambiguous} if it has at most one accepting run on
each given input tree.

\item
A language $K \subseteq \bbT_\xi\Sigma$ is called \emph{bi-unam\-bigu\-ous} if
both $K$ and $\bbT_\xi\Sigma \setminus K$ are recognised by unambiguous automata. \qedhere
\end{enuma}
\end{Def}
\begin{Thm}[Bilkowski, Skrzypczak~\cite{BilkowskiSkrzypczak13}]%
\label{Thm: characterisation of bi-unambiguous languages}
A language $K \subseteq \bbT_\xi\Sigma$ is bi-unam\-bigu\-ous if, and only if, it is recognised
by a morphism $\varphi : \bbT\Sigma \to \frakA$ of\/ $\bbT^\thin$-algebras
to a finitary unambiguous\/ $\bbT^\thin$-algebra~$\frakA$.
\end{Thm}

Unfortunately, the question of whether all trees have strongly $\bbT^\thin$-consistent labellings
is still an open problem, one which turns out to be equivalent to the existence of the
following kind of choice functions.
\begin{Def}
The \emph{Thin Choice Conjecture} states that there does \emph{not} exist an
$\MSO$-formula~$\varphi(x;Z)$ such that, for every thin (unlabelled) tree~$t$
and every non-empty set $P \subseteq \dom(t)$ of parameters,
the formula $\varphi(x;P)$ defines a unique element of~$P$.
\end{Def}

\begin{Thm}[Bilkowski, Skrzypczak~\cite{BilkowskiSkrzypczak13}]%
\label{Thm: characterisations of Thin Choice}
The following statements are equivalent.
\begin{enumn}
\item The Thin Choice Conjecture holds.
\item All trees have strongly\/ $\bbT^\thin$-consistent labellings,
  for every finitary\/ $\bbT^\thin$-algebra\/~$\frakA$.
\item The unique tree in\/ $\bbT_\emptyset\{a\}$ has a strongly\/ $\bbT^\thin$-consistent
  labelling, for every\/ $\bbT^\thin$-algebra\/~$\frakA$ and every $a \in A$.
\item For every morphism $\varphi : \frakA \to \frakB$ of\/ $\bbT^\thin$-algebras and
  every strongly\/ $\bbT^\thin$-consistent labelling~$\beta$ of some tree $t \in \bbT B$,
  there exists a strongly\/ $\bbT^\thin$-consistent labelling~$\alpha$ with
  $\varphi \circ \alpha = \beta$.
\end{enumn}
\end{Thm}
\noindent
(A~full proof can be found in Section~8.3 of~\cite{Skrzypczak16}, in particular
Theorem~8.74 and Proposition~8.22.)

The main applications of the Thin Choice Conjecture proved in~\cite{BilkowskiSkrzypczak13}
are as follows (cf.~Theorem~5 and Section~4 of~\cite{BilkowskiSkrzypczak13}).
For the definition of a pseudo-variety and a syntactic algebra
we refer the reader to standard accounts on algebraic language theory.
In the context of the monadic framework, this theory is worked out
in~\cite{Bojanczyk20,Blumensath21,BlumensathLN2}.
\begin{Thm}[Bilkowski, Skrzypczak]
Suppose that the Thin Choice Conjecture holds.
    \begin{enuma}
\item The class of unambiguous\/ $\bbT^\thin$-algebras forms a pseudo-variety.
\item A language $K \subseteq \bbT\Sigma$ is bi-unambiguous if, and only if,
  (the $\bbT^\thin$-reduct of) its syntactic algebra
  is unambiguous.
\item Bi-unambiguity of a language is decidable.
\end{enuma}
\end{Thm}

\section{Branch-continuous algebras}   
\label{Sect:branch-continuous}

In this final section, we take a look at a few other natural classes of $\bbT^\thin$-algebras
where unique $\bbT$-expansions exist. Originally, these classes were introduced
in~\cite{BlumensathZZ,Blumensath20}. A~unified account can be found in~\cite{BlumensathLN2}.
The simplest example consists of algebras that are constructed from an $\omega$-semigroup
as follows.
For the definition it is convenient to introduce a variant~$\bbT^?$ of the monad~$\bbT$
for trees where we are allowed to omit variables.
\begin{Def}
We set $\bbT^? X := (\bbT^?_\xi X)_{\xi \in \Xi}$ where
\begin{align*}
  \bbT^?_\xi X := \sum_{\zeta \subseteq \xi} \bbT_\zeta X\,.
\end{align*}
The functor~$\bbT^{?\thin}$ is defined analogously from~$\bbT^\thin$.

The corresponding flattening operation $\Flat : \bbT^?\bbT^? \Rightarrow \bbT^?$ is defined
in the obvious way\?: if the label at a vertex~$v$ is missing some variables, we omit
the corresponding subtrees and then flatten the resulting tree.
\end{Def}
It follows that a $\bbT^?$-algebra is just a $\bbT$-algebra that is equipped with additional
functions $A_\zeta \to A_\xi$, for all $\zeta \subseteq \xi$ (satisfying some natural laws).
We start with algebras arising from an $\omega$-semigroup in the following way.
\begin{Def}
    \begin{enuma}
\item Let $\bbT^{?\thin} \subseteq \bbT^0 \subseteq \bbT^?$.
A $\bbT^0$-algebra~$\frakA$ is \emph{semigroup-like} if it is generated by
$A_\emptyset \cup A_{\{z\}}$.

\item
Let $\frakS = \langle S,S_\omega\rangle$ be an $\omega$-semigroup.
We denote by $\TA(\frakS)$ the $\bbT^?$-algebra $\langle A,\pi\rangle$ with domains
\begin{align*}
  A_\xi := S_\omega + S \times \xi\,, \quad\text{for } \xi \in \Xi\,.
\end{align*}
For elements $\langle a,x\rangle \in S \times \xi$, we will use the more suggestive
notation $a(x)$.
The product is defined as follows. Given $t \in \bbT^?_\xi A$, let $\beta = (v_i)_i$ be the path
defined as follows. We start with the root~$v_0$ of~$t$.
Having chosen~$v_i$, we take a look at its label~$t(v_i)$.
If $t(v_i) = a_i(z_i) \in S \times \zeta_i$, we choose $v_{i+1} := \suc_{z_i}(v_i)$.
Otherwise, the path ends at~$v_i$. Let $(a_i)_i$ be the corresponding sequence of labels.
(If the path is finite, the last label~$a_n$ is either an element of~$S_\omega$ or a variable.)
We set
\begin{align*}
  \pi(t) := \prod_i a_i\,.
\end{align*}
Note that this product can be of one the following forms\?:
\begin{itemize}
\item an infinite product $a_0\cdot a_1\cdot\cdots \in S_\omega$ with $a_i \in S$,
\item a finite product $a_0 \cdot\cdots\cdot a_n \in S_\omega$ with $a_0,\dots,a_{n-1} \in S$
  and $a_n \in S_\omega$,
\item a finite product $\langle a_0 \cdot\cdots\cdot a_{n-1}, a_n\rangle \in S \times \xi$
  with $a_0,\dots,a_{n-1} \in S$ and $a_n \in \xi$~is a variable.
  \qedhere
\end{itemize}
\end{enuma}
\end{Def}
The following characterisation of semigroup-like algebras is from
\cite{BlumensathZZ,BlumensathLN2} (Proposition~4.21 of the former,
Proposition~\smaller{VIII}.2.7 of the latter\?; these propositions are only formulated for
$\bbT^?$-algebras, but the proof also works for $\bbT^{?\thin}$-algebras).
\begin{Prop}\label{Prop: adjunction for TA}
A $\bbT^?$-algebra~$\frakA$ is semigroup-like if, and only if, there exists a surjective
$\bbT^?$-morphism $\TA(\frakS) \to \frakA$, for some $\omega$-semigroup~$\frakS$.

Similarly, a $\bbT^{?\thin}$-algebra~$\frakA$ is semigroup-like if, and only if,
there exists a surjective $\bbT^{?\thin}$-morphism $\TA(\frakS) \to \frakA$,
for some $\omega$-semigroup~$\frakS$.
\end{Prop}

\begin{Lem}\label{Lem: semigroup-like algebras have unique expansions}
Every semigroup-like $\bbT^{?\thin}$-algebra is unambiguous and has a unique $\bbT^?$-expansion.
This expansion is again semigroup-like.
\end{Lem}
\begin{proof}
Let $\frakA$~be a semigroup-like $\bbT^{?\thin}$-algebra.
By Proposition~\ref{Prop: adjunction for TA}, $\frakA$~is a quotient of the
$\bbT^{?\thin}$-reduct of~$\TA(\frakS)$.
The corresponding quotient of~$\TA(\frakS)$ is a $\bbT^?$-expansion of~$\frakA$.
In particular, $\frakA$~has a semigroup-like $\bbT^?$-expansion.

For uniqueness, let us first prove that there exists at most one
strong labelling scheme for~$\frakA$.
Fix a tree $t \in \bbT^? A$ and let $\lambda$~and~$\mu$ be two strongly
$\bbT^{?\thin}$-consistent labellings of~$t$.
Given a vertex $v \in \dom(t)$, let $\beta$~be the path starting at~$v$
that we constructed in the definition of $\pi_+(t|_v)$.
We choose a thin factor~$p$ of~$t|_v$ containing this path.
Then $\bbT^{?\thin}$-consistency implies that $\lambda(v) = \pi(p) = \mu(v)$.
Hence, $\lambda = \mu$.

To conclude the proof, consider two $\bbT^?$-expansions $\frakA_1$~and~$\frakA_2$ of~$\frakA$.
By Proposition~\ref{Prop: expansions and labelling schemes}\,(b)
and the claim we have proved above, it follows that the $\bbT^\thin$-reduct of~$\frakA$
has at most one~$\bbT$-expansion. Hence, the $\bbT$-reducts of $\frakA_1$~and~$\frakA_2$
coincide.
Furthermore, the product of a $\bbT^?$-algebra is uniquely determined by its $\bbT$-reduct
and the by functions $A_\zeta \to A_\xi$ it induces.
Since the latter functions are given by products of finite trees and $\frakA_1$~and~$\frakA_2$
have the same $\bbT^{?\thin}$-reduct, it follows that these functions are also the same
for $\frakA_1$~and~$\frakA_2$. Consequently, the products of $\frakA_1$~and~$\frakA_2$ coincide.
\end{proof}

This lemma is hardly surprising, since the product of a semigroup-like algebra only depends
on a single branch of the given tree.
We can extend this result to more complicated classes of algebras as follows.
So far, we have mostly ignored the fact that our algebras are ordered.
(The ordering is needed to characterise logics that are not closed under negation,
something we are not concerned with in the present article.)
The next two classes of examples on the other hand make essential use of the ordering.
We start by introducing some notation concerning meets and joins.
\begin{Def}
Let $A$~be a sorted set.

    \begin{enuma}
\item For $X \subseteq A$, we set
\begin{align*}
  \Aboveseg X &:= \set{ a \in A }{ a \geq x \text{ for some } x \in X }\,, \\
  \Belowseg X &:= \set{ a \in A }{ a \leq x \text{ for some } x \in X }\,.
\end{align*}
The set~$X$ is \emph{upwards closed} if $\Aboveseg X = X$, and it is \emph{downwards closed}
if $\Belowseg X = X$.

\item We define two functors $\bbU$~and~$\bbD$ as follows. For sorted sets~$A$,
we set $\bbU A := (\bbU_\xi A)_\xi$ and $\bbD A := (\bbD_\xi A)_\xi$ where
\begin{align*}
  \bbU_\xi A &:= \set{ I \subseteq A_\xi }{ I \text{ is upwards closed} }\,, \\
  \bbD_\xi A &:= \set{ I \subseteq A_\xi }{ I \text{ is downwards closed} }\,.
\end{align*}
We order elements of~$\bbD A$ by inclusion, and those of~$\bbU A$ by inverse inclusion.
For functions $f : A \to B$, we define
\begin{align*}
  \bbU f(I) &:= \Aboveseg\set{ f(a) }{ a \in I }\,, \\
  \bbD f(I) &:= \Belowseg\set{ f(a) }{ a \in I }\,.
\end{align*}

\item For $\bbT^{?\thin} \subseteq \bbT^0 \subseteq \bbT^?$, $t \in \bbT^0 A$, and
$T \in \bbT^0\bbU A$ or $T \in \bbT^0\bbD A$, we write
\begin{align*}
  t \in^{\bbT^0} T \quad\defiff\quad
  &\text{$t$~and~$T$ have the same domain and if} \\
  &t(v) \in T(v)\,, \quad\text{for all vertices } v \in \dom_0(t)\,, \\
  &t(v) = T(v)\,, \quad\text{for all vertices } v \text{ labelled by a variable}\,.
\end{align*}

\item
Let $C \subseteq A$. We denote by $\llangle C\rrangle_{\inf}$ the closure of~$C$ under
(possibly infinite) meets and by $\llangle C\rrangle_{\sup}$ its closure under
(possibly infinite) joins.
$C$~is a set of \emph{meet-generators} if $\llangle C\rrangle_{\inf} = A$
and a set of \emph{join-generators} if $\llangle C\rrangle_{\sup} = A$.

\item We say that $A$ is \emph{completely ordered} if every subset $C \subseteq A$
has an infimum and a supremum. \qedhere
\end{enuma}
\end{Def}

The next, more interesting class of algebras we take a look at is the class of
\emph{deterministic} algebras, which was introduced in~\cite{Blumensath20} to give an algebraic
characterisation of the class of $\MSO$-definable $\bbT^?$-algebras.
Here, we are interested in the fact that their product is determined by its
$\bbT^{?\thin}$-reduct. The definition is as follows.
\begin{Def}
Let $\bbT^{?\thin} \subseteq \bbT^0 \subseteq \bbT^?$.

    \begin{enuma}
\item
We define a function $\dist : \bbT^0\bbU A \to \bbU\bbT^0 A$ by
\begin{align*}
  \dist(t) := \set{ s \in \bbT^0 A }{ s \in^{\bbT^0} t }\,.
\end{align*}

\item
A function $g : \frakA \to B$ from a $\bbT^0$-algebra $\frakA = \langle A,\pi\rangle$ to
a completely ordered sorted set~$B$ is \emph{meet-distributive} if $g$~preserves meets
and there exists a function $\sigma : \bbT^0\llangle\rng g\rrangle_{\inf} \to B$ such that
\begin{align*}
  \sigma \circ \bbT^0({\inf} \circ \bbU g) = {\inf} \circ \bbU(g \circ \pi) \circ \dist\,.
\end{align*}
We call the function~$\sigma$ the \emph{product of~$B$ induced} by~$g$.

A completely ordered $\bbT^0$-algebra~$\frakA$ is \emph{meet-distributive} if the identity
$\id : \frakA \to \frakA$ is meet-distributive.
\emph{Join-distributivity} is defined analogously with $\bbD$~and~$\sup$ instead of
$\bbU$~and~$\inf$.

\item A $\bbT^0$-algebra~$\frakA$ is \emph{deterministic} if it is meet-distributive and
it has a semi\-group-like subalgebra~$\frakC$ such that $C$~forms a set of meet-generators
of~$\frakA$. \qedhere
\end{enuma}
\end{Def}

We will show in the next lemma that, for a meet-distributive algebra, the
induced product coincides with the actual product.
Hence, meet-distributive algebras are those where the product commutes with meets.
More generally, it will follow by the results below that
an injective morphism $e : \frakA \to \frakB$ of $\bbT^?$-algebras
is meet-distributive if the restriction of~$\frakB$ to $\rng e$~is meet-distributive.
The name `deterministic algebra' stems from the fact that such algebras correspond to
deterministic automata.
A~typical example of a deterministic algebra is one where every element is of the form
\begin{align*}
  a_0(x_0) \sqcap\cdots\sqcap a_{m-1}(x_{m-1}) \sqcap b_0 \sqcap\cdots\sqcap b_{n-1}\,,
\end{align*}
where $a_0,\dots,a_{m-1} \in S$ and $b_0,\dots,b_{n-1} \in S_\omega$
are elements of some $\omega$-semigroup $\frakS = \langle S,S_\omega\rangle$
and the $x_0,\dots,x_{m-1}$ are variables.

We start with two technical lemmas. The first one is trivial.
\begin{Lem}\label{Lem: condition for meet-distributivity of an algebra}
Let $\bbT^{?\thin} \subseteq \bbT^0 \subseteq \bbT^?$.
A~completely ordered\/ $\bbT^0$-algebra\/~$\frakA$ is meet-distributive if, and only if,
\begin{align*}
  \pi \circ \bbT^0{\inf} = {\inf} \circ \bbU\pi \circ \dist\,.
\end{align*}
\end{Lem}
\begin{proof}
$(\Leftarrow)$ In the definition of meet-distributivity, we can take $\sigma := \pi$.

$(\Rightarrow)$ Let $\sigma$~be the function from the definition of meet-distributivity.
Given a tree $t \in \bbT^0 A$, let $T \in \bbT^0\bbU A$ be the tree with labels
$T(v) = \set{ a \in A }{ a \geq t(v) }$. Then
\begin{align*}
  \sigma(t) = \sigma(\bbT^0\inf(T))
            = \inf \Aboveseg{\set{ \pi(s) }{ s \in^{\bbT^0} T }}
            = \inf \Aboveseg{\{ \pi(t) \}}
            = \pi(t)\,.
\end{align*}
Hence, we have
\begin{align*}
  \pi \circ \bbT^0{\inf}
  &= \sigma \circ \bbT^0({\inf} \circ \bbU\id) \\
  &= {\inf} \circ \bbU(\id \circ \pi) \circ \dist
   = {\inf} \circ \bbU\pi \circ \dist\,.
\end{align*}
\upqed
\end{proof}

Meet-distributive functions can be used to transfer a $\bbT^?$-algebra product
from their domain to their codomain.
\begin{Lem}\label{Lem: extending a product to the meet closure}
Let $\bbT^{?\thin} \subseteq \bbT^0 \subseteq \bbT^?$.
Let $\varphi : \frakC \to A$ be a meet-distributive function such that\/
$\rng \varphi$~is a set of meet-generators of~$A$.
There exists a unique function $\sigma : \bbT^0 A \to A$ such that
$\langle A,\sigma\rangle$ is a meet-distributive\/ $\bbT^0$-algebra
and $\varphi$~a morphism of\/ $\bbT^0$-algebras.
\end{Lem}
\begin{proof}
To make our proof more concise, we use some properties of the function
$\dist : \bbT^0\bbU \Rightarrow \bbU\bbT^0$.
We have shown in~\cite{Blumensath23} that $\dist$~is what is called a \emph{distributive law,}
which means it is a natural transformation satisfying the equations
\begin{alignat*}{-1}
   \dist \circ \Flat      &= \bbU\Flat \circ \dist \circ \bbT^0\dist\,, \qquad\quad
  &\dist \circ \sing      &= \bbU\sing\,, \\
   \dist \circ \bbT^0\union &= \union \circ \bbU\dist \circ \dist\,, \qquad\quad
  &\dist \circ \bbT^0\pt    &= \pt\,,
\end{alignat*}
where $\union : \bbU\bbU A \to \bbU A$ maps a set of sets to its union and
$\pt : A \to \bbU A$ is defined by $\pt(a) := \Aboveseg\{ a \}$.

Let $\sigma : \bbT^0 A \to A$ be the product induced by~$\varphi$ as in the definition of
meet-distributivity.
To see that $\langle A,\sigma\rangle$ is a $\bbT^0$-algebra, note that
\begin{alignat*}{-1}
       & \sigma \circ \sing \circ ({\inf} \circ \bbU\varphi) \\
  {}={}& \sigma \circ \bbT^0({\inf} \circ \pt) \circ \sing \circ ({\inf} \circ \bbU\varphi)
         &&\qquad\text{[${\inf} \circ \pt = \id$]}\\
  {}={}& \sigma \circ \bbT^0({\inf} \circ \pt) \circ \bbT^0({\inf} \circ \bbU\varphi) \circ \sing
         &&\qquad\text{[$\sing$ nat. trans.]}\\
  {}={}& \sigma \circ \bbT^0{\inf} \circ \bbT^0\bbU({\inf} \circ \bbU\varphi) \circ \bbT^0\pt
           \circ \sing
         &&\qquad\text{[$\pt$ nat. trans.]}\\
  {}={}& \sigma \circ \bbT^0{\inf} \circ \bbT^0({\union} \circ \bbU\bbU\varphi) \circ \bbT^0\pt
           \circ \sing
         &&\qquad\text{[${\inf} \circ \bbU{\inf} = {\inf} \circ \union$]}\\
  {}={}& \sigma \circ \bbT^0{\inf} \circ \bbT^0(\bbU\varphi \circ {\union}) \circ \bbT^0\pt
           \circ \sing
         &&\qquad\text{[$\union$ nat. trans.]}\\
  {}={}& {\inf} \circ \bbU(\varphi \circ \pi) \circ \dist \circ \bbT^0\union \circ \bbT^0\pt
           \circ \sing
         &&\qquad\text{[$\varphi$ meet-dist.]}\\
  {}={}& {\inf} \circ \bbU(\varphi \circ \pi) \circ \dist \circ \sing
         &&\qquad\text{[$\union \circ \pt = \id$]}\\
  {}={}& {\inf} \circ \bbU(\varphi \circ \pi) \circ \bbU\sing
         &&\qquad\text{[$\dist$ dist. law]}\\
  {}={}& {\inf} \circ \bbU\varphi\,,
         &&\qquad\text{[unit law for $\pi$]} \displaybreak[0]\\[0.5em]
       & \sigma \circ \bbT^0\sigma \circ \bbT^0\bbT^0({\inf} \circ \bbU\varphi) \\
  {}={}& \sigma \circ \bbT^0({\inf} \circ \bbU(\varphi \circ \pi) \circ \dist)
         &&\qquad\text{[$\varphi$ meet-dist.]}\\
  {}={}& {\inf} \circ \bbU(\varphi \circ \pi) \circ \dist \circ \bbT^0\bbU\pi \circ \bbT^0\dist
         &&\qquad\text{[$\varphi$ meet-dist.]} \\
  {}={}& {\inf} \circ \bbU(\varphi \circ \pi) \circ \bbU\bbT^0\pi \circ \dist \circ \bbT^0\dist
         &&\qquad\text{[$\dist$ nat. trans.]} \\
  {}={}& {\inf} \circ \bbU(\varphi \circ \pi) \circ \bbU\Flat \circ \dist \circ \bbT^0\dist
         &&\qquad\text{[$\frakC$ $\bbT^0$-algebra]} \\
  {}={}& {\inf} \circ \bbU(\varphi \circ \pi) \circ \dist \circ \Flat
         &&\qquad\text{[$\dist$ dist. law]} \\
  {}={}& \sigma \circ \bbT^0{\inf} \circ \bbT^0\bbU\varphi \circ \Flat
         &&\qquad\text{[$\varphi$ meet-dist.]} \\
  {}={}& \sigma \circ \Flat \circ \bbT^0\bbT^0({\inf} \circ \bbU\varphi)\,.
         &&\qquad\text{[$\Flat$ nat. trans.]}
\end{alignat*}
Since ${\inf} \circ \bbU\varphi$ is surjective and $\bbT^0$~preserves surjectivity,
it follows that
\begin{align*}
  \sigma \circ \sing = \id
  \qtextq{and}
  \sigma \circ \bbT^0\sigma = \sigma \circ \Flat\,.
\end{align*}

To see that $\langle A,\sigma\rangle$~is meet-distributive, note that
\begin{align*}
      &\sigma \circ \bbT^0{\inf} \circ \bbT^0\bbU({\inf} \circ \bbU\varphi) \\
{}={} &\sigma \circ \bbT^0({\inf} \circ {\union} \circ \bbU\bbU\varphi)
        &&\qquad\text{[${\inf} \circ \bbU{\inf} = {\inf} \circ \union$]}\\
{}={} &\sigma \circ \bbT^0({\inf} \circ \bbU\varphi \circ {\union})
        &&\qquad\text{[$\union$ nat. trans.]}\\
{}={} &{\inf} \circ \bbU(\varphi \circ \pi) \circ \dist \circ \bbT^0\union
        &&\qquad\text{[$\varphi$ meet-dist.]}\\
{}={} &{\inf} \circ \bbU(\varphi \circ \pi) \circ \union \circ \bbU\dist \circ \dist
        &&\qquad\text{[$\dist$ dist. law]}\\
{}={} &{\inf} \circ \union \circ \bbU\bbU(\varphi \circ \pi) \circ \bbU\dist \circ \dist
        &&\qquad\text{[$\union$ nat. trans.]}\\
{}={} &{\inf} \circ \bbU{\inf} \circ \bbU\bbU(\varphi \circ \pi) \circ \bbU\dist \circ \dist
        &&\qquad\text{[${\inf} \circ \bbU{\inf} = {\inf} \circ \union$]}\\
{}={} &{\inf} \circ \bbU(\sigma \circ \bbT^0({\inf} \circ \bbU\varphi)) \circ \dist
        &&\qquad\text{[$\varphi$ meet-dist.]}\\
{}={} &{\inf} \circ \bbU\sigma \circ \dist \circ \bbT^0\bbU({\inf} \circ \bbU\varphi)\,.
        &&\qquad\text{[$\dist$ nat. trans.]}
\end{align*}
By surjectivity of $\bbT^0\bbU({\inf} \circ \bbU\varphi)$, this implies that
\begin{align*}
  \sigma \circ \bbT^0{\inf}
  = {\inf} \circ \bbU\sigma \circ \dist\,.
\end{align*}
Hence, the claim follows by Lemma~\ref{Lem: condition for meet-distributivity of an algebra}.

To see that $\varphi$~is a morphism of $\bbT^0$-algebras, note that
\begin{alignat*}{-1}
  \sigma \circ \bbT^0\varphi
  &= \sigma \circ \bbT^0({\inf} \circ \pt \circ \varphi)
       &&\qquad\text{[${\inf} \circ \pt = \id$]}\\
  &= \sigma \circ \bbT^0({\inf} \circ \bbU\varphi \circ \pt)
       &&\qquad\text{[$\pt$ nat. trans.]}\\
  &= {\inf} \circ \bbU(\varphi \circ \pi) \circ \dist \circ \bbT^0\pt
       &&\qquad\text{[$\varphi$ meet-dist.]}\\
  &= {\inf} \circ \bbU(\varphi \circ \pi) \circ \pt
       &&\qquad\text{[$\dist$ dist. law]}\\
  &= {\inf} \circ \pt \circ \varphi \circ \pi
       &&\qquad\text{[$\pt$ nat. trans.]}\\
  &= \varphi \circ \pi\,.
       &&\qquad\text{[${\inf} \circ \pt = \id$]}
\end{alignat*}

Finally, for uniqueness suppose that $\sigma' : \bbT^0 A \to A$ is another function such
that $\langle A,\sigma'\rangle$ is meet-distributive and
$\varphi : \frakC \to \langle A,\sigma'\rangle$ is a morphism of $\bbT^0$-algebras.
Then it follows that
\begin{align*}
  \sigma \circ \bbT^0({\inf} \circ \bbU\varphi)
  &= {\inf} \circ \bbU(\varphi \circ \pi) \circ \dist
       &&\qquad\text{[choice of $\sigma$]}\\
  &= {\inf} \circ \bbU(\sigma' \circ \bbT^0\varphi) \circ \dist
       &&\qquad\text{[$\varphi$ morphism]}\\
  &= {\inf} \circ \bbU\sigma' \circ \dist \circ \bbT^0\bbU\varphi
       &&\qquad\text{[$\dist$ nat. trans.]}\\
  &= \sigma' \circ \bbT^0{\inf} \circ \bbT^0\bbU\varphi\,.
       &&\qquad\text{[Lemma~\ref{Lem: condition for meet-distributivity of an algebra}]}
\end{align*}
Hence, the fact that $\bbT^0({\inf} \circ \bbU\varphi)$ is surjective implies that
$\sigma = \sigma'$.
\end{proof}

\begin{Lem}\label{Lem: inclusion of semigroup-like subalgebra is meet-distributive}
Let $A$~be a sort-wise finite, completely ordered set and\/ $\frakC$~a semigroup-like
$\bbT^?$-algera whose universe $C \subseteq A$ is contained in~$A$.
If the inclusion\/ $\frakC|_{\bbT^{?\thin}} \to A$ is meet-distributive, so is the
inclusion\/ $\frakC \to A$.
\end{Lem}
\begin{proof}
Suppose that the inclusion $i_0 : \frakC|_{\bbT^{?\thin}} \to A$ is meet-distributive
and let $\sigma_0 : \bbT^{?\thin} \llangle C\rrangle_{\inf} \to A$ be the corresponding function.
To prove that $i : \frakC \to A$ is also meet-distributive, it is sufficient to show that
\begin{align*}
  \bbT^?\inf(t) \leq \bbT^?\inf(t')
  \!\quad\Rightarrow\quad\!
  \inf {\set{ \pi(s) }{ s \in^{\bbT^?} t }}
  \leq \inf {\set{ \pi(s') }{ s' \in^{\bbT^?} t' }}\,,
\end{align*}
for all $t,t' \in \bbT^?\bbU C$,
that is, the \emph{kernel} of the function $\bbT^?\inf$ is included in the kernel of
${\inf} \circ \bbU\pi \circ \dist$. It then follows that
there exists a unique function $\sigma : \bbT^?\llangle C\rrangle_{\inf} \to A$ with
$\sigma \circ \bbT^?{\inf} = {\inf} \circ \bbU\pi \circ \dist$.

Hence, fix two trees $t,t' \in \bbT^?_\xi\bbU C$ with $\bbT^?\inf(t) \leq \bbT^?\inf(t')$.
We distinguish two cases. Let $\top_\zeta$~be the top element of~$A_\zeta$.
First, suppose that $\top_\emptyset \in C_\emptyset$.
We claim that, in this case, we have
\begin{align*}
  C_\zeta = \{\top_\zeta\}\,, \quad\text{for all } \zeta \in \Xi\,.
\end{align*}
Fix an element $a \in C_\zeta$ and let $z_0,\dots,z_{n-1}$ be an enumeration of~$\zeta$.
We have to show that $a = \top_\zeta$.
Since $\frakC$~is semigroup like, the element~$a$ is of the form $a(\bar z) = a'(z_i)$,
for some $a' \in C_{\{x\}}$ and some $i < n$.
We consider the tree $s \in \bbT^{?\thin}_\zeta\bbU C$ given by
\begin{center}
\includegraphics{Expansion-8.mps}
\end{center}
(The vertex $\Aboveseg\{a\}$ is considered to be an element of sort $\zeta + \{y\}$.)
By meet-distributivity of $\frakC|_{\bbT^{?\thin}} \to A$, we have
\begin{align*}
  a = a'(z_i)
  = \pi(\bbT^{?\thin}\inf(s))
  &= \sigma_0(\bbT^{?\thin}\inf(s)) \\
  &= \inf {\set{ \pi(r) }{ r \in^{\bbT^{?\thin}} s }}
  = \inf \emptyset
  = \top_\zeta\,,
\end{align*}
as desired. (By~(the proof of) Lemma~\ref{Lem: condition for meet-distributivity of an algebra}
the restriction of~$\sigma_0$ to~$\bbT^{?\thin} C$ coincides with~$\pi$.)

To prove meet-distributivity of $\frakC \to A$, it is now sufficient to note that,
\begin{alignat*}{-1}
  \inf {\set{ \pi(r) }{ r \in^{\bbT^{?\thin}} s }} &= \inf \emptyset &&= \top_\xi \\
\prefixtext{or}
  \inf {\set{ \pi(r) }{ r \in^{\bbT^{?\thin}} s }} &= \inf {\{\top_\xi\}} &&= \top_\xi\,,
\end{alignat*}
for every tree $s \in \bbT^{?\thin}_\xi\bbU C$. In particular, this holds for $t$~and~$t'$.

It remains to consider the case where $\top_\emptyset \notin C_\emptyset$.
Note that, by definition of the product of $\TA(\frakS)$ and
Proposition~\ref{Prop: adjunction for TA},
there exists, for every $s \in \bbT^? C$, some branch~$\beta$ such that
\begin{align*}
  \pi(s|_\beta) = \pi(s)\,,
\end{align*}
where $s|_\beta$~is the tree obtained from~$s$ by removing all vertices that do not belong
to~$\beta$. (This changes the sorts of the remaining vertices\?: every vertex of sort~$\xi$
has a label of the form $a(x)$ or~$c$, for some $a \in C_{\{z\}}$, $c \in C_\emptyset$,
and $x \in \xi$. We change the sort of such a vertex to $\{x\}$ while keeping the label.
In the latter case, the variable~$x$ is arbitrary.)
For each tree~$s$, we pick one such branch~$\beta$ and call it the \emph{main branch} of~$s$.

As $A$~is sort-wise finite, every infimum in~$A$ can be written as the infimum of
a finite subset. We can therefore find a finite set~$B$ of branches of~$t$ such that
\begin{align*}
  \inf {\bigset{ \pi(s) }{ s \in^{\bbT^?} t }}
  &= \inf {\bigset{ \pi(s|_\beta) }{ \beta \in B,\ s \in^{\bbT^?} t,\ \beta \text{ main branch of } s }}\,, \\
  \llap{\inf} {\bigset{ \pi(s') }{ s' \in^{\bbT^?} t' }}
  &= \inf {\bigset{ \pi(s'|_\beta) }{ \beta \in B,\ s' \in^{\bbT^?} t',\ \beta \text{ main branch of } s' }}\,.
\end{align*}
(Note that $t$~and~$t'$ have the same domain and, hence, the same set of branches.)

For each $v \in \dom(t)$, set
\begin{align*}
  X_v := \bigset{ c \in C }{ c \geq \pi(s),\ s \in^{\bbT^?} t|_v }\,.
\end{align*}
Let $r$~and~$r'$ be the trees obtained from, respectively, $t$~and~$t'$ by replacing
every subtree whose root~$v$ does not lie on a branch from~$B$ by a leaf with label~$X_v$
(the same label in $r$~and~$r'$).
As~$B$~is finite, the trees $r$~and~$r'$ are thin.
By meet-distributivity of~$i_0$, it therefore follows that
\begin{align*}
  \sigma_0(\bbT^{?\thin}\inf(r))  &= \inf {\bigset{ \pi(s) }{ s \in^{\bbT^{?\thin}} r }}\,, \\
  \sigma_0(\bbT^{?\thin}\inf(r')) &= \inf {\bigset{ \pi(s') }{ s' \in^{\bbT^{?\thin}} r' }}\,.
\end{align*}
Furthermore, by construction of $r$~and~$r'$,
\begin{align*}
  \bbT^?\inf(t) \leq \bbT^?\inf(t')
  \qtextq{implies}
  \bbT^{?\thin}\inf(r) \leq \bbT^{?\thin}\inf(r')\,.
\end{align*}
It follows that
\begin{align*}
  & \inf {\set{ \pi(s) }{ s \in^{\bbT^?} t }} \\
  &\quad{}= \inf {\set{ \pi(s) }{ s \in^{\bbT^{?\thin}} r }} \\
  &\quad{}= \sigma_0(\bbT^{?\thin}\inf(r)) \\
  &\quad{}\leq \sigma_0(\bbT^{?\thin}\inf(r')) \\
  &\quad{}= \inf {\set{ \pi(s') }{ s' \in^{\bbT^{?\thin}} r' }} \\
  &\quad{}\leq \inf {\bigset{ \pi(s'|_\beta) }{ \beta \in B,\ s' \in^{\bbT^{?\thin}} r',\ \beta \text{ main branch of } s' }} \\
  &\quad{}\leq \inf {\bigset{ \pi(s'|_\beta) }{ \beta \in B,\ s' \in^{\bbT^?} t',\ \beta \text{ main branch of } s' }} \\
  &\quad{}= \inf {\set{ \pi(s') }{ s' \in^{\bbT^?} t' }}\,.
\end{align*}
For the first step above, note that associativity of the product implies that
\begin{align*}
  &\inf {\bigset{ \pi(s) }{ s \in^{\bbT^?} t,\ \text{the main branch of } s \text{ contains } v }} \\
  {}={} &\inf {\bigset{ \pi(s) }{ s \in^{\bbT^?} r,\ \text{the main branch of } s \text{ contains } v }}\,,
\end{align*}
for every vertex~$v$ that is replaced in~$r$ by a constant.
For the sixth step, we have to show that every $s' \in^{\bbT^?} t'$ induces some
$s'' \in^{\bbT^{?\thin}} r'$. For a contradiction, suppose otherwise.
Then there must be some leaf~$v$ of~$r'$ with $X_v = \emptyset$.
By definition of~$X_v$, it follows that we can find a vertex
$u \in \dom(t) \setminus \dom(r)$ in the subtree attached at~$v$ such that $t(u) = \emptyset$.
Hence, $\inf t(u) \leq \inf t'(u)$ implies that $\inf t'(u) = \top$.
Since $\top \notin C$, it follows that $t'(u) = \emptyset$.
A~contradiction to the fact that $s' \in^{\bbT^?} t'$.
\end{proof}

\begin{Thm}\label{Thm: deterministic algebras have unique expansions}
Every deterministic\/ $\bbT^{?\thin}$-algebra has a unique meet-dis\-tribu\-tive\/
$\bbT^?$-expansion.
\end{Thm}
\begin{proof}
Let $\frakA = \langle A,\pi\rangle$ be a deterministic $\bbT^{?\thin}$-algebra and
let $\frakC \subseteq \frakA$ be the corresponding semigroup-like subalgebra.
We can use Lemma~\ref{Lem: semigroup-like algebras have unique expansions} to find a
unique $\bbT^?$-expansion~$\frakC_+$ of~$\frakC$.
By Lemma~\ref{Lem: inclusion of semigroup-like subalgebra is meet-distributive},
the inclusion $\frakC_+ \to A$ is meet-distributive.
Consequently, we can use Lemma~\ref{Lem: extending a product to the meet closure}
to find a unique meet-distributive algebra $\frakA_+ = \langle A,\pi_+\rangle$ with
universe~$A$ that contains~$\frakC_+$ as a subalgebra.

It therefore remains to prove that $\frakA$~is
the $\bbT^{?\thin}$-reduct of this algebra~$\frakA_+$.
Hence, let $t \in \bbT^{?\thin} A$ and fix a tree $T \in \bbT^{?\thin}\bbU C$ such that
$t = \bbT^{?\thin}\inf(T)$. By meet-distributivity and the fact that the products
$\pi$~and~$\pi_+$ agree on trees in $\bbT^{?\thin} C$, it follows that
\begin{align*}
  \pi_+(t) = \inf \Aboveseg{\set{ \pi_+(s) }{ s \in^{\bbT^?} T  }}
           = \inf \Aboveseg{\set{ \pi(s) }{ s \in^{\bbT^?} T  }}
           = \pi(t)\,,
\end{align*}
as desired.
\end{proof}

\begin{Cor}
Every deterministic $\bbT^?$-algebra is uniquely determined by its $\bbT^{?\thin}$-reduct.
\end{Cor}

We can generalise deterministic algebras by also allowing joins.
The resulting algebras are called \emph{branch-continuous.}
They were introduced in~\cite{BlumensathZZ} as an algebraic analogue to tree automata.
\begin{Def}
Let $\bbT^{?\thin} \subseteq \bbT^0 \subseteq \bbT^?$.
A $\bbT^0$-algebra~$\frakA$ is \emph{branch-continuous} if it is join-distributive and
it has a deterministic subalgebra~$\frakC$ such that $C$~forms a set of join-generators
of~$\frakA$ and the inclusion $\frakC \to A$ is meet-distributive.
\end{Def}
\begin{Exam}
A typical example of a branch-continuous algebra consists of an algebra of (profiles of) certain
games. A~\emph{regular game} is played on a directed graph the form
$\frakG = \langle V_\rmI,V_{\rmI\rmI},E,\lambda,\mu,v_0\rangle$
where the set $V = V_\rmI + V_{\rmI\rmI}$ of vertices is divided into two parts,
one for each player, and the vertices and edges are labelled by elements of
some finite $\omega$-semigroup $\frakS = \langle S,S_\omega\rangle$.
This labelling is given by the functions
\begin{align*}
  \lambda : E \to S
  \qtextq{and}
  \mu : V \to \PSet(S_\omega)\,.
\end{align*}
We assume that $\mu(v) \neq \emptyset$, for every leaf~$v$.
The vertex $v_0 \in V$ denotes the initial position of the game.

Since we want to compose games we allow some of the leaves of~$\frakG$ to be labelled
by (distinct) variables. For such leaves~$v$, we assume that $\mu(v) = \emptyset$.

The game is played between two players, Player~I and Player~II, and proceeds as follows.
It starts in the position~$v_0$.
If the game has reached some position $v \in V$,
the player to whom~$v$ belongs chooses either some semigroup element $c \in \mu(v)$ or
some outgoing edge $v \to u$.
In the first case the game terminates, otherwise it continues in position~$u$.
It follows that each play of the game produces a finite or infinite path starting at the root.
The labelling of this path is of one of the following forms.
\begin{align*}
  a_0a_1a_2\cdots\,,\quad
  a_0\cdots a_{n-1}c\,,\quad
  a_0\cdots a_{n-1}x\,,
\end{align*}
where $a_i \in S$, $c \in S_\omega$, and $x$~is one of the variables.
Each such sequence can be multiplied to either an element of~$S_\omega$ or an element
of the form $a(x)$ with $a \in S$ and $x$~a variable.
We call this product the \emph{outcome} of the play.

The set of all games (over some fixed $\omega$-semigroup~$\frakS$) forms a $\bbT^?$-algebra.
But here we are interested in a finitary quotient of this algebra.
When we are only interested in the possible outcomes of a game and not in its game graph,
we can represent each game~$\frakG$ as a term of the form
\begin{align*}
  \sup_{\sigma_\rmI} \inf_{\sigma_{\rmI\rmI}} a_{\sigma_\rmI,\sigma_{\rmI\rmI}}\,,
\end{align*}
where $\sigma_\rmI$ and $\sigma_{\rmI\rmI}$ range over all strategies for the respective
player and $a_{\sigma_\rmI,\sigma_{\rmI\rmI}}$ is the outcome of the game
when both players play according to the indicated strategies.
Let us call such a term the \emph{profile} of the game~$\frakG$.

The set of all possible profiles (again, for some fixed $\omega$-semigroup~$\frakS$),
forms a $\bbT^?$-algebra~$\frakA$ which is branch-continuous.
The elements of the form $a_{\sigma_\rmI,\sigma_{\rmI\rmI}}$ form a semigroup-like subalgebra,
those of the form $\inf_{\sigma_{\rmI\rmI}} a_{\sigma_\rmI,\sigma_{\rmI\rmI}}$
form a deterministic subalgebra, and every element of~$A$ is a join of such elements.
\end{Exam}

Using join-distributivity and meet-distributivity, one can show that a product $\pi(t)$
in a branch-continuous algebra can be computed by taking a join over meets over
products along single branches of~$t$ (see~\cite{BlumensathZZ} for details).
In particular, a product of this form is $\MSO$-definable.
Together with the translation of automata into branch-continuous $\bbT^?$-algebras,
this leads to the following two results from~\cite{BlumensathZZ}
(Proposition~4.37 and Theorem~4.42, respectively).
\begin{Prop}
Every finitary branch-continuous $\bbT^?$-algebra is $\MSO$-defin\-able.
\end{Prop}
\begin{Thm}
A language $K \subseteq \bbT^?\Sigma$ is regular if, and only if, it is recognised by
a morphism into a finitary branch-continuous $\bbT^?$-algebra.
\end{Thm}
The class of branch-continuous algebras is a proper subclass of the one of $\MSO$-definable
algebras. Both classes can play a similar role in language theory.
The reason we usually work with $\MSO$-definable algebras instead of branch-continuous ones
is that the latter do not form a pseudo-variety\?: the class of branch-continuous algebras
is not closed under finitely-generated subalgebras.
In particular, syntactic algebras are usually not branch-continuous.
Here, we are more interested in the fact that branch-continuous algebras have
unique branch-continuous expansions, although we can only prove uniqueness, not existance,
since we are missing an analogue of
Lemma~\ref{Lem: inclusion of semigroup-like subalgebra is meet-distributive}.
The proof makes use of the following observation.
\begin{Lem}\label{Lem: minimal deterministic subalgebra}
Let $\bbT^{?\thin} \subseteq \bbT^0 \subseteq \bbT^?$.
Every finitary branch-continuous $\bbT^0$-algebra~$\frakA$ has a least deterministic
subalgebra~$\frakC$ such that $C$~forms a set of join-generators of~$\frakA$
and the inclusion $\frakC \to A$ is meet-distributive.
\end{Lem}
\begin{proof}
Let $I \subseteq A$ be the set of all \emph{join-irreducible} elements of~$A$,
i.e., elements that cannot be expressed as a (possibly infinite) join of strictly
smaller elements. Then $I$~is contained in every set of join-generators of~$A$.
Furthermore, since $A$~is sort-wise finite, $I$~forms a set of join-generators of~$A$.
(Given an element $a \in A$, one can show by induction on the number of elements strictly
smaller than~$a$ that $a$~is a join of elements of~$I$.)

As $\frakA$~is branch-continuous, there exists a deterministic subalgebra
$\frakD \subseteq \frakA$ such that $D$~forms a set of join-generators of~$A$ and
the inclusion $D \to A$ is meet-distributive. Note that $I \subseteq D$ by the above remark.
Let $U$~be the closure of $I_{<2} := I_\emptyset \cup I_{\{z\}}$ under the product of~$\frakA$
and let $C$~be the closure of~$U$ under meets.
We claim that the set~$C$ induces the desired subalgebra of~$\frakA$.

We start by proving that $C$~is a set of join-generators of~$\frakA$.
To do so it is sufficient to show that $I \subseteq C$. Hence, let $a \in I$.
Since $I \subseteq D$, it follows that $a \in D$.
As~$\frakD$~is deterministic, we can write~$a$ as a meet of elements of some
semigroup-like subalgebra of~$\frakD$. Furthermore, each element of this subalgebra
can be written as a product of elements of arity at most~$1$. Hence,
\begin{align*}
  a = \inf_{i<n} \pi(t_i)\,, \quad\text{for some } t_i \in \bbT^0D_{<2}\,.
\end{align*}
As $D_{<2} \subseteq \llangle I_{<2}\rrangle_{\sup}$,
we can find trees $T_i \in \bbT^0\PSet(I_{<2})$ such that
\begin{align*}
  t_i = \bbT^0\sup(T_i)\,.
\end{align*}
Hence,
\begin{align*}
  a &= \inf_{i<n} \pi(\bbT^0\sup(T_i)) \\
    &= \inf_{i<n} \sup {\set{ \pi(s) }{ s \in^{\bbT^0} T_i }} \\
    &= \sup_f \inf_{i<n} \pi(f(i))\,,
\end{align*}
where $f$ ranges over all functions $f : [n] \to \bbT^0 I_{<2}$ such that
$f(i) \in^{\bbT^0} T_i$, for all $i<n$.
By join-irreducibility of $a \in I$, it follows that
\begin{alignat*}{-1}
  a &= \inf_{i<n} \pi(f(i))\,, &&\quad\text{for some } f\,, \\
    &= \inf_{i<n} \pi(s_i)\,,  &&\quad\text{for some } s_i \in^{\bbT^0} T_i\,.
\end{alignat*}
Since $s_i \in \bbT^0I_{<2}$ and $C$~is closed under meets, it follows that $a \in C$.

Next, let us prove that $C$~does indeed induce a subalgebra~$\frakC$ of~$\frakA$.
Hence, let $t \in \bbT^0 C$. Then there exists a tree $T \in \bbT^0\PSet(U)$ such that
\begin{align*}
  t = \bbT^0\inf(T)\,.
\end{align*}
Since $T \in \bbT^0\PSet(D)$ and $\frakD$~is meet-distributive,
it follows that
\begin{align*}
  \pi(t)
   = \pi(\bbT^0\inf(T))
   = \inf \set{ \pi(s) }{ s \in^{\bbT^0} T }\,.
\end{align*}
Since $U$~induces a subalgebra, it follows that $\pi(s) \in U$, for all $s \in^{\bbT^0} T$.
By closure of~$C$ under meets, we therefore have $\pi(t) \in C$.

We claim that the subalgebra~$\frakC$ is deterministic and that the inclusion $\frakC \to A$
is meet-distributive. The latter holds since $C \subseteq D$ and the inclusion $\frakD \to A$
is meet-distributive.
For the former, note that $\frakC$~has a semigroup-like subalgebra with domain~$U$.
Furthermore, $U$~forms a set of meet-generators of~$\frakC$.
Finally, meet-distributivity of~$\frakC$ follows from the facts that $\frakC \subseteq \frakD$
and that $\frakD$~is meet-distributive.

It remains to show $\frakC$~is the least subalgebra with the above properties.
Since $I$~is contained in every set of join-generators,
it follows that $U$~is contained in every subalgebra of~$\frakA$ that forms a set of
join-generators.
Hence, $C$~is contained in every subalgebra of~$\frakA$ that is closed under meets
and that forms a set of join-generators of~$\frakA$.
\end{proof}

\begin{Thm}\label{Thm: branch-continuous algebras have unique expansions}
Every finitary branch-continuous\/ $\bbT^?$-algebra is uniquely determined by its\/
$\bbT^{?\thin}$-reduct.
\end{Thm}
\begin{proof}
Let $\frakA = \langle A,\pi\rangle$ and $\frakA' = \langle A,\pi'\rangle$
be two branch-continuous $\bbT^?$-algebras with the same $\bbT^{?\thin}$-reduct.
Let $\frakC$~and~$\frakC'$ be the $\bbT^{?\thin}$-reduct of the
corresponding deterministic subalgebras.
Then $\frakC$~and~$\frakC'$ are deterministic subalgebras of~$\frakA|_{\bbT^{?\thin}}$
such that the universes $C$~and~$C'$ form sets of join-generators of~$A$ and the inclusions
$\frakC \to A$ and $\frakC' \to A$ are meet-distributive.
By Lemma~\ref{Lem: minimal deterministic subalgebra},
there exists a deterministic subalgebra $\frakC_0 \subseteq \frakA$ such that $C_0$~forms
a set of join-generators of~$\frakA$, the inclusion $\frakC_0 \to A$ is meet-distributive, and
such that $C_0 \subseteq C,C'$.
Note that, by Theorem~\ref{Thm: deterministic algebras have unique expansions},
the restrictions of $\pi$~and~$\pi'$ to $\bbT^? C_0$ coincide.
For $t \in \bbT^? A$, it therefore follows by join-distributivity of $\frakA$~and~$\frakA'$ that
\begin{align*}
  \pi(t)
  &= \sup {\bigset{ \pi(s) }{ s \leq t,\ s \in \bbT^? C_0 }} \\
  &= \sup {\bigset{ \pi'(s) }{ s \leq t,\ s \in \bbT^? C_0 }}
   = \pi'(t)\,.
\end{align*}
\upqed
\end{proof}
\begin{Exam}
In Lemma~\ref{Lem: non-unique definable T-expansions}, we have constructed an
$\MSO$-definable $\bbT^{?\thin}$-algebra~$\frakA$ with two different
$\MSO$-definable $\bbT$-expansions $\frakA_0$~and~$\frakA_1$.
It is straightforward to check that $\frakA$~and~$\frakA_0$ are branch-continuous.
The corresponding $\omega$-semigroup is $\frakS := \langle S,S_\omega\rangle$
where $S = \{0,1\}$, $S_\omega = \{0,1\}$ and every product is just the minimum.
The algebra~$\frakA_1$ on the other hand is not branch-continuous. While it is meet-distributive
and join-distributive, it is not generated (as in the definition of branch-continuity)
by a semigroup-like subalgebra.
\end{Exam}

\section{Conclusion}   

We have presented several approaches to the expansion problem for tree algebras.
Our results suggest that there exists a dividing line between thin trees and non-thin
ones. For classes of thin trees, we can use the existing combinatorial theory
for $\omega$-semigroups. As~a consequence we were able to obtain the results
in Section~\ref{Sect:thin trees}, which can be considered to completely solve
the expansion problem for such classes.
For non-thin trees on the other hand, our results are much more fragmentary.
We were able to solve the problem for the inclusion $\bbT^\reg \subseteq \bbT$
(at least for $\MSO$-definable algebra),
but the more important inclusions $\bbT^\thin \subseteq \bbT$ and
$\bbT^\wilke \subseteq \bbT^\reg$ had to be left open.
Theorems \ref{Thm: reduction to Txthin}~and~\ref{Thm: branch-continuous algebras have unique expansions}
might be seen as an indication the $\MSO$-definable algebras are, in a certain sense,
`controlled' by their $\bbT^\thin$-reducts. But note that we have shown in
Lemma~\ref{Lem: non-unique definable T-expansions} that, in general, $\MSO$-definable algebras
are not uniquely determined by their $\bbT^\thin$-reduct.
An important task that we have to leave open is
a~classification of all $\bbT$-extensions of a given $\MSO$-definable $\bbT^\thin$-algebra.

In general, the methods we have developed seem to work somewhat well if there exists
a unique expansion (or at least a unique expansion with a certain property, like a unique
$\MSO$-definable expansion, or a unique branch-continuous one),
but there is currently no approach to prove the existence of several expansions.

Promising next steps towards further progress seem to include
\begin{itemize}
\item trying to generalise some of our existing tools from thin trees to general ones\?; and/or
\item finding counterexamples delineating the parameter space where such generalisations
  do not exist any more.
\end{itemize}
As a problem to work on let us mention a generalisation of Simon's Factorisation Tree Theorem
to trees. But this seems to be a very hard problem. There are two technical frameworks that
might be of help here\?: \textsc{(i)}~one can try to flesh out the theory of Green's relations
for tree algebras, and \textsc{(ii)}~one can try to make use of
Tame Congruence Theory~\cite{HobbyMcKenzie88}.
While developing these two theories for tree algebras is not that difficult,
it is not at all obvious how to apply them to concrete problems, like the one mentioned above.

{\small\raggedright
\bibliographystyle{alphaurl}
\bibliography{Expansion}}

\end{document}